%% file: ms_final.tex
\documentclass[12pt,twocolumn]{aastex62} 
\usepackage[T1]{fontenc}
\usepackage{ae,aecompl}
\usepackage{amsmath}
\usepackage{rotating}
\usepackage{graphics}
\usepackage{color}
\usepackage{color}
\newcommand{\degree}{\hbox{$^\circ$}}

\newcommand{\ltsimeq}{\la}
\newcommand{\gtsimeq}{\ga}

\newcommand{\msun}{M$_{\odot}$}

\newcommand{\HI}{H{\sc i}}
\newcommand{\HII}{H{\sc ii}}

\newcommand{\ha}{H$\alpha$}
\newcommand{\inch}{$^{\prime\prime}$}

\shortauthors{McQuinn et al.}
\shorttitle{Galactic Winds in Low-Mass Galaxies}

\begin{document}
\title{Galactic Winds in Low-Mass Galaxies}

\author{Kristen.~B.~W. McQuinn}
\affiliation{Rutgers University, Department of Physics and Astronomy, 136 Frelinghuysen Road, Piscataway, NJ 08854, USA} 
\email{kristen.mcquinn@rutgers.edu}

\author{Liese van~Zee}
\affiliation{Department of Astronomy, Indiana University, 727 East 3rd Street, Bloomington, IN 47405, USA} 

\author{Evan D. Skillman}
\affiliation{Minnesota Institute for Astrophysics, School of Physics and Astronomy, 116 Church Street, S.E., University of Minnesota, Minneapolis, MN 55455, USA} 

\begin{abstract}

Stellar-feedback driven outflows are predicted to play a fundamental role in the baryon cycle of low-mass galaxies. However, observational constraints of winds in nearby dwarf galaxies are limited as outflows are transient, intrinsically low-surface brightness features, and, thus, difficult to detect. Using deep \ha\ observations, we search for winds in a sample of twelve nearby dwarfs (M$_* \sim 10^7 - 10^{9.3}$ \msun) which host on-going or recent starbursts. We detect features which we classify as winds in 6 galaxies, fountain candidates in 5 galaxies, and diffuse ISM in 1 system. Winds are found preferentially in galaxies with centrally concentrated star formation, while fountains are found in galaxies with spatially distributed star formation. We suggest that the concentration of star formation is a predictor for whether a low-mass galaxy will develop a wind. The spatial extent of {\em all} detected ionized gas is limited ($<\frac{1}{10}$ virial radius) and would still be considered the ISM by cosmological simulations. Our observations suggest that the majority of material expelled from dwarfs does not escape to the intergalactic medium but remains in the halo and may be recycled to the galaxies. Derived mass-loading factors range from 0.2$-$7 (with only a weak dependency on circular velocity or stellar mass), in tension with higher values in simulations needed to reproduce realistic low-mass galaxies and resolve discrepancies with $\Lambda$CDM. The sample is part of the panchromatic STARBurst IRegular Dwarf Survey $-$ STARBIRDS $-$ designed to characterize the starburst phenomenon in dwarf galaxies. We also report a previously uncatalogued nearby galaxy (J1118$+$7913).
\end{abstract} 


\section{Introduction}\label{sec:intro}
\subsection{On the Impact of Galactic Winds}
Stellar feedback-driven galactic winds are a major driver of the baryon cycle, impacting the evolution of galaxies. This is especially true for low-mass systems where feedback can more easily launch winds given potential wells that are much shallower than in massive galaxies. Early galaxy evolution models predicted that winds should have a dominant impact in the dwarf galaxy regime, but the detailed scope of the predictions were limited \citep[e.g.,][]{Dekel1986, MacLow1999, Ferrara2000}. Recent hydrodynamical simulations are able to implement complex stellar feedback physics, taking into account the interconnected effects of ISM heating, gas removal from star-forming disks, possible re-accretion of ejected material, and the impediment of future gas accretion due to the heating of gas surrounding a galaxy \citep[e.g.,][and references therein]{Oppenheimer2009, Christensen2016, Lu2017}. Thus, we now have sophisticated, quantitative predictions on the expected impact of stellar feedback and winds, including the role winds have in: regulating the growth of stellar mass in low-mass galaxies, reducing the baryon fraction in dwarfs compared to more massive systems \citep{Dave2009}, governing the metal content in dwarfs \citep[i.e., the mass-metallicity scaling relation;][]{Brooks2007}, and enriching the circumgalactic and intergalactic mediums (CGM and IGM respectively) with expelled metals \citep{Christensen2018}.

Stellar feedback and winds may be the key ingredients for resolving discrepancies between properties of low-mass galaxies simulated with the standard $\Lambda$ Cold Dark Matter ($\Lambda$CDM) paradigm and low-mass galaxies observed in the nearby Universe \citep[i.e., the ``small-scale crisis of $\Lambda$CDM; see][for a review and references therein]{Bullock2017}. Focusing on just one of the discrepancies, $\Lambda$CDM simulations predict a steeply rising ``cuspy'' density profile in dwarfs whereas observations show a flatter ``cored'' density profile \citep[see][for a review]{deBlok2010}. The transport of gas to larger radii in dwarf galaxies via stellar-feedback processes has been proposed as a possible solution to this long-standing core-cusp debate \citep[e.g.,][]{Navarro1996, Governato2010, Governato2012, Teyssier2013, Onorbe2015}. In this scenario, repeated expansion of gas to larger radii can alter the gravitational potential and may be a mechanism to transform the cuspy mass profiles seen in $\Lambda$CDM simulations to centrally cored profiles observed in some dwarfs. The same repeated radial expansion of the gas is also predicted to cause stellar migration of older populations \citep{Stinson2009, El-Badry2016}, and may account for the lack of observed intermediate-age stellar gradients in low-mass galaxies supporting an inside-out galaxy formation model \citep{McQuinn2017a}. 

\subsection{The Complexity of Observing Galactic Winds}\label{sec:complexity}
Detailed observations of winds in nearby, low-mass galaxies are essential to testing the stellar feedback processes, wind models, and baryon cycle predictions from hydrodynamical simulations. Yet, somewhat surprisingly, there are relatively few empirical constraints on galactic winds in nearby, low-mass galaxies. Existing observations are limited, in part, because measuring stellar feedback processes is challenging. Detecting feedback-driven outflows in low-mass galaxies is difficult due to the intrinsic low-surface brightness nature and transient nature of winds. Furthermore, galactic outflows have multiple phases (i.e., hot, warm, cold) and each phase requires a different observation. Hot winds ($T\sim10^6$ K) are fast-moving, low-density gas thought to be primarily comprised of metal-enriched supernova-ejecta and observable in the X-ray. While the hot phase may be metal-enhanced, it is not expected to entrain significant amounts of gas \citep[simulations suggest only $\sim$10\% of the mass in the hot phase;][]{Kim2017}. The material hot winds do contain has the greatest likelihood to leave the disk and enrich the CGM or IGM. Many hot-phase winds have been observed, although only in a small number of low-mass systems in the nearby Universe \citep{Heckman1995, Summers2003, Summers2004, Hartwell2004, Ott2005a, McQuinn2018}. 

Warm winds ($T\sim10^4$ K) are a mix of ionized gas swept up by stellar feedback and cooler gas shock-heated and entrained by expanding hot-phase bubbles. The warm phase has lower velocities than the hot phase and simulations suggest that most of the outflow mass is contained in this phase \citep{Hopkins2012, Onorbe2015, Christensen2016, Hu2019, Nelson2019}. Similar to the hot-phase, the properties of warm winds in low-mass galaxies in the nearby Universe are fairly limited. While there are numerous detections of extraplanar ionized gas around nearby galaxies \citep[for example, the Survey for Ionization in Neutral Gas Galaxies (SINGG)][]{Meurer2004, Meurer2006}, \ha\ emission in outer disks of spiral galaxies \citep[e.g.,][]{Barnes2011}, and a plethora of detections of winds and outflows at redshifts of $z=0.5$ or greater, the primary detailed observational constraints on warm winds in low-mass galaxies come from the characterization of expanding superbubbles of ionized hydrogen in a small sample of dwarfs \citep{Marlowe1995, Martin1998, Martin1999, Heckman2015, Chisholm2017}. 

Finally, cold, neutral gas may be entrained in outflows, although the evidence is sometimes contradictory. Cold gas outflows have been reported in NGC~5253 \citep{Kobulnicky2008} and I~Zw~18 \citep{Lelli2014b} based on \HI\ observations, but no evidence of cold outflows were found in either galaxy based on Na D absorption doublet observations \citep{Schwartz2004}, although the latter study did find evidence of outflowing cold gas from 3 other low-mass systems. Radial motions of \HI\ in the dwarf galaxy UGC~6456 suggest a small outflow of cold gas based on modelling the \HI\ rotation curves \citep[][and see \S\ref{sec:wind_detections} below]{Lelli2014b}. While there is evidence of ISM blow-out in some systems based on the \HI\ observations \citep[NGC~625, DDO~165;][respectively]{Cannon2004, Cannon2011a}, the amount of atomic gas typically estimated in outflows is small relative to the warm and hot phases \citep[e.g.,][]{Schwartz2004, Lelli2014a}. \citet{Lelli2014b} estimate that, in a sample of 11 starburst dwarfs, $\ltsimeq$10\% of the atomic gas mass may be lost based on comparing \HI\ maps with measured star formation activity from \citet{McQuinn2010a}. There is also evidence of molecular gas entrained in outflows in some systems \citep[i.e., M82;][]{Walter2002} and in more massive systems from ALMA CO observations \citep[i.e., NGC~253;][]{Bolatto2013a}. While the current detections are limited, future observations of low-mass systems with ALMA may reveal previously undetected molecular gas entrained in outflows. 

\subsection{How Much Gas is Expelled by Winds?}\label{sec:intro_mass_load}
Detecting winds is a first step, but to observationally characterize a wind and quantitatively test different theoretical models, a key parameter to measure is the rate at which gas mass is transported out of a system (the mass loss rate, $\dot{\rm M}$) relative to the star formation rate (SFR), referred to as the mass-loading factor ($\eta \equiv \dot{\rm M} $/ SFR). Theoretically, there is little consensus on how much gas is expected to be expelled from low-mass galaxies. Thus, observationally determined mass-loading factors not only uniquely characterize stellar feedback, such measurements can also help discern which implementations of baryon physics in simulations most closely match nature.

\subsubsection{Observed Mass-Loading Factors}
On the observational side in the nearby Universe, constraints include mass-loading factors of $0.5-5$ in ten high surface brightness dwarf galaxies \citep{Martin1999}, $\sim6-8$ in five low-mass galaxies with a higher upper limit of 63 on the extreme system I~Zw~18 \citep{Heckman2015}, and values of 11 and 19 for I~Zw~18 and the similarly extreme starburst SBS~1415$+$437 respectively \citep{Chisholm2017}. We discuss these results in detail in \S\ref{sec:discuss}. There are additional detections of stellar-feedback driven outflows of gas in low-mass systems that provide a mix of spatial constraints, kinematic measurements, and energy calculations of warm winds \citep[e.g.,][]{Meurer1992, Marlowe1995, Schwartz2004} but most studies do not calculate mass-loading factors. 

More generally, \citet{Dalcanton2007} calculated that the fraction of gas mass lost through galactic outflows is modest ($<$15\%) at galaxy masses of dwarfs and spirals, whereas the fraction of metals lost in outflows increases steadily at lower masses, reaching 50\% at galaxy rotation speeds of $\sim$ 30 km s$^{-1}$ and as much as 95$\pm$2\% at $15\pm5$ km s$^{1}$ \citep{McQuinn2015f}. 

\subsubsection{Theoretical Mass-Loading Factors}
On the theoretical side, early models of outflows in low-mass galaxies simply {\it assumed} they were driven by the thermal pressure of gas heated by supernova \citep[e.g.,][]{Larson1974, Dekel1986}. When models added supernovae energy, feedback-driven mass loss was predicted in galaxies with $M_{gas} <10^{7}$ \msun, whereas significant ($>$70\%) metal loss was expected in galaxies up to $M_{gas} \sim10^9$ \msun\ \citep{MacLow1999}. Separately, outflow rates of $10^{-5}-10^{-2}$ \msun\ yr$^{-1}$ were predicted in low-mass galaxies, peaking in systems with $M_{gas} \approx 10^8$ \msun\ \citep{Ferrara2000}. In these early models, star formation was assumed to be {\it both co-spatial and co-temporal}, which increases the impact of stellar feedback but is at odds with the larger spatial distribution and longer timescales of observed starbursts in dwarf galaxies \citep{McQuinn2012a}. 

In contrast, detailed modeling of starburst dwarf galaxies predicted that while galaxies may be efficient at transporting energy, they are inefficient at transporting mass out of a system \citep{Strickland2000a}. Later models showed that when radiative losses are included, unbound galactic winds are less likely and star formation feedback may not be energetic enough to heat the gas above the escape temperature of dark matter halos \citep{Marcolini2005}. 

In more recent hydrodynamical cosmological simulations, winds are often mass-loaded by prescription due to resolution limitations \citep[e.g.,][]{Vogelsberger2013, Ford2014}. Commonly used prescriptions are based on either a energy-driven wind \citep[i.e., an energy-conserving wind with $\eta\ \propto$ V$_{\rm circ}^{-2}$; where V$_{\rm circ}$ is the halo circular velocity;][]{Chevalier1985} or a momentum-driven wind \citep[i.e., a momentum-conserving wind with $\eta\ \propto$ V$_{\rm circ}^{-1}$;][]{Murray2005}. Energy-driven winds are expected to have a larger impact in low-mass galaxies whereas momentum-driven winds may play a larger role in higher mass galaxies \citep[i.e.,][]{Murray2011, Hopkins2012}. Additional factors such as cosmic ray streaming are also being explored in dwarf galaxy simulations which show mass-loading factors of 5 and as much as 60\% of the ISM may be expelled through winds \citep{Uhlig2012}.

In zoom-in hydrodynamical simulations that increase the resolution on individual galaxies and include stellar feedback processes, mass-loading factors are calculated directly from the simulated star formation activity and the resulting winds \citep[e.g.,][]{Muratov2015, Christensen2016, Hu2019, Nelson2019}. The simulations necessarily still rely on formulae to cope with the sub-grid physics for star formation and supernovae feedback, which can vary between simulations. The mass-loading factors reported vary by more than two orders of magnitude for similar halo masses between the different simulations. We discuss these in detail in \S\ref{sec:discuss}. 

\subsection{New Observational Constraints on Galactic Winds}
Given the importance of winds, the range in predictions of mass-loading factors, and the limited observational constraints, additional study is warranted. We present a comprehensive analysis on the impact of stellar feedback-driven winds in twelve nearby, low-mass, actively star-forming galaxies with masses ranging from M$_* \sim 10^7 - 10^{9.3}$ \msun. We have obtained deep \ha\ imaging that is $\sim2$ orders of magnitude more sensitive in surface brightness and include galaxies an order of magnitude less massive than in a number of previous studies constraining wind properties in dwarfs \citep[e.g.,][]{Marlowe1995, Martin1998}. We use these observations to identify warm winds and smaller-scale fountain flows, map the full extent of the winds, connect the winds to the star formation properties, and calculate the mass-loading factors. 

The sample is built from low-mass galaxies with on-going or recent bursts of star formation and, thus, are excellent candidates for hosting galactic winds. Eleven galaxies are from the STARBurst IRRegular Dwarf Survey \citep[STARBIRDS][]{McQuinn2015b}; an additional galaxy (NGC~4190) was added to the sample based on starburst characteristics measured by \citet{McQuinn2015d}. 

One of the strengths of this study is the rich ancillary datasets available on the sample. All galaxies have previously derived star formation histories from HST observations of their resolved stellar populations \citep{McQuinn2009, McQuinn2010a, McQuinn2015b} and measurements of the spatial distribution of recent star formation \citep{McQuinn2012a}. All galaxies have existing \HI\ observations to compare with the ionized gas distributions, providing critical context for whether or not the ionized gas is reaching outside both the stellar and gas disk and constitutes an outflow. Preliminary ionized gas kinematics are available from an on-going IFU spectroscopic observing campaign (L. van Zee et al.\ in preparation). Three galaxies have $Chandra$ X-ray imaging of the hot phase winds \citep{McQuinn2018}, facilitating a direct comparison of the warm and hot phase ISM. 

The paper is organized as follows. In \S\ref{sec:sample}, we describe STARBIRDS, the galaxies selected for this study, and the data sets used which include ground-based \ha\ and broad band imaging from the Kitt Peak National Observation (KPNO) 4m, the University of Arizona 2.3m Bok telescopes, and the WIYN 0.9m telescope, and archival \HI\ observations from the Very Large Array (VLA). The reduced \ha\ and broadband datasets are available from the STARBIRDS multi-wavelength data archive hosted by MAST, that already includes $HST$ optical imaging, {\it Spitzer Space Telescope} infrared imaging, and GALEX Space Telescope ultraviolet imaging \citep{McQuinn2015b, McQuinn2015c}. In \S\ref{sec:final_images}, we measure the structural parameters and geometry of the stellar components in the galaxies and present the final R-band and \ha\ continuum subtracted images of the sample. 

In \S\ref{sec:outflows}, we present a comparison of the ionized and neutral gas components. We discuss our overall findings on the presence or absence of winds and fountains. In \S\ref{sec:spatial}, we quantify the spatial extent as a function of estimated virial radius and present a comparison of the warm-phase and hot-phase gas in three galaxies (NGC~4214, NGC~1569, NGC~4449) for which we have both \ha\ and X-ray imaging. In \S\ref{sec:calc_mass_load}, we calculate the mass-loading factors, in \S\ref{sec:discuss}, we discuss our overall findings with a comparison of our results to cosmological simulations, and in \S\ref{sec:conclusions}, we summarize our conclusions. 

In Appendix~\ref{sec:atlas}, we provide a galaxy atlas with maps of the stellar component, ionized gas, neutral gas, and \HI\ velocity fields for the full sample with an accompanying detailed discussion of each galaxy. In Appendix~\ref{sec:j1118_7913}, we report the serendipitous discovery of a galaxy in the nearby universe, J1118$+$7913, including imaging and an optical spectrum.

\input{tab1}

\section{The Galaxy Sample and Observations}\label{sec:sample}
\subsection{STARBIRDS Galaxies}
The galaxies selected for study are part of STARBIRDS, a panchromatic study aimed at characterizing the lifecycle and impact of starbursts on dwarf galaxies in the nearby universe. The sample spans a range of absolute magnitude, inclination angle, and metallicity. All galaxies are classified as a ``starburst'' with current or recent elevated levels of star formation activity. However, it is important to highlight that the delineation as a starburst and the starburst properties are, in part, determined by comparing the recent activity to an historical average rather than the flux output or star formation rate meeting an absolute threshold. Thus, the sample includes dwarf galaxies with a range in star formation activity that may be more typical of the low-mass, star-forming galaxy population at the present, rather than being comprised of only extreme starburst systems that are atypical of low-mass systems.

Table~\ref{tab:galaxies} lists the galaxy sample, basic properties, and structural parameters. From the larger STARBIRDS sample of twenty galaxies, we selected the eleven northern hemisphere starburst dwarf galaxies visible in the spring. In addition to the eleven STARBIRDS galaxies, the observational footprints also include seven nearby galaxies. One of these systems, NGC~4190, has been classified as a starburst based on the SFH in \citet{McQuinn2015d}. Thus, we have added this galaxy to the sample for searching for galactic outflows. Of the remaining six galaxies, five are well-known, nearby systems. For completeness, we include structural measurements of these additional galaxies from our observations (see Table~\ref{tab:galaxies} and \S\ref{sec:structure}), but do not include them in further analysis on outflows. The final system is a newly catalogued galaxy, J1118$+$7913, found in the field of view of the STARBIRDS target UGC~6456. We provide structural parameters in Table~\ref{tab:galaxies}, and \ha, B, R imaging and an optical spectrum in Appendix~\ref{sec:j1118_7913}.

Table~\ref{tab:star_formation} summarizes the star formation properties in the sample. Previous results from STARBIRDS have shown that the starburst events last 100's of Myr in the galaxies, comparable to the dynamical timescales \citep{McQuinn2009, McQuinn2010a, McQuinn2010b}. Two of the galaxies (NGC~4163, UGC~9128) are post-starburst with SFRs declining over the last 100 Myr. 

\input{tab2}

The spatial concentration of the starburst activity has important implications for the development of an outflow. Simulations show cavities or bubbles created in the ISM from stellar feedback expand and merge to form superbubbles, a precursor to an outflow \citep{Kim2017}. If star formation is highly concentrated, this increases the probability of both a superbubble forming and an outflow developing. \citet{McQuinn2012a} measured the concentration of recent star formation based on a number of variables, including the concentration ratio of young stars (traced by blue helium burning (BHeB) stars) and old stars (traced by red giant branch (RGB) stars). The spatial distribution of recent star formation, listed in Table~\ref{tab:star_formation}, ranges from being highly concentred (e.g., UGC~6456, NGC~4163) to widely distributed (e.g., NGC~4068). Using the same analysis, we report the spatial concentration of the additional starburst NGC~4190 from the data in \citet{McQuinn2015d}. Two galaxies (Ho~II, NGC~4214) do not have sufficient coverage of their stellar disk in the $HST$ images to use this approach. For these systems, we use \ha\ emission as a surrogate tracer of young stars and the R-band emission for the older stars from our new imaging. Specifically, we compare the radius encompassing 90\% of the total \ha\ flux to the radius of the stellar disk. We checked some of the individual galaxies with data suitable for both methods and found that this approach returned concentration values comparable to the BHeB / RGB ratios. We compare the detected warm outflows in \ha\ emission to these spatial extents in our analysis.

\input{tab3}

\subsection{\ha\ and Broad-Band Ground-Based Observations}
Our primary new observations are deep \ha-on and -off narrow-band imaging with complementary broad-band B and R imaging. \ha\ emission is arguably one of the most sensitive tracers of the warm ionized gas in galaxies. The broadband images are useful for measuring the spatial extent of the underlying stellar populations.

The observations were obtained on three telescopes. In the spring of 2013, we observed all but one galaxy in the sample (NGC~1569; see below) using two telescopes with wide fields of view, the KPNO Mayall 4m telescope with the Mosaic 1.1 instrument and the Steward Observatory Bok 90\inch\ telescope with the Prime Imager instrument. The  Mosaic 1.1 imager on the KPNO 4m telescope covers a field of view of 35\arcmin $\times$ 35\arcmin\ with a pixel scale of 0.26\arcsec\ pixel$^{-1}$, while the Bok 90\inch\ covers 1-square degree with a pixel scale of 0.455\arcsec\ pixel$^{-1}$. These large fields of view are more than sufficient to ensure any outflows or extended extraplanar features are fully imaged. 

Both the KPNO and the Bok observing runs occurred during dark time to avoid scattered lunar light interfering with possible detections of low-surface brightness outflow features. We used the same narrow-band filters on both telescopes, the KPNO \ha\ filter (KP1009) centered on the \ha\ wavelength (central $\lambda = $6574.74\AA, FWHM $=$80.62\AA) and the narrow band KPNO filter centered $\sim$80 \AA\ from the \ha\ wavelength (KP1011; central $\lambda = $6654.19\AA, FWHM $=$80.62\AA) as an \ha-off filter. The narrow-band \ha-off filter allows continuum subtraction from the \ha-on image from a matched, narrow-band filter shifted slightly in wavelength.

The details of the observing runs are listed in Table~\ref{tab:observations}. The observing strategy included obtaining six 20 min exposures in each narrow band filter over a 5-point dither pattern with the central pointing repeated, and five 2 min broad B-band and R-band filter exposures over the same dither pattern. Our chosen dither pattern allowed for the removal of bad pixels and cosmic rays in the final mosaic images. Airmass ranged from 1.00$-$1.49. Observations were alternated between the \ha-on and -off filters to minimize changes in seeing conditions and changes in the point spread function (PSF). Total integration goals were 2 hours on and 2 hours off in the narrow band \ha\ filters and 10 min in the B and R bands. In the narrow band filters, we achieved the 2 hour integration times in 6 of the 10 targets including one target with a slightly longer integration time of 2 hours 20 min. The remaining 4 targets have slightly shorter integration times of 1 hour 40 min due to weather conditions at the telescope. Similarly for the B and R broad band filters, 4 of the targets have slightly shorter integration times of 8 min. Seeing conditions varied between $1.3-1.9$\arcsec.

In October of 2016, we obtained similar optical imaging observations of the iconic starburst galaxy NGC 1569 using the WIYN 0.9m telescope\footnote{The 0.9m telescope is operated by WIYN Inc.\ on behalf of a Consortium of partner Universities and Organizations (see www.noao.edu/0.9m for a list of the current partners). WIYN is a joint partnership of the University of Wisconsin at Madison, Indiana University, Yale University, and the National Optical Astronomical Observatory.} with the Half-Degree Imager (HDI) detector. While not equivalent in depth to the H$\alpha$ images described above, these images illustrate the dramatic ionized gas outflows that may occur with the right conditions. The WIYN 0.9m observations include two 20 min exposures in the H$\alpha$ filter centered at 6580 \AA, one 20 min exposure in the narrowband continuum filter centered at 6660 \AA, four 15 min exposures in B and four 5 min exposures in R.  

\subsection{Data Reduction of Optical Images}
The \ha\ narrow-band and B, R broad-band data from the KPNO and Bok telescopes were uniformly processed using the data reduction tools available in the \textsc{iraf} package \textsc{mscred}, with some additional steps unique to each telescope. For example, the KPNO data includes pupil ghost artifacts while the Bok data lacked bad pixel maps and an astrometric solution. Here, we describe the basic reductions steps applied to both data sets and include a description of any additional individual processing required.
 
The images were overscan corrected and trimmed. Crosstalk corrections were applied to the KPNO data, but none were used for the Bok data. Bad pixels, identified in the KPNO Mosaic 1.1 standard masks and/or flagged by inspection from careful examination of the data, were interpolated over using \textsc{fixpix}. For both data sets, the individual amps were merged on each CCD.

Bias images were combined from each observing run and subtracted from the data and flat field frames. Dark exposures of 20 min each were taken during both observing runs. The dark current for the KPNO data was negligible; no correction was applied.  There was structure, however, in the dark exposures for the Bok telescope. We applied a correction using an appropriately scaled dark frame to all Bok data and flat field frames.

For the KPNO data, the dome flats were combined to create an initial flat field image. This flat field had a ghost pupil image in the center of both narrow band images and the B-band images. \textsc{mscred} includes the task \textsc{rmpupil} that is designed to remove the pupil ghost from a CCD with 8 amps. However, the upgraded Mosaic 1.1 imager has 16 amps. Thus, to create a smooth behavior of the gains in the central CCDs, we used the \textsc{IRAF} task \textsc{blkavg} to average the values in the CCDs and divided the dome flats by this smoothed image. Then, we used the task \textsc{rmpupil} to remove the pupil ghost and subsequently restored the original amplifier gains by multiplying by the smoothed image. The result is a final flat image free of the pupil ghost artifact. We applied the corrected dome flat to all target observations to create flattened images. While flattened, the narrow-band and B-band images still have the pupil ghost artifact that requires further processing. 

For the Bok data, a gradient was present on the dome flats which is likely a result of improper illumination of the flat field screen. Thus, we used only the twilight flats to perform an initial flat-field of the data. Twilight flats were combined into one flat field image for each filter and applied to the data. 

For both data sets, we created a super-sky flat by masking objects in all of the long exposure images using the task \textsc{objmasks} and combining them using the task \textsc{sflatcombine}. For the Bok data, the super-sky flat was used to adjust for small jumps between different CCD amps. For the KPNO data, the narrrow-band and B-band super-sky flats are used to create a separate ``image'' of the pupil ghost using \textsc{mscpupil}. This pupil image was then scaled and removed from all standard star and target observations using the pupil mask template and the task \textsc{rmpupil}. This last step was done iteratively for a number of data frames to produce pupil ghost-free images. The corrected frames were re-masked and re-combined to make a pupil ghost-free super-sky flat. Finally, the super-sky flats were normalized by the mode of the images and applied to the corrected data frames for the narrow-band and B-band images and to all other (non-pupil ghost impacted) data frames.

Calibration solutions were determined through matching aperture photometry measurements of photometric and spectrophotometric standard stars with the photometric catalogs \citep[e.g.,][]{Landolt1992, Oke1990}, and solving for zero-point and color terms. The calibration solutions were incorporated into the headers of all galaxy images, which were also scaled to the data obtained on the most photometric night at the lowest airmass. We verified our calibrations by comparing the total \ha\ fluxes within the D$_{25}$ diameter of the galaxies to values in the literature and found they were an excellent match to previously reported fluxes.

The data frames were cosmic ray cleaned using L.A. Cosmic \citep{vanDokkum2001}. Meteor trails were edited by hand using \textsc{imedit}. We registered the images to the same coordinates, merged the CCDs, and projected them onto a common frame using the tasks \textsc{msctpeak, mscsetwcs, msccmatch,} and \textsc{mscimage}. 

Sky subtractions were performed on the images using the \textsc{galphot} sky fitting program. We identified regions of each image that were relatively devoid of bright stars across each image using \textsc{marksky} and subtracted the mode of the regions from the images. 

Additional image artifacts of ghost bleed trails (i.e., echoes of bleed trails) and ``dots'' above bright stars caused by slight PSF mismatches or saturation were identified in the KPNO data frames. While these do not impact our science goals, we removed many of them from the images using the task \textsc{imedit} on the individual data frames, particularly in the areas around the galaxies, to improve the cosmetics of our mosaics.

Finally, the data frames were trimmed, aligned, and median stacked to create the final mosaics in each filter. Continuum subtracted \ha\ images were made by subtracting the narrow-band off-image from the on-image\footnote{We also created a second set of continuum subtracted \ha\ images of the fields at the rest-frame of the narrow-band off-images by subtracting the on-images from the off-images. These are available from the STARBIRDS data archive via MAST but not considered in the analysis here.}. As some of the stars were poorly subtracted due to changing seeing conditions, we also masked the residuals around the galaxies to improve the overall cosmetics of the final images.

\input{tab4}

The WIYN 0.9m data on NGC~1569 were processed in the standard manner similar to the KPNO and Bok data and the individual exposures in each filter were averaged to create the final images.  All images were spatially aligned and the final H$\alpha$ image was created by subtracting an appropriately scaled narrowband continuum 
image from the 6580 \AA~filter image.

\subsection{VLA Neutral Hydrogen Observations}\label{sec:hi}
Neutral hydrogen observations with the Very Large Array (VLA) were retrieved from the NRAO data 
archive for all of the galaxies in the sample. As indicated in Table 4, almost all of these 
observations have been published previously.  The observations included both flux and phase 
calibrators in addition to the target sources.  For the majority of the observations, the data
were processed using standard tasks in AIPS \citep[see, e.g.,][]{Richards2016}.  Phase and gain 
calibration solutions were derived based on 'Channel 0' data (the inner 75\% of the bandpass) and 
copied to the line data.  Bandpass calibrations for the line data were determined 
from observations of the flux calibrator.  Continuum emission was removed with the AIPS task 
UVLIN. Natural weight data cubes were created using the task IMAGR with a robust factor of 5.  
For datasets that included B, C, and D configuration data, an additional data cube was created that
had similar spatial resolution as C-configuration observations by limiting the $uv$ range to 0 to 20 kilolambda 
and applying a $uv$ taper at 15 kilolambda.  The data cubes were analyzed in the GIPSY package
\citep{vanderHulst1992} following standard practice including a primary beam correction for
the moment 0 maps.  While the majority of the archival data were reprocessed as described above, 
finished data products were also downloaded from the LITTLE THINGS \citep{Hunter2012} and VLA-ANGST
\citep{Ott2012} projects for UGC 9128, NGC 4163, and NGC 4190.

Estimates of the maximum circular rotation velocity (V$_{\rm circ}$) were derived from the moment 1 maps.  While
the majority of these galaxies have clear evidence of non-circular rotation or tidal disturbances,
rotation curves were derived using a tilted-ring model (\textsc{rotcur}) in GIPSY.  Given the nature of the
gas kinematics in these systems, these model fits should only be considered indicative, at best. Indeed,
for the majority of the galaxies, it was not possible to derive a robust fit based on the moment 1
map alone.  Often, the physical center and systemic velocity were estimated based on the optical
parameters and centroid of the integrated neutral hydrogen profile, respectively. Since the optical
and gaseous components were not always congruent in these galaxies, inclination angles and position angles 
were estimated based on the gas distribution and kinematics. In most cases, the maximum circular rotation velocity is recorded as the rotation velocity at the last measured point of the neutral hydrogen distribution (R$_{\rm max}$). However, in the extreme cases where the extended gas is clearly not in rotation, the maximum circular rotation velocity was measured at the edge of the region where the gas appeared to be still in regular circular rotation, well inside the tidal tails. For three galaxies which have particularly complicated kinematics, we adopt V$_{\rm circ}$ derived from detailed dynamical modelling of \citet[][NGC~1569, NGC~4214]{Lelli2014b} and \citet[][NGC~4449]{Hunter1998}. Circular velocities and the radius at which they were measured are listed in Table~\ref{tab:mass_load}.

\begin{figure*}
\includegraphics[width=0.95\textwidth]{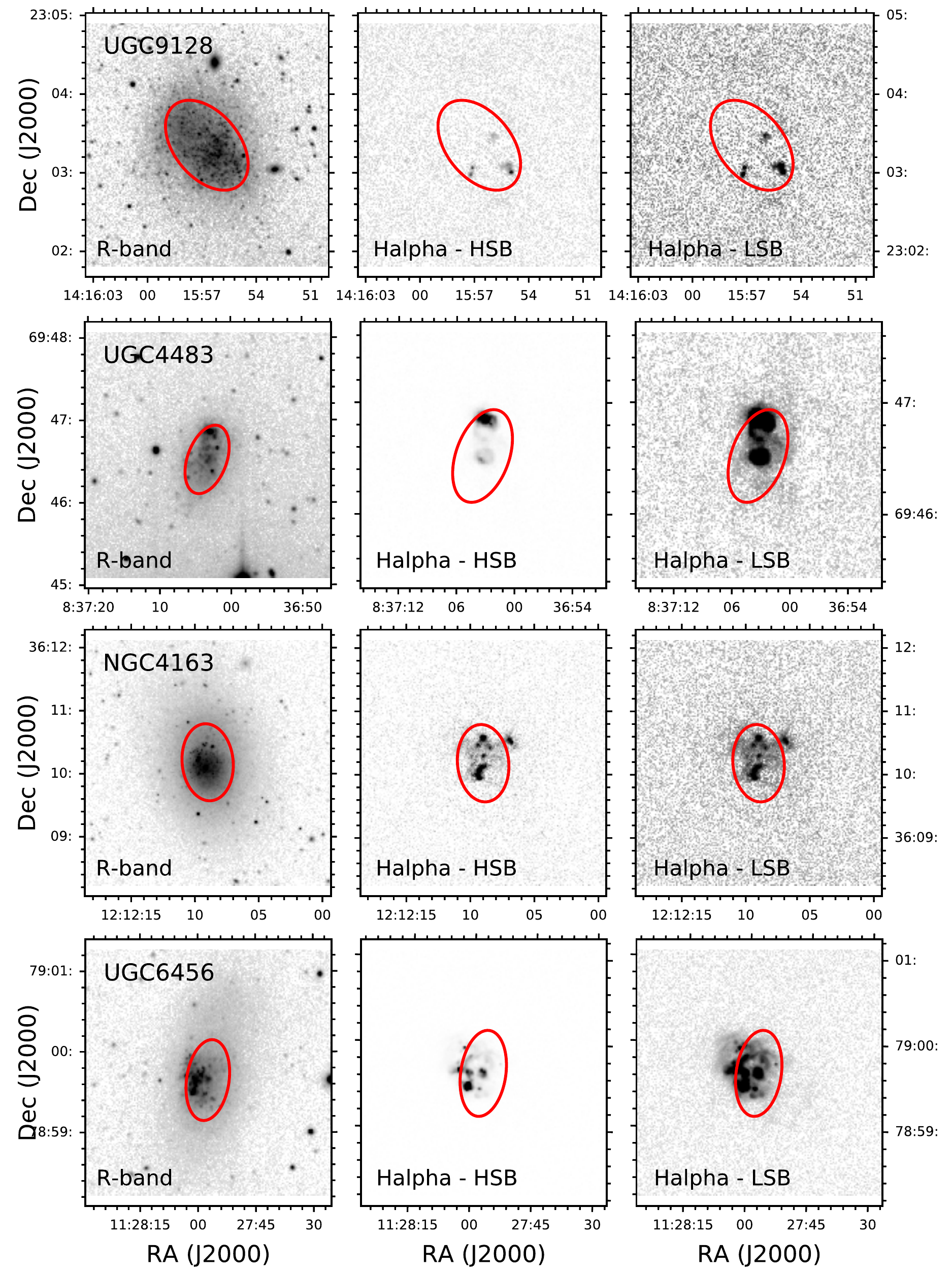}
\caption{Left: R-band image of UGC~9128, UGC~4483, NGC~4163, and UGC~6456 with North up and East left. The red ellipse outlines the stellar disk of the galaxy (2.2$\times$ scale length) based on the geometry in Table~\ref{tab:galaxies}. Middle and Right: Continuum subtracted \ha\ maps shown with a stretch that highlights the high-surface brightness and low-surface brightness \ha\ emission, respectively. All images are cropped to $\sim2\times$ the optical diameter defined by the B-band 25 mag arcsec$^{-2}$ isophote.}
\label{fig:full_fova}
\end{figure*}

\begin{figure*}
\includegraphics[width=0.95\textwidth]{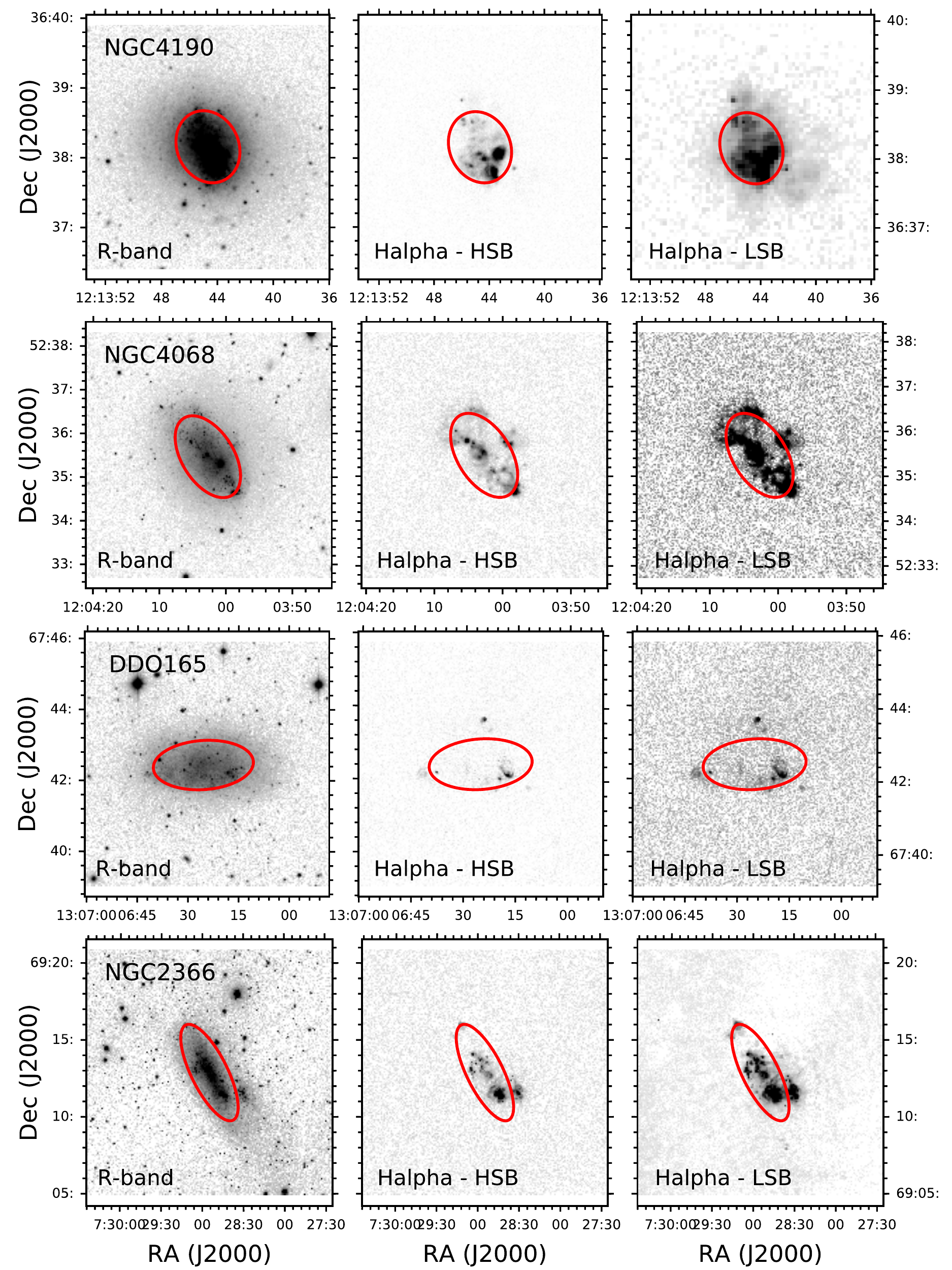}
\caption{R-band and \ha\ continuum subtracted images of NGC~4190, NGC~4068, DDO~165, and NGC~2366. See Figure~\ref{fig:full_fova} captions for details.}
\label{fig:full_fovb}
\end{figure*}

\begin{figure*}
\includegraphics[width=0.95\textwidth]{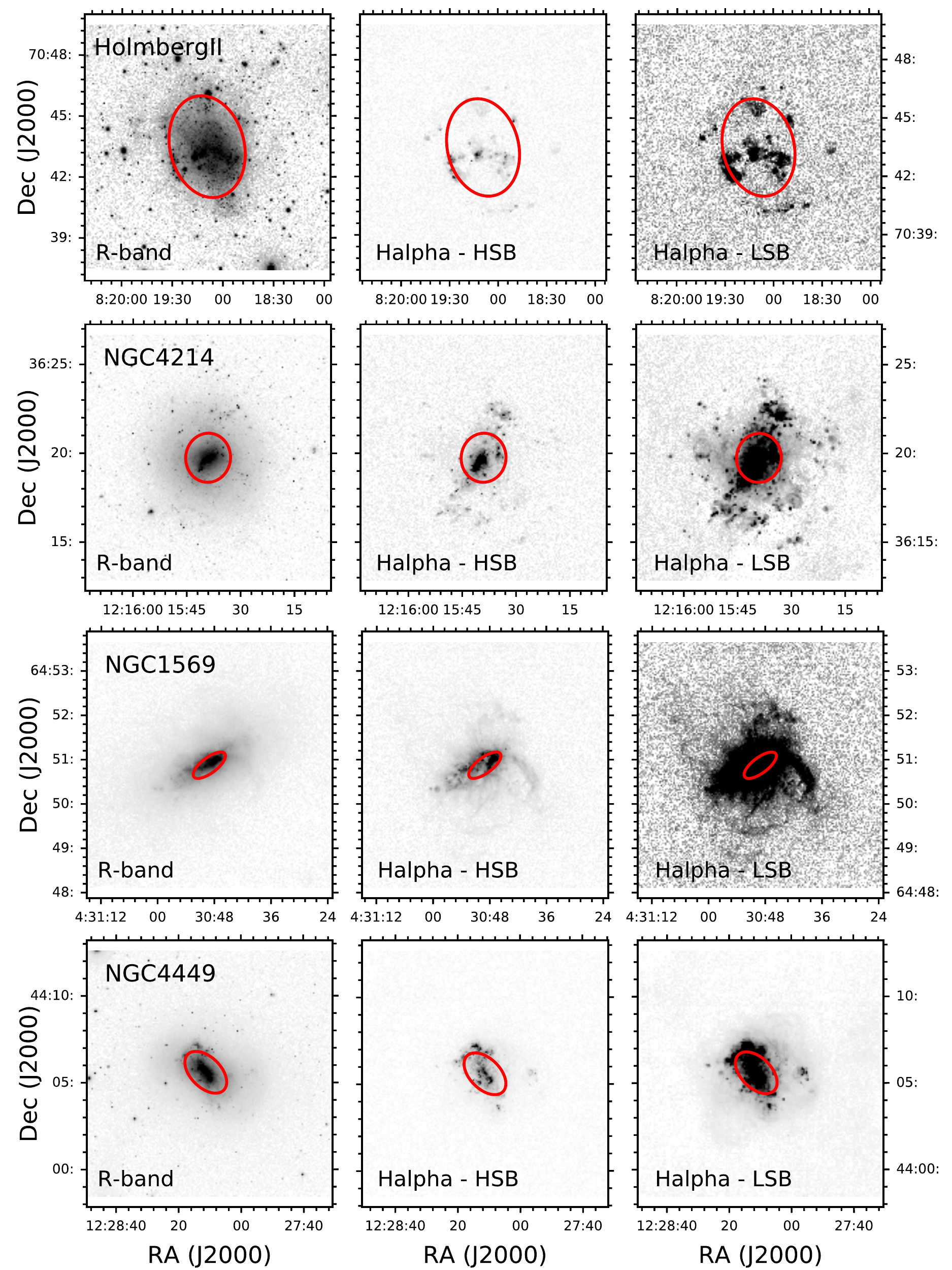}
\caption{R-band and \ha\ continuum subtracted images of Holmberg~II, NGC~4214, NGC~1569, and NGC~4449. See Figure~\ref{fig:full_fova} captions for details.}
\label{fig:full_fovc}
\end{figure*}

\section{Galaxy Geometry and Final Images}\label{sec:final_images}
\subsection{Structural Parameters}\label{sec:structure}
The geometry of the stellar components were determined by fitting elliptical isophotes to the data. Table~\ref{tab:galaxies} lists the scale length (h), ellipticity ($1 - \frac{b}{a}$), position angle, and approximate diameter of the B-band 25 mag arcsec$^{-2}$ isophote (D$_{25}$), based on fitting ellipses to our R-band images of the galaxies. These parameters are used to define the inclination angle (where cos $i = \frac{b}{a}$), the extent of the stellar disks (2.2$\times$ scale length), and provide a reference for the extent and orientation of any detected outflow of ionized gas. 

Briefly, we used the \textsc{iraf} task \textsc{ellipse} to fit isophotes to the R-band images, after interpolating over significant foreground stars and background galaxies with the task \textsc{imedit}. However, even after eliminating contamination, concentric isophotes can be skewed as the galaxies have irregular morphologies and host regions of on-going star formation. Thus, to circumvent this, we determined the ellipticity and position angle of the ellipse from the outer isophote of the galaxies, where the surface brightness is generally the least impacted by recent star formation, and used that shape to define concentric ellipses across the galaxies. 

Table~\ref{tab:galaxies} also lists the structural parameters of six additional galaxies. To take advantage of the wide fields of view of the KNPO 4m and Bok 2.3m telescopes, we designed the pointings of our observations such that the footprints included not only our target galaxies, but a number of other dwarf galaxies nearby on the sky. For completeness, we include measurements for them in Table~\ref{tab:galaxies}. Analysis of NGC~4102 will be presented in a separate study (L.~van Zee et al., in preparation). We also serendipitously detected a nearby galaxy in the field with UGC~6456, namely J1118$+$7913; we present an image, a spectrum, and a brief description of the system in Appendix~\ref{sec:j1118_7913}. 

\subsection{H$\alpha$ Fluxes and Sensitivity Limits of the Images}\label{sec:sensitivity}
Table~\ref{tab:observations} lists the total measured \ha\ flux and the \ha\ 3$\sigma$ surface brightness limit reached for each galaxy. The total \ha\ fluxes include emission detected both inside and outside the disks of the galaxies. We did not correct for nitrogen as the contribution in low metallicity systems is expected to be only a small percentage of the total flux measured in the narrow band \ha\ filter. We determined the surface brightness limits by smoothing the images to a 9\arcsec\ resolution and measuring the sky noise in the environs of each galaxy. Smoothing the images reduces the noise levels and allows lower surface brightness features around the galaxies to be identified. We experimented with different smoothing kernels and kernel sizes. Trial and error showed that a 9\arcsec\ Gaussian kernel yielded the best results, minimizing noise while retaining the ability to identify structures in the galaxies. At the typical distance of the galaxies, this corresponds to a spatial scale of 0.15 kpc. 

The \ha\ surface brightness limits are used to explore the extent of the ionized gas at high confidence levels on a case-by-case basis and determine the mass-loading factors of the winds (see Figures~\ref{fig:sblimits}$-$\ref{fig:SB_histograms}, \S\ref{sec:outflows}$-$\ref{sec:calc_mass_load}, and the galaxy atlas in Figures~\ref{fig:ngc4190}-\ref{fig:ngc4449} of Appendix~\ref{sec:atlas}). The surface brightness levels were calculated using the total counts per $\sq \arcsec$ based on $3\times$ sky sigma values of the smoothed images, our calibration zeropoints and extinction correction coefficients. For the deep observations from the KPNO and Bok telescopes, our surface brightness limits range from $1.4\times10^{-18} $ to $6.7\times10^{-18}$  erg s$^{-1}$ cm$^{-2}$ arcsec$^{-2}$ and can be compared with the expected 1$\sigma$ \ha\ surface brightness level of $5\times10^{-19}$ erg s$^{-1}$ cm$^{-2}$ arcsec$^{-2}$ from diffuse gas ionized by the metagalactic UV radiation field \citep{Bland-Hawthorn1997}. The shorter integration observations of NGC~1569 from WIYN reach a surface brightness limit of $1.8\times10^{-17}$  erg s$^{-1}$ cm$^{-2}$ arcsec$^{-2}$. 

\subsection{Final R-band and \ha\ Images of the Sample}
The left panels in Figures~\ref{fig:full_fova}$-$\ref{fig:full_fovc} present the R-band images of the observations centered on the galaxies and cover a field of view of $\sim2\times$ D$_{25}$. The red ellipses encompass the majority of the stellar disks and are based on the ellipticity, position angle, and 2.2 $\times$ scale length from Table~\ref{tab:galaxies}. The middle and right panels show the \ha\ continuum subtracted images in the same fields of view. The images in the middle panels have a square root of the flux stretch to highlight the higher surface brightness regions and structure in the \ha\ emission. The peaks in \ha\ emission are localized around star forming regions identifiable and cross-referenced in the $HST$ optical imaging presented in \citet{McQuinn2010a}. The images in the right panels are shown with a flux-squared stretch to highlight the lower surface brightness regions and diffuse ionized gas. 

\subsection{Bubbles, Shells, and Diffuse Ionized Gas}
From Figures~\ref{fig:full_fova}$-$\ref{fig:full_fovc}, it is immediately apparent there is extended, low surface brightness \ha\ emission in and around nearly all of the galaxies that would not be detected in shallower observations. In addition, the morphology of the extended \ha\ emission is varied with significant substructure. There are bubble-like features seen in nearly all of the galaxies, particularly visible in the middle panels. The \ha\ emitting bubbles and shells are often limb-brightened, indicating that much of the optically emitting gas resides on the surface of largely hollow structures \citep[cf.][]{Veilleux2005}. Filaments and arcs of ionized gas are seen in many galaxies, most notably NGC~4214, NGC~1569, and NGC~4449, that are consistent with the morphology of expanding and/or ruptured superbubbles. Complex structures such as these have long been noted in previous \ha\ studies \citep[e.g.,][]{Hunter1990, Meurer1992, Marlowe1995, Martin1998, Hunter1997, vanZee2000, GilDePaz2003, Kennicutt2003, James2004, Garrido2004, Hunter2004, Meurer2006, Kennicutt2008, Lee2016}. 

Diffuse \ha\ emission is detected throughout the interiors of the galaxies. This diffuse emission does not appear to be associated with a particular star forming region but can more generally be considered the warm-phase ISM. Despite the rich archive of \ha\ observations, much of this diffuse ionized gas constitutes new detections for these galaxies due to the high sensitivity of these observations. 

The complex bubbles and structures seen in the images show the localized impact of stellar feedback and is well-described theoretically. Briefly, the formation of a galactic outflow is predicted to be the cumulative effect of stellar feedback from star-forming regions \citep[see e.g.,][and references therein]{Veilleux2005}. Stellar winds and radiation pressure from higher mass stars can create low-density cavities within a nascent cloud over Myr timescales, while direct heating of the ISM from these same stars ionizes the surrounding neutral ISM. Multiple supernovae over longer timescales ($t>10$ Myr) inject energy into these cavities forming expanding bubbles of hot gas that shock-heat and entrain surrounding gas as they push outward. In small star-forming regions, the hot gas in a bubble can quickly cool, adding turbulence to the ISM but without a larger scale impact on the gas \citep{Kim2017}. In larger and/or adjacent star-forming regions, bubbles can merge into superbubble structures and generate an outflow if they break through the edge of the gaseous disk in a galaxy. 

\begin{figure*}
\includegraphics[width=0.98\textwidth]{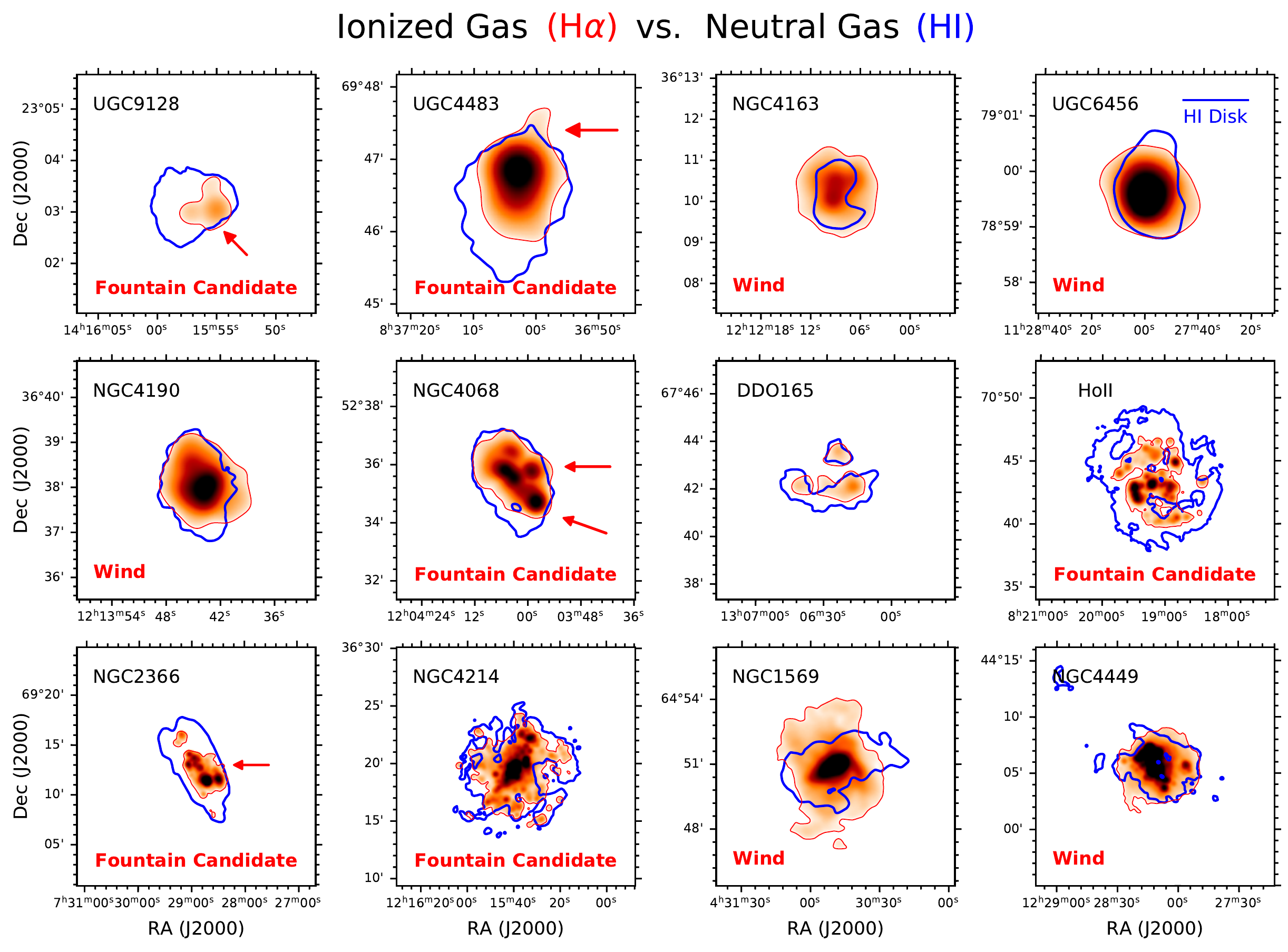}
\caption{\ha\ maps smoothed with a 9\arcsec\ Gaussian kernel, outlined in red at the 3$\sigma$ detection limit. The extent of the \HI\ disk is shown with a solid blue line. Five galaxies have clearly detected winds with ionized gas extending past the neutral hydrogen over significant sold angles. Six galaxies have fountain flow candidates with ionized gas pushing past the \HI\ disk in small regions, many of which are marked with arrows. One galaxy does not show clear evidence of a wind or fountain which may be due to confusion with cirrus contamination in the surrounding region.}
\label{fig:sblimits}
\end{figure*}

\section{Winds and Fountains}\label{sec:outflows}
\subsection{Classification Criteria and Detections}
We classify the ionized gas detected in the sample into the following 3 general categories: (i) galactic winds, (ii) local fountain flow candidates, and, in cases where the identification is ambiguous, (iii) diffuse, ionized ISM. 

The classification as a wind or fountain is based on the simple criterion that the ionized gas traced by \ha\ emission must reach farther than the ISM traced by the atomic hydrogen emission. The neutral hydrogen in dwarf irregular galaxies is generally more extended than the stellar component and has important implications for whether or not the ionized gas is breaking out of a galaxy. Typically, the size of an \HI\ disk is defined by where the \HI\ column density reaches $1.25\times10^{20}$ cm$^{-2}$, corresponding to a surface density of $\sim1$ \msun\ pc$^{-2}$. However, this column density is low enough (and most often extremely patchy, seen even in the low and medium resolution of the VLA configurations) that the drag force of the neutral gas does not significantly impede the motion of any outflowing ionized gas. Thus, we chose a subjective and slightly higher \HI\ column density of $5\times10^{20}$ cm$^{-2}$ to define the region of denser ISM where ``break-through'' of the ionized gas occurs. 

The distinction between a wind and a fountain flow is made based on whether the \ha\ emission outside the \HI\ disk is widespread or confined to small, localized regions. Although a subjective criterion, as demonstrated below, the \ha\ emission is either clearly extended over significant solid angles around a galaxy and lacks a sharp boundary consistent with a wind, or the \ha\ emission extends past the \HI\ in one or two localized regions. We conservatively label the systems with small-scale outflows as fountain ``candidates'' because whether or not the ionized gas is breaking through the \HI\ disk depends more sensitively on our chosen \HI\ column density threshold than the winds. Finally, in a case without any unambiguous extended \ha\ emission, we  classify the system as having diffuse, ionized ISM, but with significant caveats as discussed below. 

Note that while the orientation of the galaxies can impact our classifications, the majority of the galaxies have favorable viewing angles for discriminating between a wind and a small scale outflow. The exceptions are Ho~II and NGC~4214, which are nearly face-on and which also have holes in their \HI\ disks. It is possible that the \ha\ emission, which we classify as fountains along the line of sight, may be part of more significant outflow events in these two galaxies. 

Figure~\ref{fig:sblimits} shows the smoothed \ha\ emission of the sample in red (outlined at the $3\sigma$ surface brightness limits of each galaxy from Table~\ref{tab:observations}) and the extent of the \HI\ gas at a column density of $5\times10^{20}$ cm$^{-2}$ (outlined in solid blue). Based on the above criteria, we find the following:
\begin{itemize}
\item 5 Winds, 42\% of the sample: \\
\indent \indent NGC~4163, UGC~6546, NGC~4190, NGC~1569, NGC~4449
\item 6 Fountain Candidates, 50\% of the sample: \\
\indent \indent UGC~9128, UGC~4483, NGC~4068, NGC~2366, Holmberg~II, NGC~4214
\item 1 Diffuse, Ionized ISM, 8\% of the sample:\\
\indent \indent DDO~165
\end{itemize}
The galaxies with identified galactic winds and fountain candidates are labeled. For clarity, arrows mark the small-scale, fountains in UGC~9128, UGC~4483, NGC~4068, and NGC~2366. For the remaining two galaxies with fountain candidates, Ho~II and NGC~4214, close inspection of the images reveals \ha\ emission is detected in and around a couple of the ISM holes.

\subsection{Details on the Galactic Winds}\label{sec:wind_detections}
Galactic winds are detected in five galaxies: NGC~4163, UGC~6456, NGC~4190, NGC~1569, and NGC~4449. The morphologies of these winds, seen most clearly in the un-smoothed \ha\ images in the right panels of Figures~\ref{fig:full_fova}$-$\ref{fig:full_fovc}, are varied. 

In NGC~4163, the ionized gas lacks a clearly defined edge. The \ha\ emission follows three main pathways until spreading over a significant extent of the galaxy disk, consistent with the broad and clumpy warm phase wind morphologies seen in simulations \citep{Hopkins2012}, including \ha\ specific simulations \citep{Ceverino2016}. 

In UGC~6456, the \ha\ emission traces a series of superbubble structures in the center of the galaxy with ionized gas outflowing from the minor axis. Previous observations of UGC~6456 with less sensitive observations did not detect the wind \citep{Martin1998, Lozinskaya2006, Arkhipova2007}, highlighting the advantages of observations reaching very low surface brightness levels. In the western region, low column density \HI\ gas with discrepant velocities is coincident with the ionized gas outflow. Radial motions of \HI\ have been measured at $\sim5$ km s$^{-1}$ along approximately this same axis \citep{Lelli2014b}. Combined with our new observations, this suggests that there is colder gas entrained in the warm-phase wind. 

\begin{figure*}
\includegraphics[width=0.95\textwidth]{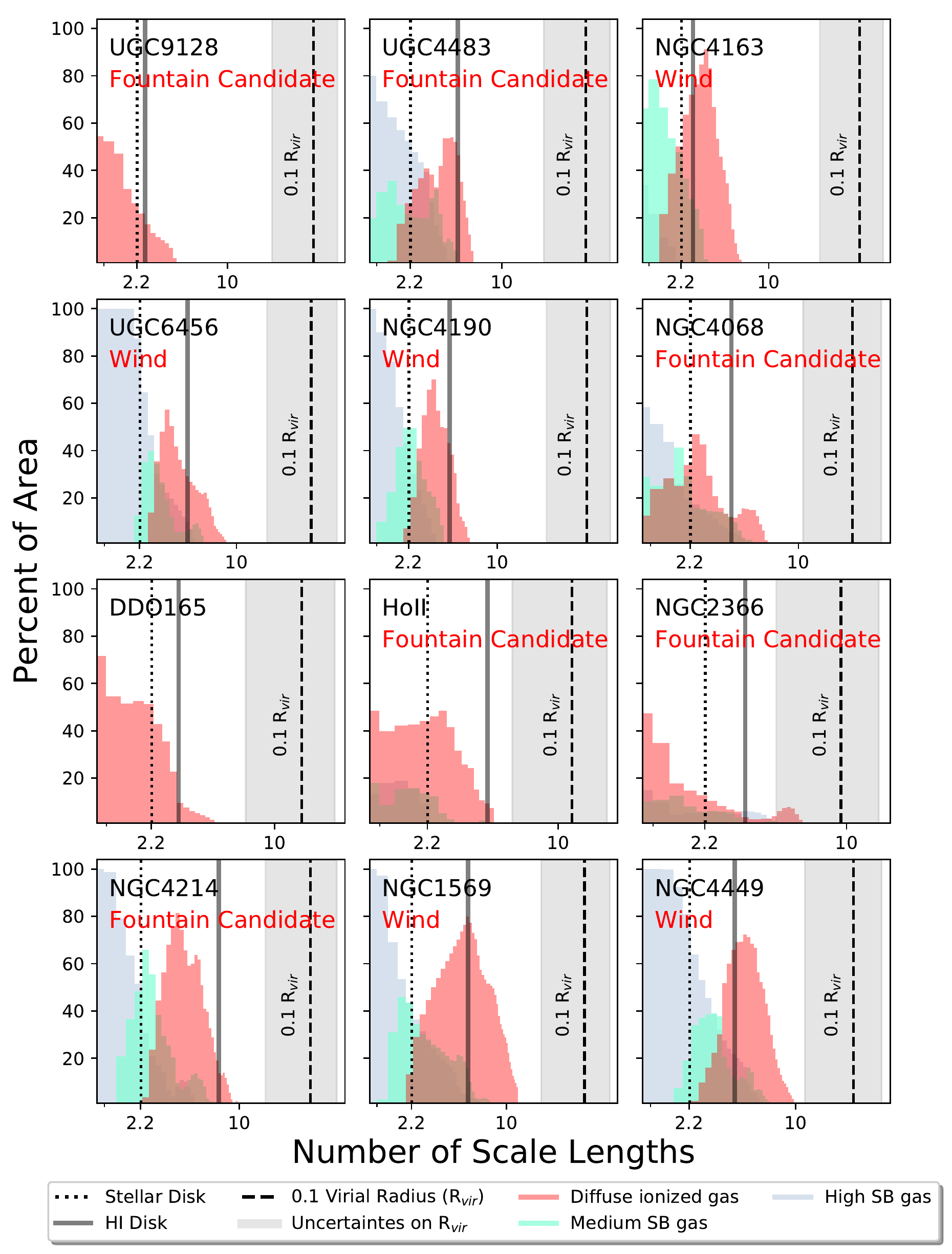} 
\caption{Distribution of high (flux$\geq30\sigma$), medium ($30\sigma> $flux$ \geq10\sigma$), and low ($10\sigma> $flux$ \geq3\sigma$) surface brightness gas as a function of the scale length in the sample. The extent of the stellar disk at $2.2\times$ the scale length is marked with a dotted line; \HI\ with a column density of $5\times10^{20}$ cm$^{-2}$ is detected out to larger distances in all galaxies, shown as a solid gray line. The location of 0.1 R$_{\rm vir}$, defining the boundary between the ISM in a galaxy and the CGM in simulations, is marked with dashed black vertical lines with assumed uncertainties of 50\% shaded in gray.} 
\label{fig:SB_histograms}
\end{figure*}

In NGC~4190, the wind is strongest to the west, lacking a clearly defined edge. Low column density and extremely patchy \HI\ extends out from the main \HI\ disk with an irregular morphology and in the same western region as the \ha\ emission (see also Figure~\ref{fig:ngc4190} in the Appendix). While not conclusive, this suggests some of the atomic gas may be entrained in the warm wind. 

In the NGC~1569, the wind extends in both directions along the minor axis of the galaxy with both substructure (i.e., arcs and filaments) and \ha\ emission without a defined edge. 

In NGC~4449, the wind morphology appears as ruptured superbubbles  (seen most clearly in the right panel of Figure~\ref{fig:full_fova}) and extends to large radii, reaching $\sim6$ kpc. Ionized gas is also detected in the larger environment away from the main galaxy (seen in the smoothed image of Figure~\ref{fig:sblimits}). The diffuse \ha\ emission has been previously studied in NGC~4449 around the main part of the galaxy \citep{Martin1998}; much of the larger spatial extent is newly detected in our deep, wide field observations.

Four of the galaxies with winds, NGC~4163, UGC~6456, NGC~4190, and NGC~1569, have the most highly concentrated recent star formation in the sample (see Table~\ref{tab:star_formation}) based on the star formation spatial properties measured in \citet{McQuinn2010b, McQuinn2012a}. This supports the argument that galaxies with highly concentrated stellar feedback are more likely to develop a galactic wind. NGC~4163 is classified as a post-starburst galaxy based on the global star formation \citep{McQuinn2009}. However, the star formation in the central region of the galaxy remains elevated \citep{McQuinn2012a}, emphasizing that the spatial concentration of the star-formation activity plays an important role in a driving galactic wind. For NGC~4449, the star formation is more distributed across the stellar disk. NGC~4449 is tidally disrupting a lower mass galaxy in the vicinity \citep{Martinez-Delgado2012} and is also thought to have undergone another recent minor merger \citep{Hunter1998}. These interactions are a major driver of ongoing starburst in NGC~4449 and complicates making a connection between the spatial concentration of star formation and the presence of a wind.

These five systems represent the clearest examples of star-formation driven galactic winds in the sample. We calculate the mass-loading factor of the winds in \S\ref{sec:calc_mass_load}. \\

\subsection{Details on the Fountain Flow Candidates}
Small-scale fountain candidates are detected in six galaxies: UGC~9128, UGC~4483, NGC~4068, NGC~2366, Ho~II, and NGC~4214. The recent star formation in the galaxies UGC~4483, NGC~4068, and NGC~2366 have some of the largest spatial distributions in the sample, with young BHeB stars being nearly as extended as the older RGB detected in the $HST$ imaging  (see Table~\ref{tab:star_formation}). The ``diluted'' impact of the stellar feedback appears to still clear paths in the ISM to vent ionized gas, but on a smaller, more localized scale. UGC~9128 is more centralized, but is classified as a ``fossil'' burst with a declining rate of star formation over the past $\sim25$ Myr. The spatial extents of the fountain flows range from being fairly modest (UGC~9128, NGC~4068, NGC~2366) to more extended (UGC~4483) and appear to be driven by pockets of star-forming regions.

As previously mentioned, Holmberg~II and NGC~4214 are nearly face-on. Both galaxies have significant holes in their neutral gas distribution, a clear indication of significant, recent stellar feedback, and lower column densities pathways in the ISM cleared by previous outflows \citep[e.g.,][]{Warren2011}. Close inspection of Figure~\ref{fig:sblimits} shows a number of regions in each galaxy with \ha\ emission detected where there are holes in the \HI\ disk seen in projection or near the edges of the neutral hydrogen. \citet{Martin1998} reported expansion velocities of the ionized gas in the center of NGC~4214 that are consistent with gas moving out of the disk. The physical extent of the ionized gas is difficult to measure due to the inclination angles. It is possible that the \ha\ emission we classify as a small-scale fountain, may be a larger outflow or wind event.

The fountain flows are a mixing agent for metals in the ISM and can transport gas from the star forming regions out past the main \HI\ disk. Given the limited spatial extent of these outflows, the material will likely to be recycled into the galaxy impacting future star formation activity on short timescales. The returning material is an important mechanism to replenish the material available to fuel star formation \citep{Oppenheimer2010} and can increase the turbulence in the ISM,  thereby maintaining low star formation efficiencies \citep[e.g.][]{Kim2017}. 

\subsection{Details on the Diffuse, Ionized ISM}
We do not detect clear evidence of a wind or fountain in DDO~165. The \HI\ disk of the galaxy is known to have a significant hole attributed to a blow-out of the ISM due to previous recent episodes of star formation \citep{Cannon2011a}. In Figure~\ref{fig:sblimits}, \ha\ emission is detected outside the remaining \HI. However, the region around DDO~165 also contains Galactic cirrus. Thus, it is unclear whether the \ha\ emission is from ionized gas expelled earlier by stellar feedback or contamination from foreground emission.

\section{Spatial Extents}\label{sec:spatial}
\subsection{\ha\ Emission vs. Stars, Gas, and Virial Radius}
Figure~\ref{fig:SB_histograms} provides a quantitative measurement of the spatial extent of the \ha\ emission. These histograms plot the radial distribution of low, medium, and high surface brightness ionized gas based on thresholds of 3, 9, and 27 $\times$ the surface brightness limits listed in Table~\ref{tab:observations} as a function of area in elliptical annuli using the geometry in Table~\ref{tab:observations}. The main stellar disk at 2.2$\times$ the scale length is marked as is the extent of the \HI\ for each galaxy. 

Supporting the wind classification from the \ha\ and \HI\ comparisons in Figure~\ref{fig:sblimits}, five galaxies (NGC~4163, UGC~6546, NGC~4190, NGC~1569, and NGC~4449) have extended low surface brightness \ha\ emission in a significant area ($\sim20-90$\% at larger radii), reaching distances outside the \HI\ disk in projection. The radial profiles of the low surface brightness \ha\ emission in these galaxies are similar, increasing in area near the edge of the stellar disk and smoothly declining as the wind drops below detection limits. The projected distances of the detected winds are $\sim2-3$ times the stellar disks of the galaxies corresponding to physical projected distances of $\gtsimeq1 \sim6$ kpc. With the exception of NGC~4190, the galaxies are highly inclined, suggesting the full physical scales of the winds are not dissimilar from their projected distances.

In the fountain flow candidates, the low surface brightness \ha\ emission has a different radial profile. The emission profiles in five galaxies (UGC~9128, UGC~4483, NGC~4068, NGC~2366, and Ho~II) decline at smaller radii before increasing again where the fountain flows are detected (seen as a `bump' in the low surface brightness \ha\ emission histogram). The exception is NGC~4214 where the radial profile shows a consistent pattern with the other fountain flows but where the extended low surface brightness \ha\ radial profile is also consistent with the profiles of the winds. It is possible that NGC~4214 has a wind, but the nearly face-on inclination angle makes it difficult to unambiguously characterize the detected outflow. We tentatively classify NGC~4214 as a fountain candidate.

Figure~\ref{fig:SB_histograms} also shows the estimated distance of a tenth of the virial radius (0.1 R$_{\rm vir}$) for each galaxy. We highlight 0.1 R$_{\rm vir}$ as this is the distance used as the ISM-CGM separation in cosmological simulations \citep[e.g.,][]{Muratov2015} and, thus, provides a useful reference point with which to compare the extent of the detected low surface brightness ionized gas. We estimate the virial radii by assuming the stellar mass-halo mass abundance matching relation from \citet{Moster2010} and the parameters for converting to virial radius from the COS-Dwarf Survey \citep{Bordoloi2014}. The virial radii calculation used in \citet{Bordoloi2014} has uncertainties of order 50\%, which we adopt and show as shaded regions in Figure~\ref{fig:SB_histograms}. Table~\ref{tab:mass_load} lists the halo mass, virial radius, and corresponding escape velocity for the sample. Note that these values depend on the stellar mass-halo mass relation assumed, which has higher uncertainties at lower galaxy masses. Results from \citet{Behroozi2013} suggest a flattening of the stellar mass-halo mass relation below M$_{\rm halo} \sim10^{11}$ \msun. If we assume this relation, the halo mass, virial radius, and escape velocities differ, but are still within the range of our adopted uncertainties. As a consistency check, we also calculated the dynamical masses of the galaxies based on the \HI\ rotation curves. We used M$_{\rm dyn} = 2.33\times10^5 \cdot$ V$_{\rm circ}^2 \cdot$ R$_{\rm max}$ (where V$_{\rm circ}$ is in units of km s$^{-1}$, R$_{\rm max}$ is in units of kpc, and M$_{\rm dyn}$ is in units of \msun). Our values of M$_{\rm dyn}$ were consistently lower than the M$_{\rm halo}$ values, as expected, given that the gas rotation curves do not plateau. We also confirmed that the stellar and halo masses agreed with the range reported for simulated low-mass galaxies in \citet{Muratov2015}. The halo masses and virial radii only represent approximate values to help place our measurements in the context of the larger galactic systems. Even with the high uncertainties, nearly all of the low surface brightness ionized gas detected lies inside 0.1 R$_{\rm vir}$ and would still be considered part of the ISM in simulations. We discuss the implications of the limited spatial extent of the winds relative to simulations in \S\ref{sec:discuss}.

\subsection{Comparison of the Warm and Hot Wind Phase}
The warm, ionized wind phase is thought to carry most of the mass out of galaxies, whereas the hot wind phase (T$>10^6$ K) is thought to have a higher velocity and be preferentially enriched with metals \citep{Veilleux2005}. Thus, it is interesting to compare the distribution and extent of the warm, ionized gas in our observations with the hot phase. 

\begin{figure}
\includegraphics[width=0.48\textwidth]{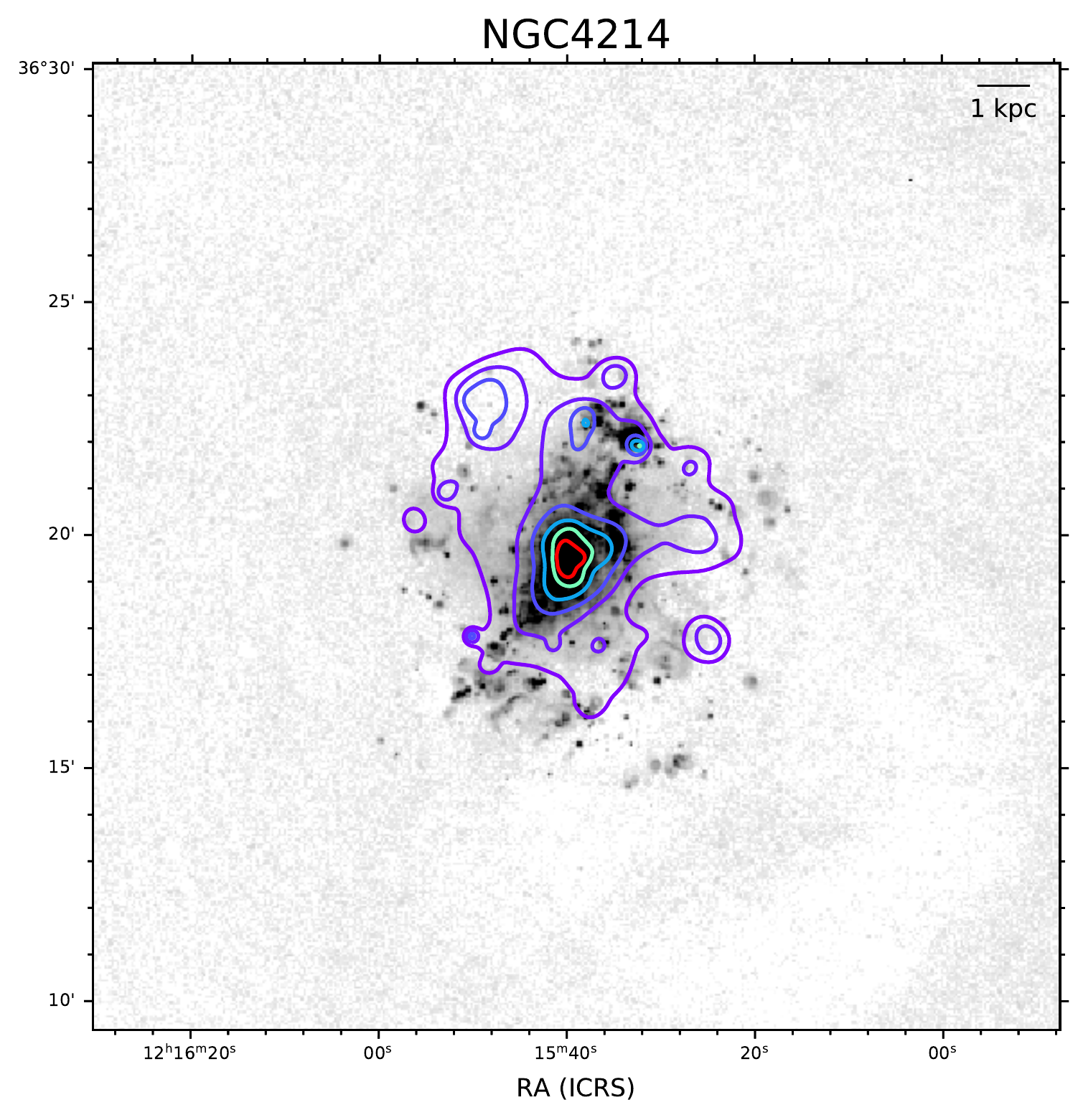}
\vspace{-15pt}
\caption{NGC~4214. \ha\ map overlaid with soft, diffuse X-ray contours from \citet{McQuinn2018}. The detected warm and hot gas phases have similar extent. \HII\ regions are found farther than the detected X-ray emission in projection. Soft X-ray contours levels are 2, 4, 8, 16, 32, and 64$\sigma$.}
\label{fig:ngc4214_xray}
\end{figure}

\begin{figure}
\includegraphics[width=0.48\textwidth]{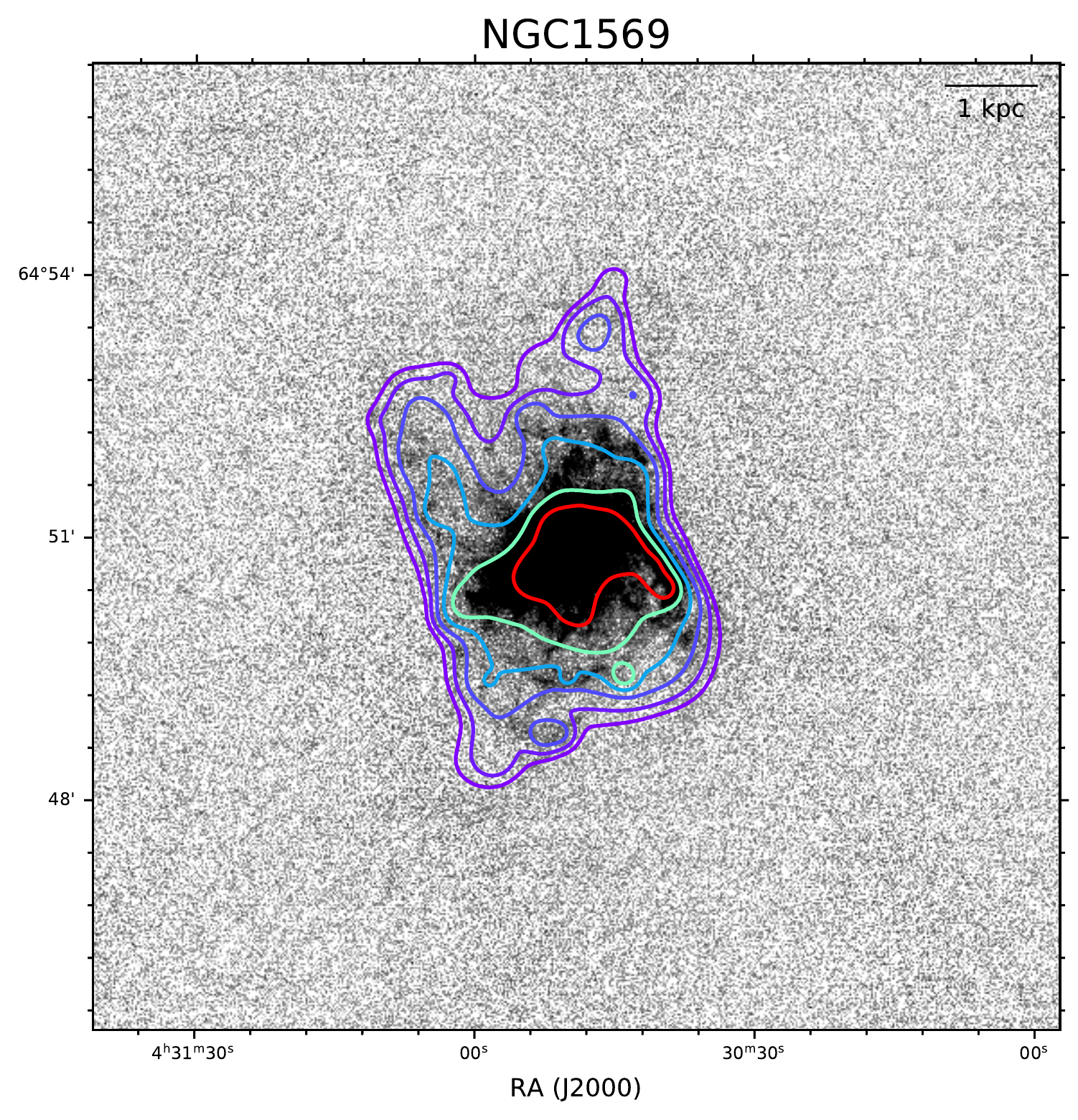}
\vspace{-15pt}
\caption{NGC~1569. \ha\ map overlaid with soft, diffuse X-ray contours from \citet{McQuinn2018}. The detected warm and hot gas phases have similar morphology and extent, although low surface brightness \ha\ emission is seen at larger radii than the detected X-ray emission. Soft X-ray contours levels are 2, 4, 8, 16, 32, and 64$\sigma$.}
\label{fig:ngc1569_xray}
\end{figure}

\begin{figure}
\includegraphics[width=0.48\textwidth]{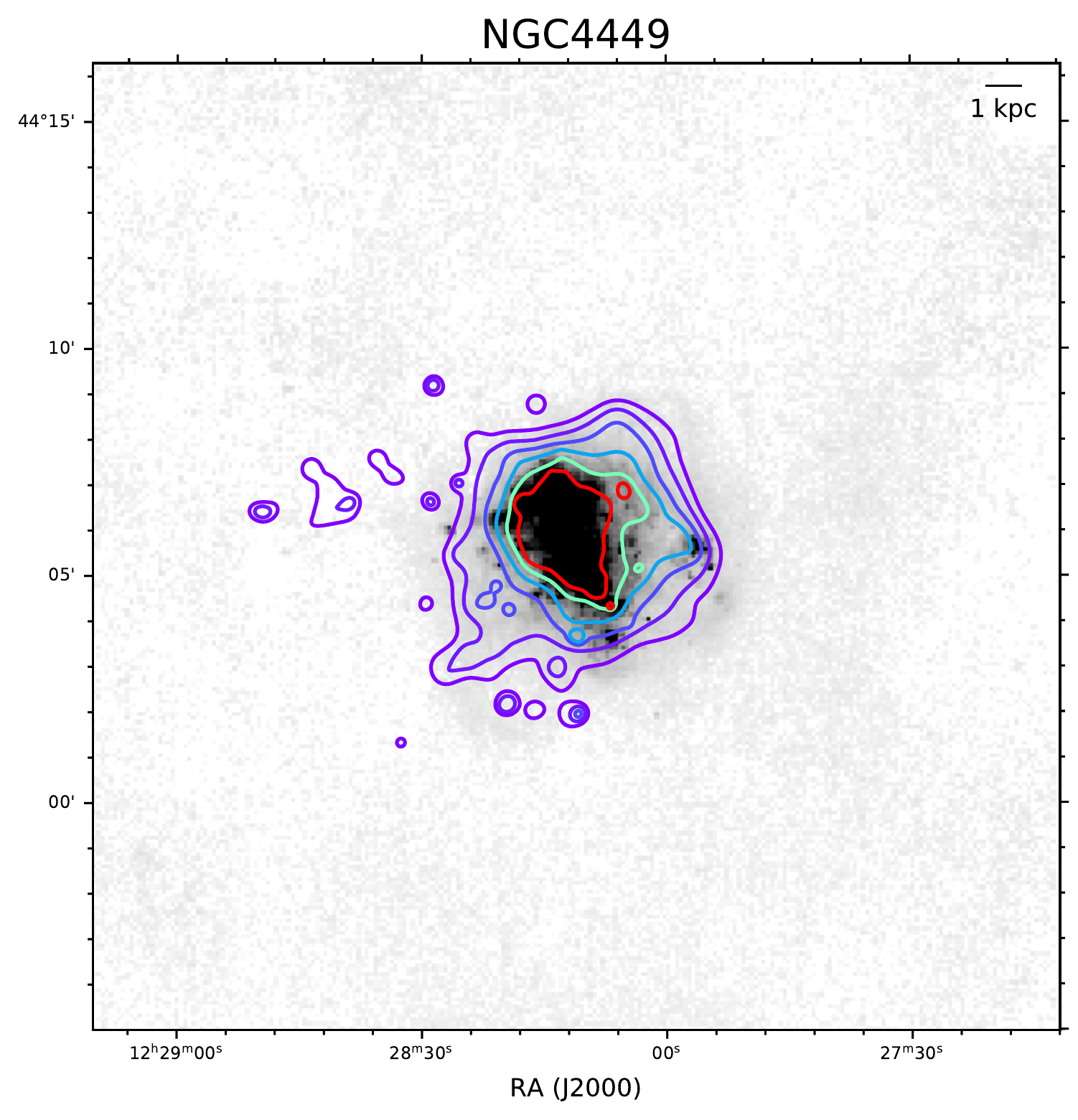}
\vspace{-15pt}
\caption{NGC~4449. \ha\ map overlaid with X-ray contours from \citet{McQuinn2018}. The detected \ha\ emission is more extended in nearly all directions than the hot wind detected in soft, diffuse X-ray emission. Soft X-ray contours levels are 2, 4, 8, 16, 32, and 64$\sigma$.}
\label{fig:ngc4449_xray}
\end{figure}

Three galaxies in the sample, NGC~4214, NGC~1569, and NGC~4449, have existing X-ray imaging from the $Chandra$ Space Telescope. In Figures~\ref{fig:ngc4214_xray}$-$\ref{fig:ngc4449_xray}, we present the \ha\ emission maps with diffuse, soft X-ray contours overlaid. The X-ray contours are from the reduced $Chandra$ data from a companion STARBIRDS paper on galactic outflows and timescales \citep{McQuinn2018}. 

The overall extent of the hot and warm gas phases are comparable in all three galaxies, but there are some notable differences. In NGC~4214, there are \ha\ knots associated with \HII\ regions outside the X-ray emission in projection. In NGC~1569 and NGC~4449, the \ha\ and X-ray emission morphologies follow each other closely, although low surface brightness \ha\ emission is found at slightly larger distances from the galaxies than the detected X-ray emission. Assuming the gas in NGC~1569 and NGC~4449 is expanding along the same low density paths in the \HI, this implies the hot wind lies inside the warm wind, possibly driving the expansion of the warm gas in both systems.

\newpage
\section{Calculating Mass-Loading Factors}\label{sec:calc_mass_load}
The mass-loading factor ($\eta$) compares the mass loss rate from outflows ($\dot{\rm M}$) to the star formation rate (SFR). Given the range in mass-loading factors of a few to more than a hundred from simulations and the relatively sparse observational constraints, the mass loading of the winds and fountains detected in the STARBIRDS sample can provide needed, quantitative constraints on the impact of stellar feedback on low-mass galaxies. 

We calculate $\dot{\rm M}$ by estimating the mass of ionized gas passing through a thin shell defined by the outer edge of the identified winds and fountain flows over timescales determined by the thickness of the shell and the velocity of the gas. This approach is similar to what is done in simulations where the instantaneous mass flux through a thin spherical shell approach is used to calculate mass-loading factors \citep[e.g.,][]{Muratov2015}. The main uncertainties in our calculated mass loss rate come from assuming a 3D geometry for each wind or outflow and the adopted velocity of the outflowing material; both are discussed in more detail below. The time averaged SFRs are well-measured from the star formation histories. 

The mass of outflowing gas is based on the number density of ionized hydrogen atoms in a given volume:
\begin{equation}
{\rm M_{outflow}} = m_H \cdot \mu \cdot n_p \cdot \delta \cdot {\rm V} \label{eq:mass}
\end{equation}

\noindent where $m_H$ is the mass of hydrogen, $\mu = 1.4$ is the mean molecular mass assumed and accounts for the mass of helium, $n_p$ is the number density of ionized hydrogen atoms, $\delta$ is the volume filling factor, and V is the volume of the thin shell through which the ionized gas is moving. We describe each variable, beginning with the assumed geometry and filling factors.

The volume of ionized gas is determined based on the measured thin shell area of \ha\ emisison at the $3\sigma$ surface brightness limits and assuming a depth along the line of sight, $l$. One of the advantages of choosing a thin shell at the outer edge of the winds is that the \ha\ emission is at a nearly constant surface brightness and there is nearly contiguous emission around the shell (see Figure~\ref{fig:sblimits}). Thus, this simplifies the geometry assumptions. In the cases of the winds, the lowest surface brightness gas is roughly circular in shape. We assumed a cylindrical 3D geometry with a depth equal to the diameter of the approximately circular shells. In the cases of the fountain flow candidates, we assume that the outflows have a cylindrical geometry with a depth equal to the diameter of the detected fountain `finger' or region. If the geometry is more fan-like in nature, this will under-estimate the mass. 

Our calculations of the ionized gas volume depend on the orientation of the galaxies. As shown below, the dependency on our geometry in the final mass-loading factor derivation reduces to the square root of the assumed length along the line of sight. Using the inclination angles of the galaxies listed in Table~\ref{tab:galaxies} as a guide, we estimate this adds $<$20\% uncertainty to our final mass-loading factors, which is significantly less than our adopted uncertainties as described below.

For the volume filling factor, we assume that the mass of ionized gas determined from the \ha\ emission represents 10\% of the volume in the outflows (i.e., a volume filling factor $\delta = 0.1$). Previous work estimated the volume filling factors for \HII\ regions to range from 0.01 to 0.1, with a mean value of 0.02 \citep{Kennicutt1984}, which can be compared with a value of 0.01 determined based on pressure arguments from hot gas pushing the expansion of warm outflows \citep{Martin1999}. We adopt the higher value of 0.1 as the outflowing, extremely low surface brightness gas is not expected to be as highly clumped as in \HII\ regions where the \ha\ emission originates from knots of higher density gas. 

$n_p$ is related to the emission measure (EM; pc cm$^{-6}$) along the line of sight defined as:

\begin{equation}
{\rm EM} \equiv \int n_e  \cdot n_p\ dl \label{eq:EM_los}
\end{equation}

\noindent where $n_e$ is the number density of electrons and $n_e \approx 0.92 \cdot n_p$; the factor of 0.92 adjusts for singly ionized helium in the gas. The integral is evaluated at the depth along the line of sight, $l$, described above. Because the \ha\ emission from ionized hydrogen depends on $n_p$, $n_e$, and a recombination rate, the EM can also be calculated from the \ha\ surface brightness assuming a case B recombination rate \citep{Brocklehurst1971}:

\begin{equation}
{\rm EM} = 4.858 \times 10^{17} \cdot ({\rm T}/10^4)^{0.924} \cdot {\rm S(H}_{\alpha}) \label{eq:EM_sb}
\end{equation}

\noindent where the surface brightness measurement S(\ha) is in units of erg s$^{-1}$ cm$^{-2}$ arcsec$^{-2}$ and the temperature, T, is assumed to be 10,000 K. Based on the $3\sigma$ \ha\ surface brightness levels found for the sample, the EMs range from $0.7-1.7$ pc cm$^{-6}$ for our deep \ha\ observations with the exception of DDO~165 whose EM is somewhat higher at 3.3 pc cm$^{-6}$ due to the presence of Galactic cirrus contamination, and 8.6 pc cm$^{-6}$ from the shorter integration observations of NGC~1569. Final numbers are listed in Table~\ref{tab:observations}. For comparison, EMs of the warm interstellar medium of the Milky Way are of order 0.1 pc cm$^{-6}$ from the Wisconsin \ha\ Mapper (WHAM) Survey \citep{Hill2008}, and EMs of \HII\ regions range from $\sim10^3-10^5$ \citep{Kennicutt1984}. 

\input{tab5}

The mass of ionized gas can be found by simultaneously solving Eqs.~\ref{eq:EM_los} and \ref{eq:EM_sb} for $n_p$, and substituting the resulting expression for $n_p$ into Eq.~\ref{eq:mass}: 

\begin{equation}
{\rm M_{outflow}} = 1.02 \times 10^9\ \cdot m_H \cdot \delta \cdot {\rm area}\ \cdot [{\rm S(H}_{\alpha}) \cdot {\it l}]^{\frac{1}{2}}
\label{eq:mass_ha}
\end{equation}

\noindent where the volume occupied by the ionized gas is now expressed as the area of the thin shell of ionized gas $\times$ the depth $l$. 

$\dot{\rm M}$ is then determined using the mass calculated above divided by the time it takes for the gas to cross the thin shell. This crossing time is simply the thickness of the thin shell divided by the velocity of the outflowing gas. The thin shell thicknesses are measured from the smoothed \ha\ images and converted to physical scales using the distances in Table~\ref{tab:galaxies}. Velocities of the diffuse ionized gas are of order 25$-50$ km s$^{-1}$ based on the velocity dispersion measured from the ionized gas kinematics using WIYN Sparsepak IFU observations in the sample (van Zee et al.\ in preparation). For comparison, previously measured expansion velocities in dwarf galaxies range from 25$-$100 km s$^{-1}$, with a mean of 50 km s$^{-1}$ \citep{Marlowe1995, Martin1998, Schwartz2004}. Simulations of galaxies with comparable halo masses report velocities of ejected gas to be up to 5 times the circular velocity \citep[e.g., V$_{\rm circ} \ltsimeq$ V$_{\rm ejected} \ltsimeq 5\times$V$_{\rm circ}$;][]{Christensen2016}. 

We use the range in velocity of $25-50$ km s$^{-1}$ in our mass loading calculations, consistent with the ionized gas kinematics. Based on the radial extent of the ionized gas and a velocity range of $25-50$ km s $^{-1}$, the crossing time is of order a few $\times10^5$ yr. Assuming the lower velocity of 25 km s$^{-1}$, mass loss rates range from $0.01-0.6$ \msun\ yr$^{-1}$ for the winds and $0.0004-0.06$  \msun\ yr$^{-1}$ for the fountains; these values double for a higher velocity of 50 km s$^{-1}$ is assumed. We list the average of these mass loss rates for each galaxy in Table~\ref{tab:mass_load}. 

The SFRs for the final mass-loading calculation come from the SFHs, which have SFRs as a function of time. Since the winds and fountains have different spatial extents and therefore were likely launched by star formation activity on different timescales, we estimate a dynamical timescale for the outflows and use the SFR from the most closely matched time bin in the SFHs. In some galaxies where the SFRs are relatively constant, this will not impact the calculation. However, in a few systems, the SFRs show significant variations over the last $\sim100$ Myr; selecting the most closely matched timescale provides the appropriate SFR is used in the mass-loading calculation. 

The dynamical timescales are estimated using the radial distance of the thin shell perpendicular to the major axis of the disk and by assuming velocities of 25 and 50 km s$^{-1}$ for the outflowing gas (i.e., $\tau_{dyn} =$ radius / velocity). If the gas has decelerated as it has expanded outward through the ISM, these dynamical timescales will be a lower limit. The outflow timescales range from a few Myr to $\sim$35 Myr, broadly consistent with previously estimated timescales of $10-20$ Myr derived for warm-phase outflows \citep{Martin1998} and 60 Myr timescales reported by simulations to launch a wind \citep{Muratov2015}. 

For the galaxies with winds, we used the globally averaged SFRs \citep{McQuinn2009, McQuinn2010a, McQuinn2015d}. For the galaxies with fountain flows, we use the SFR derived from the central regions which are spatially adjacent to the detected outflows \citep{McQuinn2012a}. Table~\ref{tab:mass_load} lists the SFRs and corresponding time frames; SFRs from the central regions are marked with a ``(c)''. Finally, the mass-loading factor, $\eta$, is calculated for each galaxy based on the estimated $\dot{\rm M}$ and SFRs. Table~\ref{tab:mass_load} lists the average mass-loading factors assuming the range in velocities and corresponding SFRs.

\begin{figure*}
\includegraphics[width=0.98\textwidth]{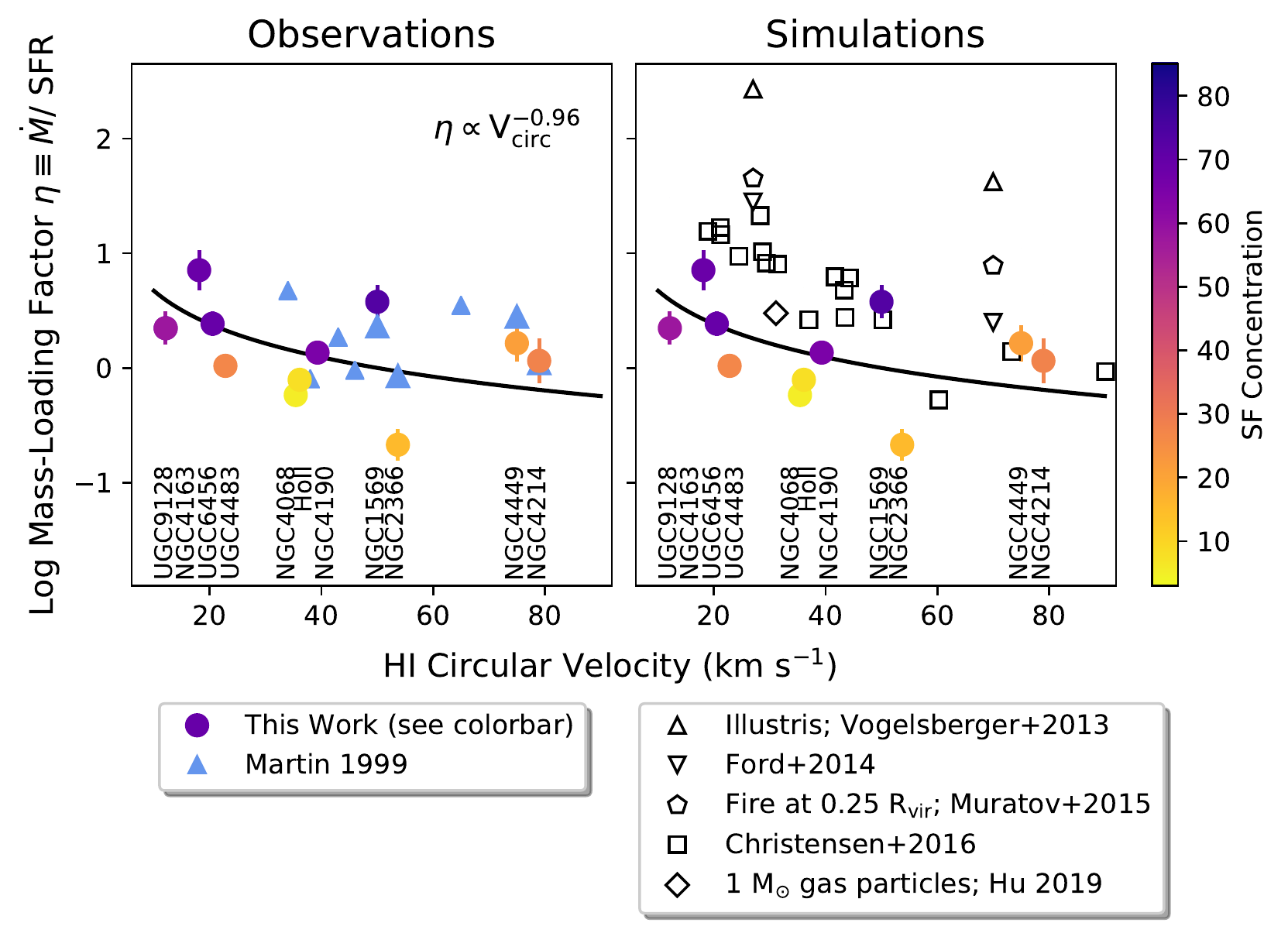}
\caption{mass-loading factors as a function circular velocity; points are color-coded by the concentration of star formation as shown in the side colorbar. Overplotted in the left panel are observationally measured mass-loading factors from \citet[][blue triangles]{Martin1999}. The best-fitting relation $\eta \propto$ V$_{\rm circ}^{-0.96}$ is shown as a solid black line. Overplotted in the right panel are predicted mass-loading factors from hydrodynamical simulations from \citet[][triangles]{Vogelsberger2013}, \citet[][upside down triangles]{Ford2014}, \citet[][pentagons]{Muratov2015} \citet[][squares]{Christensen2016}, and \citet[][diamonds]{Hu2019}. The lowest predicted values overlap with our highest values, but overall the simulations are offset to mass-loading factors significantly higher than empirically derived values.}
\label{fig:mass_load_vcirc}
\end{figure*}

Our mass-loading factors are based on the amount of {\em ionized} gas detected in the winds and fountain flows, representing the warm phase wind. However, winds are multi-phase and can include hot ($T\sim10^6$ K) and cold (atomic or molecular) gas. As described in \S\ref{sec:intro}, hot winds are fast-moving, low-density gas thought to be primarily comprised of metal-enriched supernova-ejecta and observable in the X-ray. The hot phase is expected to contribute little to the total mass budget of winds. Cold, neutral gas may be entrained in the outflows, but there is not strong, quantitative evidence that the mass budget in winds is dominated by the neutral gas phase. Thus, we conclude that while the mass-loading factors we calculate are lower limits for the {\it total} amount of gas in all phases in the winds and outflow, they likely represent the majority of material being expelled from the disks of the galaxies. 

\section{Discussion on Mass-Loading Factors}\label{sec:discuss}
The mass-loading factors ($\eta$) as a function of circular velocity measured from the \HI\ rotation curves (V$_{\rm circ}$)  are presented in Figure~\ref{fig:mass_load_vcirc}, color-coded by the concentration of the recent star formation and with accompanying galaxy labels. We show the average mass-loading factors, which range from 0.2 $\sim 7$; the uncertainties on the mass-loading factors span the values calculated using the 25-50 km s$^{-1}$ range in wind velocities we adopted based on IFU spectroscopy of the ionized gas in the sample and the varying SFRs (i.e., if the wind is traveling faster, it will have been launched by  more recent SF). The best-fitting line to our mass-loading calculations, overplotted as a solid black line, follows the relation $\eta \propto$ V$_{\rm circ}^{-0.96}$, consistent with a momentum-conserving wind.

The left panel in Figure~\ref{fig:mass_load_vcirc} also shows mass-loading factors previously derived for low-mass galaxies  from \citet[][blue triangles]{Martin1999}. There are a few galaxies that overlap with our study; for these systems, we adopt our V$_{\rm circ}$ values. For the systems unique to \citet{Martin1999}, we adopt their V$_{\rm circ}$. 

The right panel in Figure~\ref{fig:mass_load_vcirc} compares our derived mass-loading factors from values predicted by a number of hydrodynamic simulations. The studies shown range from large cosmological simulations including \citet[][Illustris, triangles]{Vogelsberger2013} and \citet[][upside down triangles]{Ford2014}, to  zoom-in simulations including \citet[][FIRE, pentagons]{Muratov2015} and \citet[][squares]{Christensen2016}, to a high-resolution simulation of an individual galaxy with gas particle masses of 1 \msun\ from \citet[][diamond]{Hu2019}. The mass-loading factors from simulations vary widely, spanning more than two orders of magnitude; the lower of these values overlap with the highest mass-loading factors derived observationally. 

Seen in Figure~\ref{fig:mass_load_vcirc}, the mass-loading factors do not exhibit a strong dependency on the circular velocity of the galaxies. Note, however, the circular velocities should be interpreted with care. As discussed in \S\ref{sec:hi}, the circular velocities are only indicative tracers of the gravitational potential as the gas kinematics are often complicated by non-circular rotation, the gaseous disks do not always extend far enough for gas rotation curves flatten, and the inclination angle of the gaseous component is difficult to estimate. In addition, V$_{\rm circ}$ in simulations is often determined at the virial radius based on the halo mass, whereas V$_{\rm circ}$ is measured observationally at the outermost radius from the \HI\ gas. Thus, not only are the observational V$_{\rm circ}$ values uncertain, they can be difficult to directly compare to V$_{\rm circ}$ from simulations at low galaxy masses \citep{Brooks2017}. 

\begin{figure*}
\includegraphics[width=0.98\textwidth]{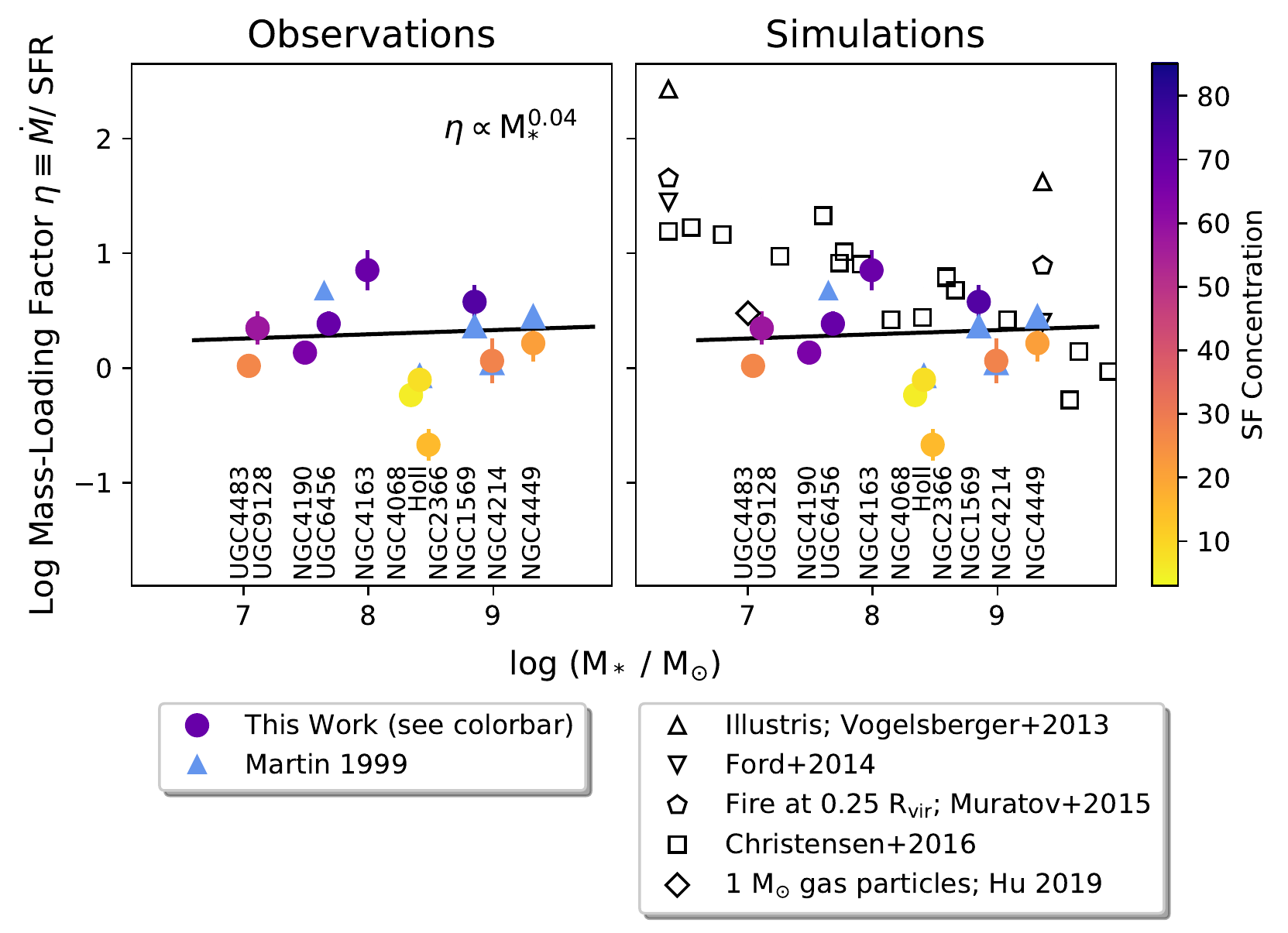}
\caption{mass-loading factors as a function stellar mass; similarly to Figure~\ref{fig:mass_load_vcirc}, points are color-coded by the concentration of star formation as shown in the side colorbar. Overplotted in the left panel are observationally measured mass-loading factors from \citet[][blue triangles]{Martin1999}. The best-fitting relation $\eta \propto$ M$_{*}^{0.04}$ is shown as a solid black line. Overplotted in the right panel are predicted mass-loading factors from hydrodynamical simulations from \citet[][triangles]{Vogelsberger2013}, \citet[][upside down triangles]{Ford2014}, \citet[][pentagons]{Muratov2015} \citet[][squares]{Christensen2016}, and \citet[][diamonds]{Hu2019}. In contrast to predictions, the mass-loading factors do not exhibit a strong dependency on stellar mass. Instead, higher mass-loading factors are associated with galaxies with centrally concentrated star formation.}
\label{fig:mass_load_mstar}
\end{figure*}

Therefore, for additional context, we present the mass-loading factors as a function of stellar mass in Figure~\ref{fig:mass_load_mstar}, with the same symbols, colors, and samples shown in Figure~\ref{fig:mass_load_vcirc}.\footnote{Stellar masses were not available for four galaxies from \citet{Martin1999}.} The observationally-derived mass-loading factors show very little dependence on stellar mass ($\eta \propto$ M$_*^{0.04}$), with a high degree of scatter. The flat distribution of mass-loading factors as a function of stellar mass is very different from simulation results which consistently show a steep dependency with stellar mass. We discuss this in detail in \S\ref{sec:predictions}. From Figures~\ref{fig:mass_load_vcirc} \& \ref{fig:mass_load_mstar}, we find a correlation with the concentration of star formation, where galaxies with more centrally concentrated star formation have higher mass-loading factors than those with star formation that is distributed across the stellar disk of the galaxy. 

\subsection{Detailed Comparison with Previous Observations}
The full range in observed mass-loading factors shown in Figure~\ref{fig:mass_load_vcirc} is modest, from 0.2 for a fountain flow in NGC~2366 to 7.1 for a wind in NGC~4163. Our values are comparable to those derived by \citet{Martin1999} over the narrow mass range probed by both studies. For three of the four galaxies that overlap between the two studies (NGC~1569, NGC~4214, NGC~4449), the mass-loading factors are in good agreement; for the final system (NGC~2366) our value is considerably lower. Our results extend the derivation of mass-loading factors to galaxies an order of magnitude less massive, with circular velocities below 20 km s$^{-1}$ and M$_*\sim10^7$ \msun. 

Other observationally determined mass-loading factors ranging from 6$-$19 have been reported in the literature for more extreme starbursts and derived using different methods \citep{Heckman2015, Chisholm2017}. Considering these additional studies, the full range of $\eta$ values spans two orders of magnitude from 0.2 to 19. In order to better understand what factors drive such a large range, we explore differences in the methods used to derive $\eta$ in \S\ref{sec:techniques} and differences in the samples studied in \S\ref{sec:samples}.

\subsubsection{Differences in Deriving $\eta$}\label{sec:techniques}
Our method of deriving the mass-loading factors is similar to that of \citet{Martin1999} in that we use \ha\ emission to trace the mass loss from galaxies. There are, however, some important differences in techniques. Our mass loss rates are determined at the edge of the detected outflows whereas \citet{Martin1999} scale the \ha\ flux summed from the low surface brightness expanding bubbles and the high surface brightness areas around the star forming regions powering the expanding bubbles. It is unclear how much of this ionized material will actually leave the galaxy, particularly the higher surface brightness gas at smaller radii. Thus, some of the mass-loading factors in \citet{Martin1999} may be overestimated. Our SFRs are based on the temporally resolved star formation histories from CMDs, whereas \citet{Martin1999} scale the \ha\ flux as a measure of the current star-formation activity. \ha-based SFRs are tied to star formation timescales of $\sim5$ Myr. With advancements in simulations, it is now clear that the galactic winds detected outside of low-mass galaxies were launched much earlier than 5 Myr \citep[i.e.,][]{Hopkins2012, Kim2017}, bringing into question the suitability of a 5 Myr star formation tracer given the known variability of SFR in dwarfs. Since the SFRs measured from the CMD-fitting technique shows variability on timescales of 5$-$30 Myr, the SFR adopted from a 5 Myr timescale may not accurately represent the star formation activity in the mass loading calculation. Given the relatively small range in $\eta$ between the two studies and the general agreement between the galaxies overlapping in the samples, these differences likely impact the mass-loading factors by a factor of a few at most.

The mass-loading factors in \citet{Heckman2015} and \citet{Chisholm2017} were derived using a different approach than that employed here or by \citet{Martin1999}. Mass loss rates were based on unsaturated, far-ultraviolet absorption lines including \ion{O}{1}, \ion{Si}{2}, \ion{S}{2}, and \ion{Si}{4} and depend on a number of different factors including extinction corrections, observationally motivated photoionization models, absorption line profile models, and, in the case of \citet{Heckman2015}, also employing simplifying assumptions and stacking spectra to increase the strength of the detected absorption lines. The SFRs were determined by scaling the ultraviolet (UV) flux, which is tied to a $\sim100$ Myr timescale. As cooling timescales of the warm gas are much shorter than 100 Myr and SFRs can vary by a factor of a few on 100 Myr timescales \citep{McQuinn2010a}, $\eta$ values are uncertain by a similar factor. Furthermore, UV-based SFRs calibrated from theoretical scaling relations \citep[e.g.,][]{Hao2011, Murphy2011, delosReyes2019} are known to be $\sim$50\% {\em lower} than UV SFRs calibrated empirically from CMDs \citep{McQuinn2015c}. Using the theoretical calibration for UV-based SFRs translates to mass-loading factors that are systematically 50\% higher.

\subsubsection{Differences in the Samples Studied}\label{sec:samples}
As described in \S\ref{sec:sample}, our sample is comprised primarily of galaxies that, while actively star-forming and classified as starbursts, are not extreme systems. The galaxies studied in \citet{Martin1999} have similar overall star formation properties, with the exception of the extreme system I~Zw~18. In contrast, the low-mass galaxies studied in \citet{Heckman2015} and \citet{Chisholm2017} are primarily extreme starbursts, such as  I~Zw~18, SBS~0335$-$52, Haro~3, and SBS~1415$+$437. These galaxies have SFRs per unit mass \textit{two orders of magnitude higher} than our sample. The differences in star formation activity per unit mass have a profound impact on the amount of gas expelled from a galaxy, directly reflected in the mass-loading factors. The extreme starbursts are also highly centralized events with spatially and temporally coincident star formation, analogous to the star formation events assumed in models of low-mass galaxies that drive strong outflows \citep[e.g.,][]{Dekel1986}. Such extreme starbursts are rare in the nearby universe and are thought to have star formation conditions similar to systems at higher redshift. 
 
\vspace{10pt}
In summary, mass-loading factors derived observationally span two orders of magnitude. We note that precise comparisons between the mass-loading factors are challenging as different data analysis techniques introduce uncertainties that may impact the values of $\eta$ up to a factor of a few. Yet, since the mass-loading factors across the four studies vary by 100, analysis techniques alone do not account for the differences. Instead, differences in the star formation conditions in the galaxies appear to be the biggest driver of the range in $\eta$. At low redshift, the typical dwarf does not produce the star formation density of an extreme burst. Thus, even though the smaller potential wells of these galaxies make it easier for gas to escape, the mass-loss rates from these galaxies are modest relative to their star formation activity, with $\eta$ values $<1$ to several. The rare extreme starburst found at low redshift, which is more characteristic of a starburst at high redshift, is capable of driving a stronger wind, with $\eta$ values reaching up to 20.

\subsection{Detailed Comparison with Theoretical Predictions}\label{sec:predictions}
Similar to the observations, the mass-loading factors predicted from hydrodynamical simulations shown in Figures~\ref{fig:mass_load_vcirc}  \& \ref{fig:mass_load_mstar} vary by two orders of magnitude over a similar range in circular velocity and stellar mass. The predicted mass-loading factors, however, are offset to higher values than the empirically derived values, with the exception of the upper end of our stellar mass range (i.e., for M$_*\gtsimeq10^9$ \msun).

The highest predicted values are for the galaxies in the Illustris simulations \citep{Vogelsberger2013}, the simulations of \citet{Ford2014}, and FIRE simulations \citep{Muratov2015}. For Illustris, the values represent the input to the simulations using an assumed energy-driven wind (i.e., $\eta \propto$ V$_{\rm circ}^{-2}$); these are more than 2 orders of magnitude higher than our measurements. \citet{Ford2014} also used an energy-driven scaling factor for the mass-loading factors but are systematically lower than \citet{Vogelsberger2013}. The Next Generation of the Illustris simulations \citep[TNG50;][not shown]{Nelson2019} inject outflows using a scaling relation and find the mass-loading factors produced are $\sim10$ for systems with M$_* \sim10^9$ \msun\ and $\sim60$ for 10$^7$ \msun, somewhat lower than Illustris but still higher than our empircal values.

For the FIRE results \citep{Muratov2015}, the values are calculated directly from the high-resolution zoom-in simulations, rather than being based on a sub-grid prescription for mass loading. The mass-loading factors are measured at a distance of 0.25 R$_{\rm vir}$ from the center of the galaxies, significantly farther than the winds detected in our observations. At smaller radii, the mass-loading factors are reported to be larger as not all the material leaving the galaxy disk will reach the CGM at 0.25 R$_{\rm vir}$, although it is not clear if the winds measured at 0.25 R$_{\rm vir}$ include mass entrained from the CGM itself. The median wind velocities for the two low-mass halos are between $\sim15-30$ km s$^{-1}$, in general agreement with our adopted range of 25$-$50 km s$^{-1}$. One possible factor driving the higher mass-loading factors in FIRE is that the initial baryon fraction assumed for low mass galaxies may be too high; if the galaxies are formed with high gas masses, more gas must be removed from the systems to prevent high stellar masses at $z=0$.

\citet{Christensen2016} looked at gas flows in a suite of high-resolution, zoom-in simulations that included 17 galaxies in the same stellar mass range of $10^6\sim10^9$ \msun\ of the above simulations. In these simulations, V$_{\rm circ}$ values were found to be similar to what would be observed via gas rotation curves for lower galaxy masses. The mass-loading factors ranged from $\sim1-20$ at $z=0$, with a power-law fit of $\eta\ \propto$ V$_{\rm circ}^{-2.2}$; roughly consistent with an energy-driven wind. \citet{Hu2019} reports a mass-loading factor of 3 from a high-resolution (particle masses of $m_{gas} = 1$ \msun) smoothed-particle hydrodynamic simulation of an isolated dwarf galaxy (i.e., not a cosmological simulation) that resolves individual supernovae. Fountain flows comprised of ionized and cooler gas develop at small radii from the galaxy; winds containing predominantly ionized gas develop at distance of 2 kpc from the galaxy. As seen in Figures~\ref{fig:mass_load_vcirc}, the lower values of these two studies overlap with the higher empirically determined mass-loading factors. When the mass-loading factors are shown as a function of stellar mass in Figure~\ref{fig:mass_load_mstar}, the values from \citet{Christensen2016} agree for M$_* \gtsimeq 10^9$ \msun, but are an order of magnitude higher than our derived mass-loading factors for lower mass galaxies. The weaker winds produced in the simulation by \citet{Hu2019} are in good agreement with our values for a $\sim10^7$ \msun\ galaxy. 

As observationally determined mass-loading factors are based on a number of assumptions, we consider whether the differences between observations and predictions could be due to inputs in our calculations.  As described above, the main assumptions include the velocity of the outflowing gas, the volume filling factor, and the 3D geometry of the outflows. Figure~\ref{fig:mass_load_vcirc} explicitly includes a range of possible velocities. We have conservatively estimated the filling factor to be 0.1; reducing this value would lower our mass-loading factors, increasing the gap between observations and predictions. Finally, the mass-loading factors depend on the square root of the line of sight depth of the \ha\ emission, which we have estimated based on an assumed geometry (see above). If the winds and outflows reach farther distances along the line of sight, our mass-loading factors would be higher by a factor of $\sqrt l$; the line of sight depth would need to be more than 100$\times$ greater to close the gap with simulations. 

\vspace{10pt}
In summary, the mass-loading factors from simulations span two orders of magnitude and, while they overlap with observations, they are generally offset to higher values for systems with comparable circular velocities. When viewed as a function of stellar mass, empirical mass-loading factors have a flat distribution while theoretical values show a steep dependency with the lowest mass galaxies exhibiting the highest mass-loading factors. The difference in slopes means that  the observations and simulations are in general agreement with each other for galaxies with M$_*\sim10^9$ \msun, but diverge as the predicted mass-loading factors sharply increase for lower mass galaxies. 

In contrast to the weak dependence on stellar mass, the empirically-derived mass-loading factors correlate with the degree to which star formation is concentrated in the galaxies. Thus, unsurprisingly, Figures~\ref{fig:mass_load_vcirc} \& \ref{fig:mass_load_mstar} show a general trend that predicted mass-loading factors are higher when the resolution of the simulations is lower. At lower resolutions, the star particles are more massive, and, thus, feedback from the particles occurs all at once and in one location. This overly concentrated injection of energy and mass in both space and time results in a wind that is too strong. SNe occuring over a larger area and a more realistic time frame, which is only possible in higher resolution simulations, drive weaker winds. As previously noted in \citet{Heckman2015}, simulations are able to better reproduce galaxies with strong winds, but prescriptions severely disagree with weak winds.

\subsection{Likelihood of Escape to the IGM}
The final impact of the detected winds and fountain flows depends on whether or not the gas is lost from the host galaxy into the IGM. Using the values of M$_{\rm halo}$ and the virial radii from above, we calculate the escape velocity from the galaxies using V$_{\rm esc} = [2 \cdot$ G $\cdot$ M$_{\rm halo}/$R$_{\rm vir}]^{1/2}$ and list the values for the sample in Table~\ref{tab:mass_load}. Focusing on the five galaxies with winds, the escape velocities are comparable to or greater than the upper end of the 25$-$50 km s$^{-1}$ wind velocities used in the mass loading calculations and consistent with spectroscopically obtained velocities of the outflowing ionized gas. Given these velocities and the limited spatial extent of the ionized gas, it is probable that much of the detected material will remain in the CGM around the galaxies. 

The high likelihood that the gas will remain bound is consistent with results from numerical calculations and simulations. Looking specifically at dwarf galaxies, \citet{Silich1998} found that superbubbles that break through the gaseous disks, with dimensions smaller than a few kiloparsec and velocities of order 50 km s$^{-1}$, fragment into a freely flowing wind, but are dispersed by turbulent motions in the galactic halos. The morphology, sizes, and velocities described by \citet{Silich1998} are similar to what we observe in a number of galaxies, including UGC~4483, NGC~4163, UGC~6456, NGC4190, NGC~1569, and NGC~4449). 

\citet{Christensen2016}, with mass-loading factors that are some of the closest to our measurements at the upper mass range of our sample, note that feedback-driven winds are effective at driving gas out of the disks of galaxies but at lower velocities, and therefore with a lower probability of escaping the halo. The FIRE simulations show the majority of ejected material remains in the halo and is recycled to the galaxy on timescales of a few 100 Myr \citep{Angles-Alcazar2016}. In contrast, \citet{Hu2019} shows wind velocities increase in the halo due to ram pressure from more recent ejecta. As a result, more material escapes from the gravitational potential of the galaxy.  

\subsection{What About the Metals?}
The derived mass-loading factors describe the amount of {\it gas mass} in the outflows from the galaxies. These values do not directly translate to the amount of {\it metals} that may also be expelled via stellar-feedback processes. As discussed briefly in \S\ref{sec:complexity}, metals are thought to be preferentially ejected in the faster-moving hot-phase ($T\sim10^6$ K) of the winds. The hot-phase of galactic winds is even lower density than the warm winds that are the focus of the present work, entraining a much smaller amount of gas \citep[simulations suggest only $\sim$10\% of the mass in the hot phase;][]{Kim2017}. X-ray observations of this phase in nearby low-mass galaxies confirm that the gas is diffuse \citep{Heckman1995, Summers2003, Summers2004, Hartwell2004, Ott2005a, McQuinn2018}. 

Yet, despite the low gas-densities, the hot-phase wind is thought to be highly metal-enriched \citep{Dalcanton2007} and, thus, can have a profound impact on the metal content of low-mass galaxies. Metal-enriched outflows are believed to be a major driver of the mass-metallicity relation which shows a tight correlation between low-mass galaxies and low gas-phase oxygen abundances \citep[e.g.,][]{Berg2012}. Supporting this framework, recent work has measured the fraction of metals retained in the extremely low-mass (M$_* = 5.6\times10^5$ \msun) galaxy Leo~P to be only 5$\pm2$\% of the metals produced by nucleosynthesis \citep{McQuinn2015f}. As Leo~P is outside of a group environment and isolated, the interpretation is that the metals have been expelled by secular stellar feedback processes. There are no available measurements of winds or mass outflow rates in Leo~P, but, considering our low mass-loading factors, the galaxy is unlikely to have expelled 95\% of its gas mass.

Thus, despite low mass-loading factors, observational evidence suggests signficant metal loss in low-mass galaxies.
Given the qualitatively different velocities, densities, and compositions of the hot- and warm-phase winds, it is clearly not appropriate to extrapolate metal loss from galaxies based on gas mass loss. Investigating such a possible correlation requires independent measurements of the mass-loading of winds and the metal loss from winds in the same galaxies. On-going work to measure the metal retention fraction in a larger sample of nearby, gas-rich dwarf galaxies, including the STARBIRDS galaxies, is underway (K.~B.~W. McQuinn et al. in preparation). This future analysis will compare the {\it gas mass lost} with the {\it metals lost} for the current sample.

\section{Conclusions}\label{sec:conclusions}
Observationally constraining the impact of stellar feedback and the prevalence of galactic winds in low-mass galaxies is fundamental to understanding the baryon cycle in galaxies. The recent implementation of additional stellar feedback mechanisms in hydrodynamical simulations has shown that galactic winds are essential for reproducing the observed properties of galaxies and may be key to solving discrepancies between predictions from simulations assuming a $\Lambda$CDM cosmology and observations. Most of the mass in the winds is predicted to be in the warm phase \citep{Hopkins2012}, with as much as 60\% of baryons expelled from dwarfs and the majority of the mass reaching the IGM \citep{Bertone2007}. Yet, there are relatively few detections of strong winds in low-mass galaxies that reach the CGM in the present-day Universe. 

We have searched for the presence or absence of ionized winds in 12 starburst dwarf galaxies from STARBIRDS, using very deep \ha\ narrow-band imaging. The galaxies are all gas-rich, low-mass, with elevated levels of recent star formation activity, making the sample biased towards detecting stellar-feedback driven galactic mass-loss. We find evidence of galactic winds in five galaxies, fountain flows in six systems, and no clear evidence of outflows in the remaining galaxy. The winds are preferentially found in galaxies with centrally concentrated star formation whereas fountains occur in systems with spatially distributed star formation. 

In the three galaxies with complementary X-ray data from $Chandra$ (NGC~4214, NGC~1569, NGC~4449), the projected extent of the hot-phase gas is comparable to that of the ionized gas. The general expectation is that hot-phase outflows are moving faster than the warm-phase ISM and can reach farther distances, and, thus, are the mechanism for transporting metals from a galaxy to the IGM. At the sensitivity levels of the data sets, this is not the case for these three galaxies. Understanding whether this is a completeness effect or if the hot-phase outflows in these galaxies are more spatially confined will require the next generation of X-ray telescope with significantly higher sensitivity. 

Despite the prevalence of occurrence of winds and outflows, the projected spatial extent of the warm gas is modest in all cases (i.e., within 0.1 R$_{\rm vir}$) which is in conflict with the larger spatial extent of ionized winds predicted by many cosmological simulations. The physical distances may be somewhat larger due to projection effects, but given the highly inclined viewing angles of the majority of the sample, the winds would have to be preferentially mis-aligned with the minor-axes of the galaxies to be reaching the CGM as defined by simulations. The small-scale fountain flows extend just past the \HI\ disks. The orientation of the fountains depends more heavily on local ISM density variations and, thus, may have a wider range of physical sizes. Regardless, the small, local nature of these fountains suggest they are venting gas that will be recycled on short timescales rather than representing a significant expulsion of the ISM from the host galaxies. 

The mass-loading factors of the winds are modest, ranging from 0.2$-$7. Previously reported mass-loading factors include values from $\sim0.5$ to 5 for low-mass galaxies with similarly active star formation \citep{Martin1999}, and $6-19$ for extreme starbursts and star formation events driven by mergers in the local universe \citep{Heckman2015, Chisholm2017}. 

When viewed as a function of circular velocity, the highest of our mass-loading factors overlap with the lower values predicted by two simulations  \citep[e.g.,][]{Christensen2016, Hu2019}, but are one to two orders of magnitude lower than predictions from Illustris, FIRE, and TNG predictions \citep[e.g.,][]{Vogelsberger2013, Ford2014, Hopkins2012, Muratov2015, Nelson2019}. When viewed as a function of the more robustly measured quantity stellar mass, the mass-loading factors from simulations tend to agree for the higher mass galaxies (M$_* \gtsimeq 10^9$ \msun), but remain significantly higher than our values for lower mass galaxies. The exception is the result from a high-resolution simulation of an isolated dwarf with a weak wind whose mass-loading factor is consistent with our values for a M$_* \sim 10^7$ \msun\ galaxy \citep{Hu2019}.

The mass-loading factors scale as $\eta \propto$ V$_{\rm circ}^{-0.96}$, consistent with a momentum driven wind, and, contrary to simulations, have a flat distribution with stellar mass, $\eta \propto$ M$_*^{0.04}$. Instead of the gravitational potential being the dominant determinant in mass-loading factors, we suggest galaxies whose star formation in centrally concentrated will have higher mass-loading factors and galaxies with star formation distributed across the disk will have lower values.

The findings of warm winds and fountain flows in STARBIRDS galaxies support the general result that star formation in dwarfs can drive mass-loss from galaxies. However, the observed limited spatial extent and low mass-loading factors conflicts with the high mass-loading factors in winds that reach large distances predicted by many simulations. Rather than being expelled to the IGM, it is more likely that mass lost via winds and outflows will be recycled into the host systems.

Historically, strong galactic winds in dwarf galaxies have been seen as a convenient mechanism to explain their observed low metallicities, low gas mass fractions, low baryonic fractions \citep[e.g.,][]{Governato2007, Hopkins2012, Shen2014, Ma2016}, and the conversion from gas-rich (i.e., dIrr) to gas-poor (i.e., dSph) galaxies. The implementation of strong feedback and outflows in simulations also eases discrepancies with $\Lambda$CDM predictions. However, the simplest versions of these wind models often are in direct conflict with other observables \citep[e.g.,][]{Dekel1986}.  Clearly, the most extreme starbursts (i.e., centralized events with spatially and temporally coincident star formation) can result in galactic winds capable of significantly altering the global properties of a dwarf galaxy \citep[cf.][]{Veilleux2005, Heckman2015}, but the majority of star formation events in dwarf galaxies (at least at low redshift where the observations are best) are much lower impact affairs. Even in the sample studied here, which is specifically chosen based on strong star formation events, we see that the evidence of material leaving the gravitational potential well of the galaxy is scarce. These observations indicate that consideration of alternative solutions to explain some of the common properties of the dwarf galaxy population is very appropriate.

\section*{Acknowledgments}
K.B.W.M. would like to thank Eve Ostriker for providing insight on stellar feedback, the development of bubbles and superbubbles, and the mass-loading of winds, and Phil Hopkins for discussions of the implementation of feedback in the FIRE hydrodynamical simulations. We would also like to thank the referee for helpful comments that improved the manuscript. Observations reported here were obtained with the 2.3m Bok telescope operated by the Steward Observatory of University of Arizona. The Bok telescope observations were obtained as 
part of the University of Minnesota's guaranteed time on Steward Observatory facilities through membership in the  Research Corporation and its support for the Large Binocular Telescope. This material is based upon work supported by the National Science Foundation under Grant Numbers 1806926 and 1806522. Any opinions, findings, and conclusions or recommendations expressed in this material are those of the authors and do not necessarily reflect the views of the National Science Foundation. This research made use of NASA's Astrophysical Data System and the NASA/IPAC Extragalactic Database (NED) which is operated by the Jet Propulsion Laboratory, California Institute of Technology, under contract with the National Aeronautics and Space Administration. \\

This work is dedicated in loving memory to Reb Wingfield.\\

\facility{KPNO Mayall 4m telescope; Steward Bok 2.3m telescope; the WIYN Observatory, the Very Large Array.}

\renewcommand\bibname{{References}}
\bibliography{../../bibliography}

\appendix 
\section{Galaxy Atlas: Comparison of the Ionized Hydrogen with the Stars and Neutral Hydrogen}\label{sec:atlas}
\begin{figure*}
\includegraphics[width=\textwidth]{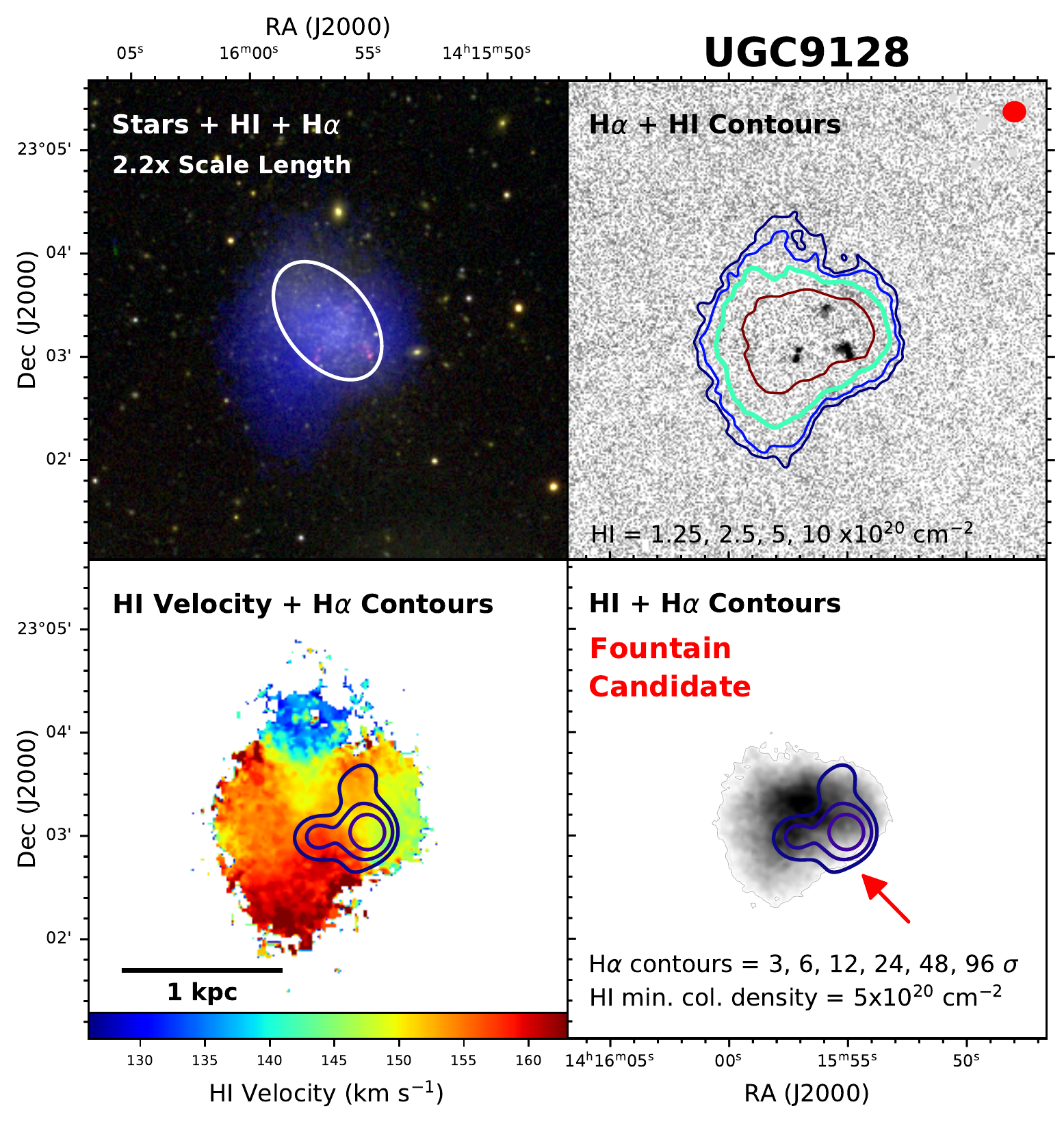}
\caption{UGC~9128. Clockwise from the top left: combined \HI, B-band, \ha, R-band images of the galaxy, \ha\ map with \HI\ contours, \HI\ map with \ha\ contours, \HI\ velocity field with \ha\ contours. See description at the beginning of Appendix~\ref{sec:atlas} for details.}
\label{fig:ugc9128}
\end{figure*}

\begin{figure*}
\includegraphics[width=\textwidth]{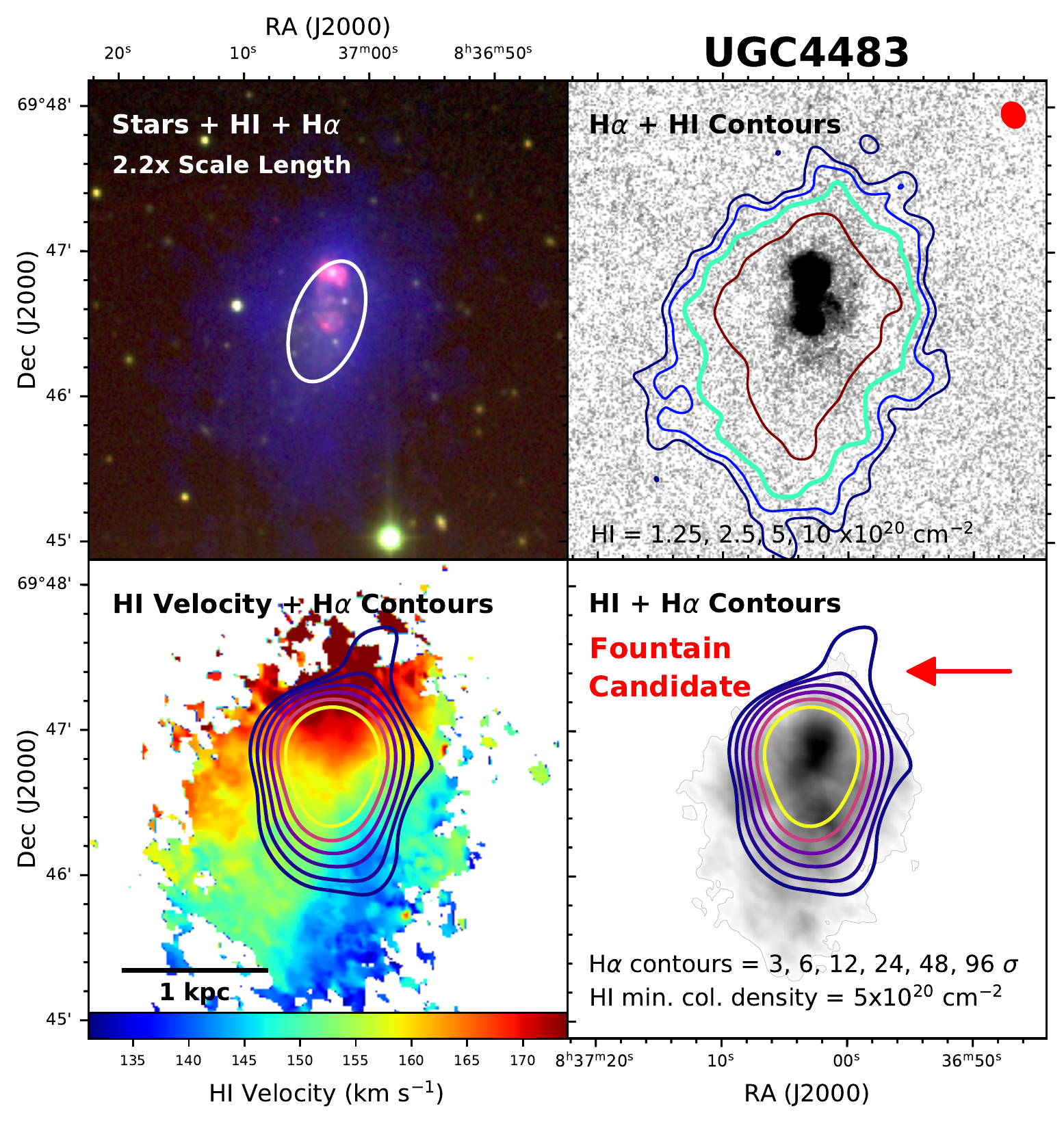}
\caption{UGC~4483. Clockwise from the top left: combined \HI, B-band, \ha, R-band images of the galaxy, \ha\ map with \HI\ contours, \HI\ map with \ha\ contours, \HI\ velocity field with \ha\ contours. See description at the beginning of Appendix~\ref{sec:atlas} for details.}
\label{fig:ugc4483}
\end{figure*}

The galaxy atlas shown in Figures~\ref{fig:ugc9128}$-$\ref{fig:ngc4449} is comprised of four panel montages for each galaxy and compares the \ha\ emission with the stellar disk and the neutral hydrogen in each system, with galaxies ordered by $M_B$ magnitude from faintest to brightest. We describe the content of each panel here and include detailed discussions of each system below.

Each the four panels in Figures~\ref{fig:ugc9128}$-$\ref{fig:ngc4449} covers the same field of view for an individual galaxy, namely $3\times$ the major-axis diameter. Physical scales are shown in the lower left panel based on the distances listed in Table~\ref{tab:galaxies}. 

\noindent\textbf{Top Left:} Color images that highlight the stellar, ionized gas, and neutral gas components. We combined the \HI\ maps and the B-band images (blue), R-band (green) and \ha\ narrow band before continuum subtraction (red). The white ellipses encompass 2.2 $\times$ scale length of each system using the geometry in Table~\ref{tab:galaxies}. 

\noindent\textbf{Top Right:} Gray-scale image of the \ha\ continuum subtracted maps with a power stretch and \HI\ contours corresponding to column densities of 1.25, 2.5, 5, 10 $\times10^{20}$ cm$^{-2}$. The \HI\ column density of $5\times10^{20}$ cm$^{-2}$ used to define the edge of the denser ISM where ``break though'' of ionized gas occurs is plotted as a thicker cyan contour line. The \HI\ beam size is shown as a solid red ellipse in the top right corner.

\noindent\textbf{Lower Right:} \HI\ column density map truncated at the slightly higher column density of $5\times10^{20}$ cm$^{-2}$ that we use to define the edge of the denser ISM. \ha\ contours smoothed with a 9\arcsec\ gaussian are overlaid and correspond to surface brightness levels of 3, 6, 12, 24, 48, and 96 $\sigma$. Galaxies with winds or fountain candidates are labelled; arrows identify where the ionized gas is breaking through from neutral hydrogen forming the fountain flows.

\noindent\textbf{Lower Left:} \HI\ velocity field out to the full extent of the detected \HI\ (i.e., below the $1.25\times10^{20}$ cm$^{-2}$ column density shown in the top right panel), with the same \ha\ contours shown in the lower right panels. The patchiness of the low density \HI\ is readily apparent in the outer edges of the \HI\ velocity maps. 

\subsection{UGC~9128} In Figure~\ref{fig:ugc9128}, the \HII\ regions are very faint in this galaxy. There are four small regions of \ha\ emission, three of which have multiple, resolved \HII\ regions seen more easily in the stretch used in Figure~\ref{fig:full_fova} above. The \ha\ emission is confined to these areas with little identifiable diffuse emission. The region to the southwest reaches outside the edge of the \HI\ disk, marked with an arrow. The \HI\ moment 2 maps (not shown) reveal that this overall region has the greatest velocity dispersion of $\sim15$ km s$^{-1}$. Based on the resolved stellar populations, UGC~9128 has had a highly variable star formation activity over the most recent few 100 Myr. The galaxy is a post-starburst galaxy whose elevated star formation activity ended $\sim15$ Myr ago. 

\subsection{UGC~4483} 
In Figure~\ref{fig:ugc4483}, UGC~4483 has a comet-like morphology and the \ha\ emission is dominated by two large, high surface brightness bubble structures. The largest bubble is located in the north coincident with the compact ``head'' of the comet structure \citep{Izotov2002}; the diffuse \ha\ emission traces a free-flowing ionized gas from both the top and the bottom right of the bubble structure. Diffuse ionized gas is detected between the two bubbles, including an arc feature that may be a ruptured bubble inside the galaxy (see Figure~\ref{fig:full_fova}). A previous \ha\ map of UGC~4483 shows the two high surface brightness knots of emission, connected by a lower surface brightness bridge \citep{GilDePaz2003}, however, these images do not detect the extension in diffuse \ha\ emission to the south seen in Figure~\ref{fig:ugc4483}. The ionized gas appears to break through the \HI\ disk in the north with a fountain indicated by the arrow. From the star formation history and BHeB stars, the starburst is spatially distributed with the young stars covering 71\% of the area of the old stars \citep{McQuinn2012a}. The \ha\ emission is coincident with a population of BHeB stars less than 100 Myr old. 

\begin{figure*}
\includegraphics[width=\textwidth]{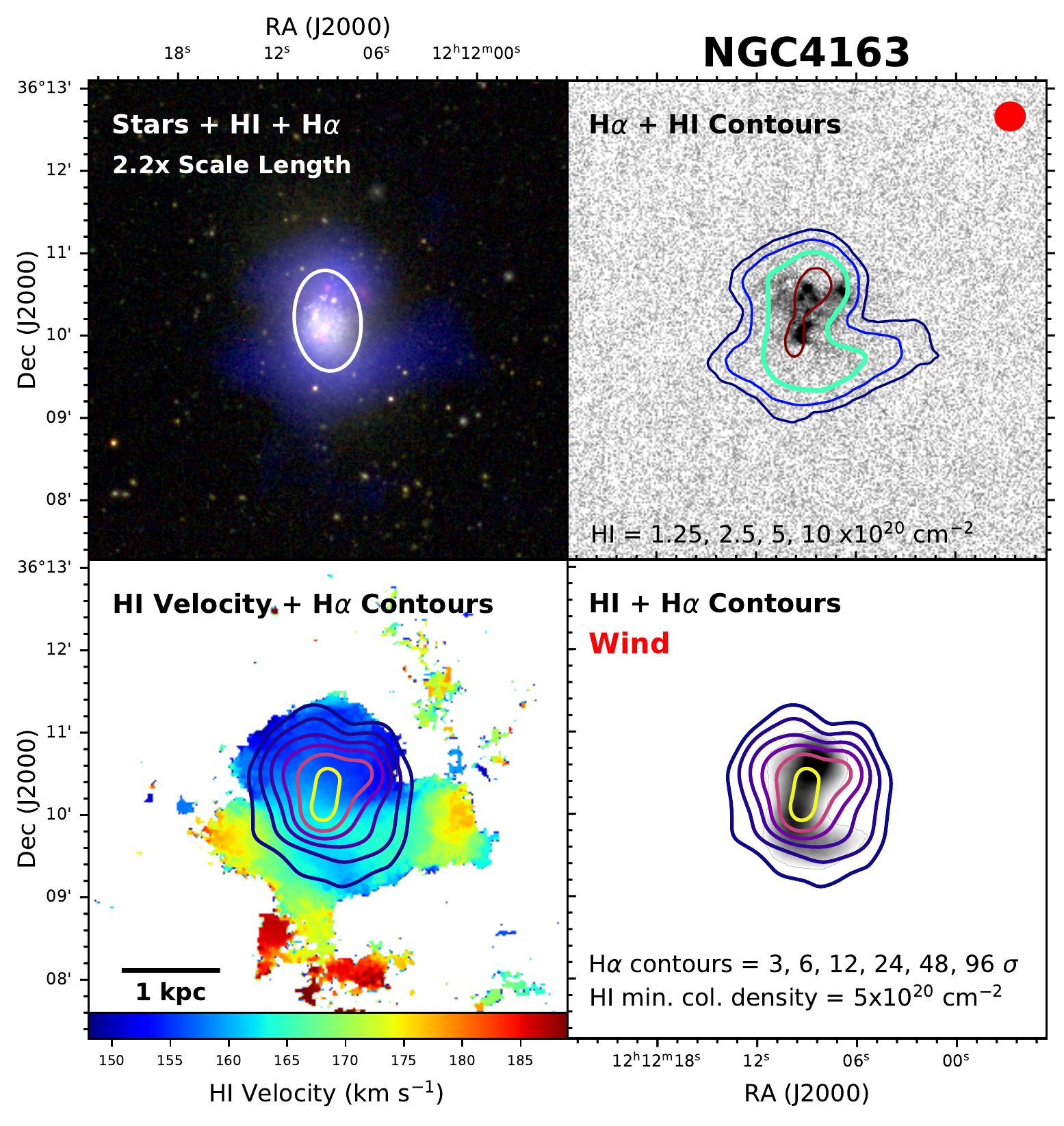}
\caption{NGC~4163. Clockwise from the top left: combined \HI, B-band, \ha, R-band images of the galaxy, \ha\ map with \HI\ contours, \HI\ map with \ha\ contours, \HI\ velocity field with \ha\ contours. See description at the beginning of Appendix~\ref{sec:atlas} for details.}
\label{fig:ngc4163}
\end{figure*}

\subsection{NGC~4163} 
In Figure~\ref{fig:ngc4163}, there is a clear detection of galactic wind in NGC~4163. We find ionized gas around the galaxy in nearly all directions, reaching the farthest distance in the west. Traced by the \ha\ emission, this gas is well outside the \HI\ disk with a morphology that suggests a free-flowing wind. The projected physical scale of the outflowing gas is much less than a tenth of the virial radius. Interestingly, the \HI\ disk is irregular and has a two opposing tails in the outskirts consistent with an interaction. There are also a number of prominent \HII\ regions and large \ha\ bubbles in the central region of the system. The central \ha\ structures are located in the same area as the young BHeB stars \citep{McQuinn2012a}, but the northern bubbles are outside of the highly concentrated, star-forming center; this is even evident from the 3-color image seen in the upper left. NGC~4163 is a post-starburst galaxy with the elevated levels of star formation ending $\sim100$ Myr ago \citep{McQuinn2010b}. 

\begin{figure*}
\includegraphics[width=\textwidth]{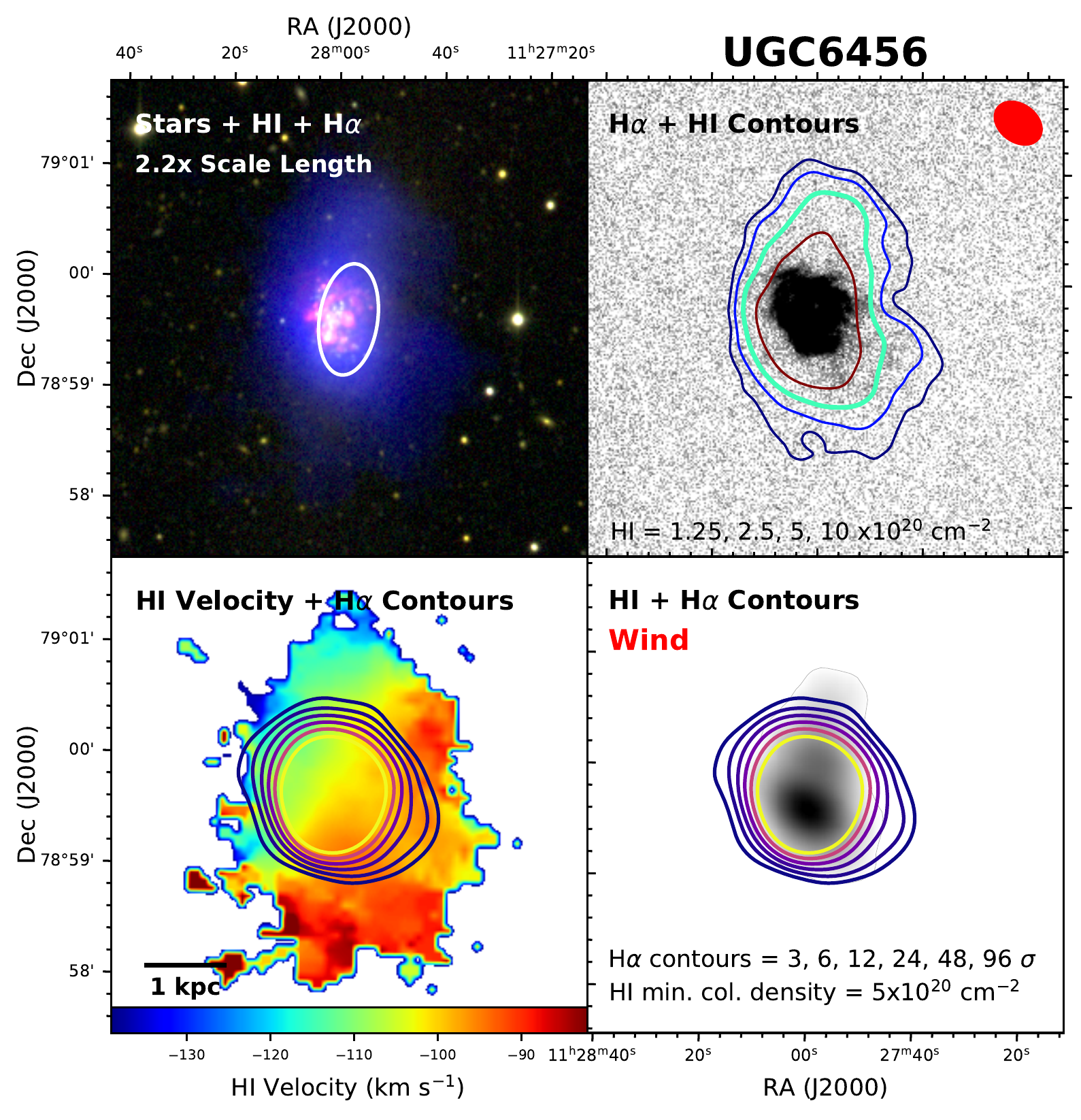}
\caption{UGC~6456. Clockwise from the top left: combined \HI, B-band, \ha, R-band images of the galaxy, \ha\ map with \HI\ contours, \HI\ map with \ha\ contours, \HI\ velocity field with \ha\ contours. See description at the beginning of Appendix~\ref{sec:atlas} for details.}
\label{fig:ugc6456}
\end{figure*}

\begin{figure*}
\includegraphics[width=\textwidth]{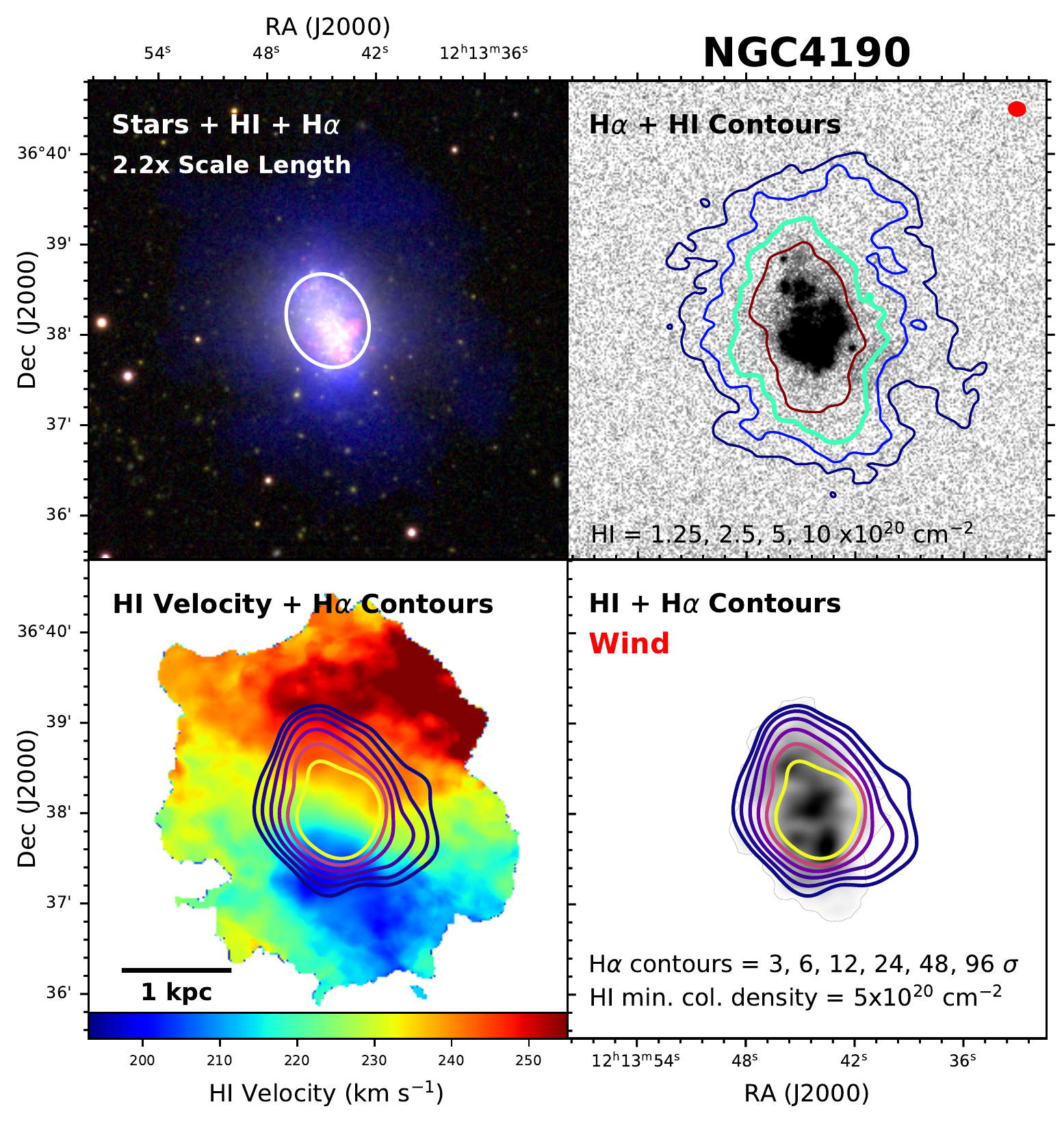}
\caption{NGC~4190. Clockwise from the top left: combined \HI, B-band, \ha, R-band images of the galaxy, \ha\ map with \HI\ contours, \HI\ map with \ha\ contours, \HI\ velocity field with \ha\ contours. See description at the beginning of Appendix~\ref{sec:atlas} for details.}\label{fig:ngc4190}
\end{figure*}

\subsection{UGC~6456} 
In Figure~\ref{fig:ugc6456}, the central $\sim1$ kpc region in UGC~6456 is dominated by higher surface brightness \ha\ emission that resolved into a number of bubbles, some of which appear to be superimposed. In this centrally concentrated starburst, the star formation activity has increased in the last 50 Myr, and the locations of the largest bubbles are also the sites of young BHeB stars  \citep{McQuinn2012a}. The diffuse \ha\ emission lies along the northeast / southwest minor axis, perpendicular to the main stellar and gas disks. A previous \ha\ map of UGC~6456 shows similar high surface brightness emission, although the low surface brightness emission is not as extended \citep{GilDePaz2003}. On the eastern and western edges of the \HI\ map, the \ha\ emission reaches past the gas and also into an area with very little detected neutral hydrogen. From the \HI\ velocity map, the gas is rotating around the minor axis and the peaks in the \HI\ velocity dispersion, reaching 15 km s$^{-1}$, are coincident with the extended ionized gas. UGC~6456 was observed previously, but no outflow was detected, presumably due to detection limits \citep{Martin1998, Lozinskaya2006, Arkhipova2007}. 

\begin{figure*}
\includegraphics[width=0.98\textwidth]{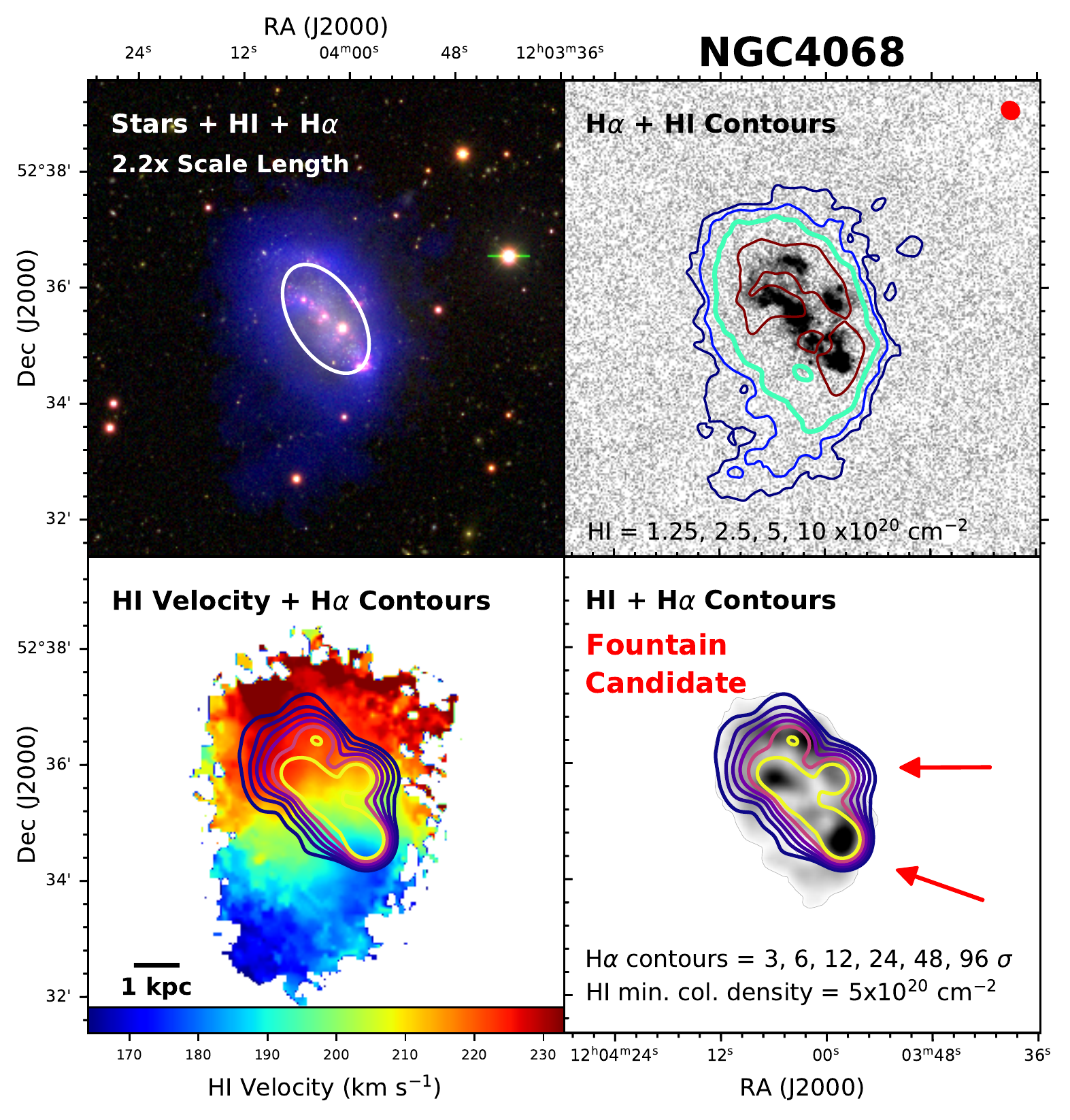}
\caption{NGC~4068. Clockwise from the top left: combined \HI, B-band, \ha, R-band images of the galaxy, \ha\ map with \HI\ contours, \HI\ map with \ha\ contours, \HI\ velocity field with \ha\ contours. See description at the beginning of Appendix~\ref{sec:atlas} for details.}
\label{fig:ngc4068}
\end{figure*}

\begin{figure*}
\includegraphics[width=\textwidth]{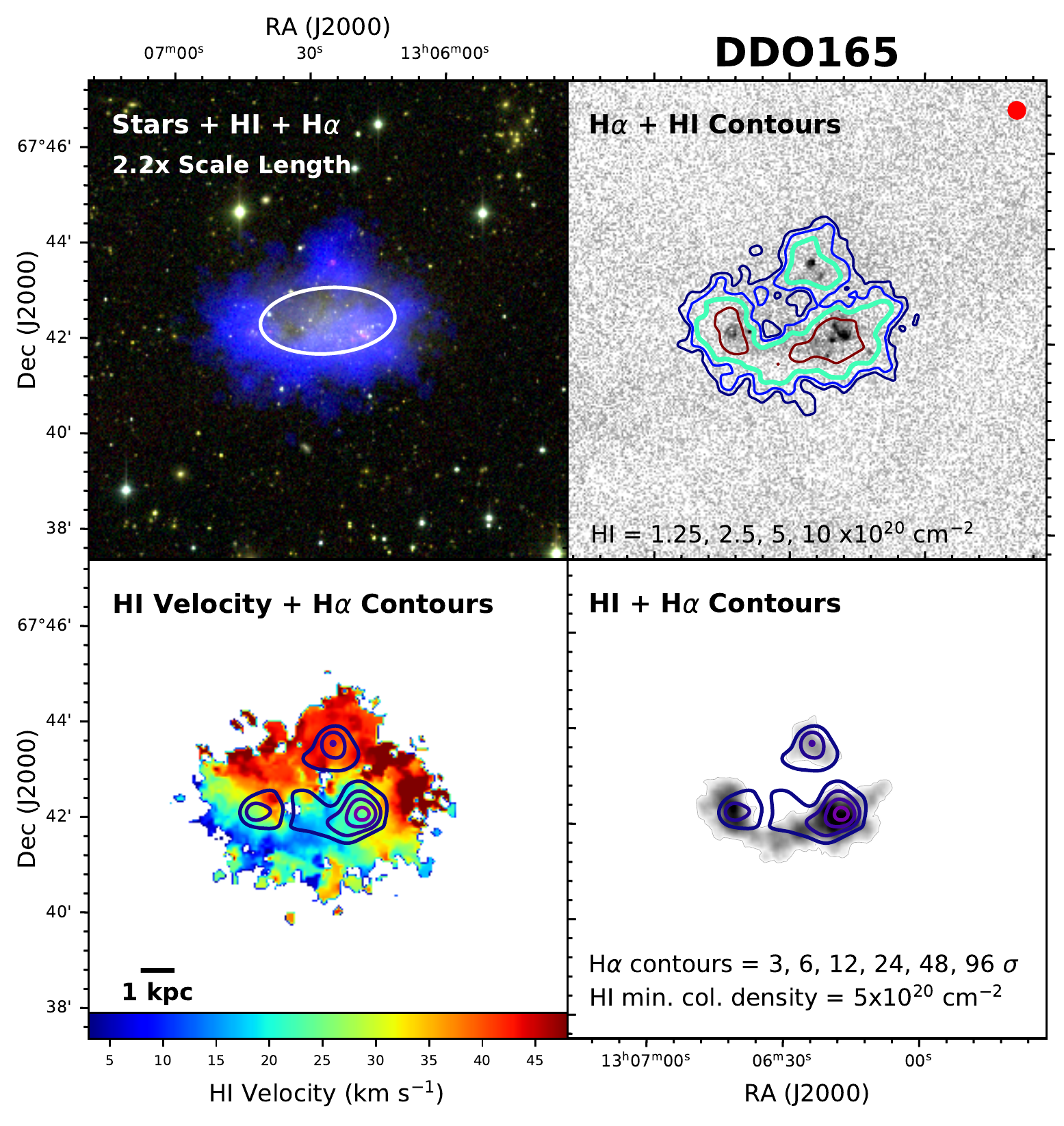}
\caption{DDO~165. Clockwise from the top left: combined \HI, B-band, \ha, R-band images of the galaxy, \ha\ map with \HI\ contours, \HI\ map with \ha\ contours, \HI\ velocity field with \ha\ contours. See description at the beginning of Appendix~\ref{sec:atlas} for details.}
\label{fig:ddo165}
\end{figure*}

\subsection{NGC~4190} 
In Figure~\ref{fig:ngc4190}, NGC~4190 has a warm wind extending roughly spherically around the galaxy with the strongest detection in two approximately orthogonal directions. The ionized gas reaches the edge of both the eastern and western sides of the \HI\ disk; this extended low surface brightness gas does not have a sharp discontinuity in the unsmoothed images (see Figure~\ref{fig:full_fovb}). There are also a number of prominent \HII\ regions and large \ha\ bubbles in the southern region of the system. 

\begin{figure*}
\includegraphics[width=\textwidth]{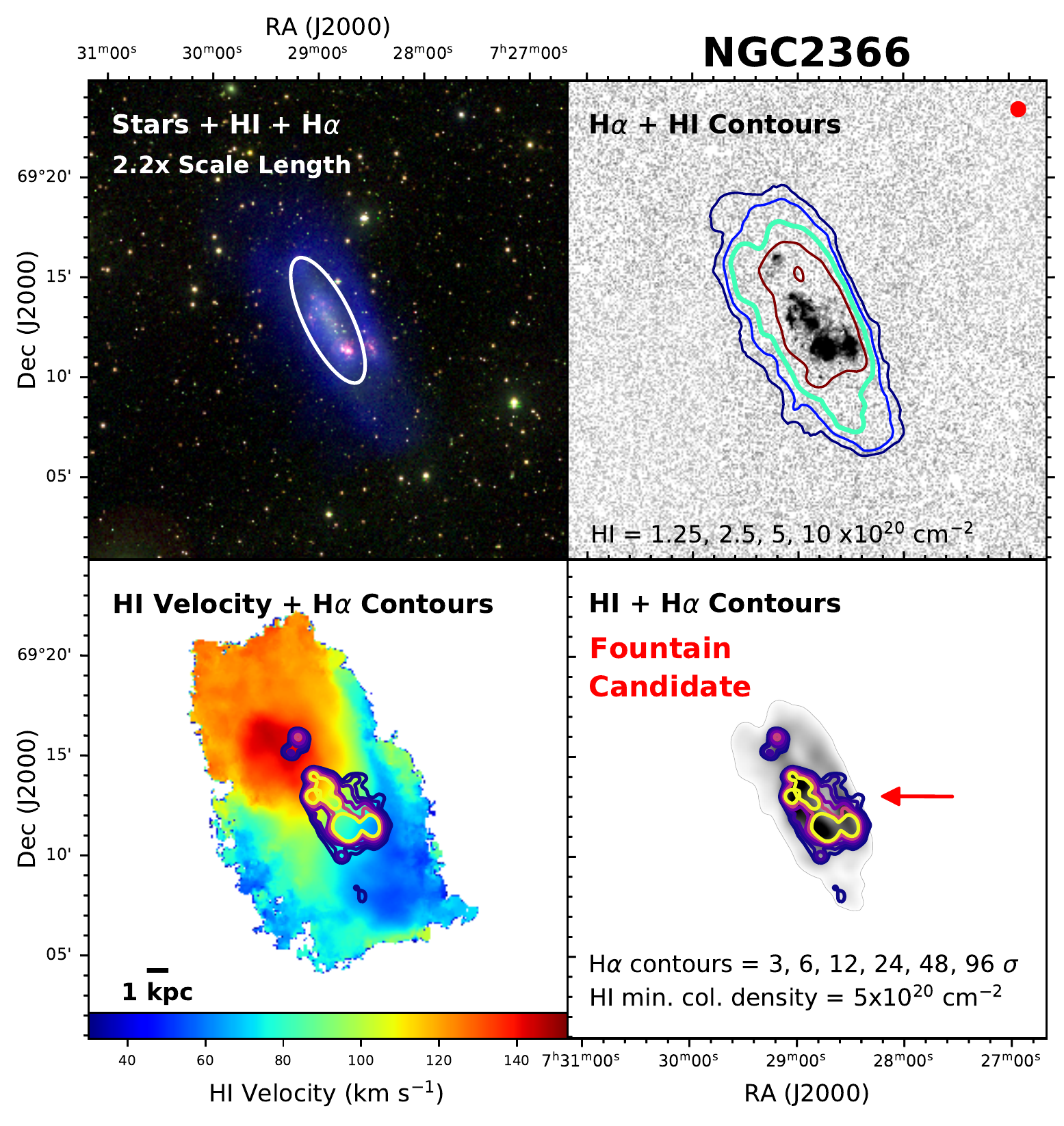}
\caption{NGC~2366. Clockwise from the top left: combined \HI, B-band, \ha, R-band images of the galaxy, \ha\ map with \HI\ contours, \HI\ map with \ha\ contours, \HI\ velocity field with \ha\ contours. See description at the beginning of Appendix~\ref{sec:atlas} for details.}\label{fig:ngc2366}
\end{figure*}

\begin{figure*}
\includegraphics[width=\textwidth]{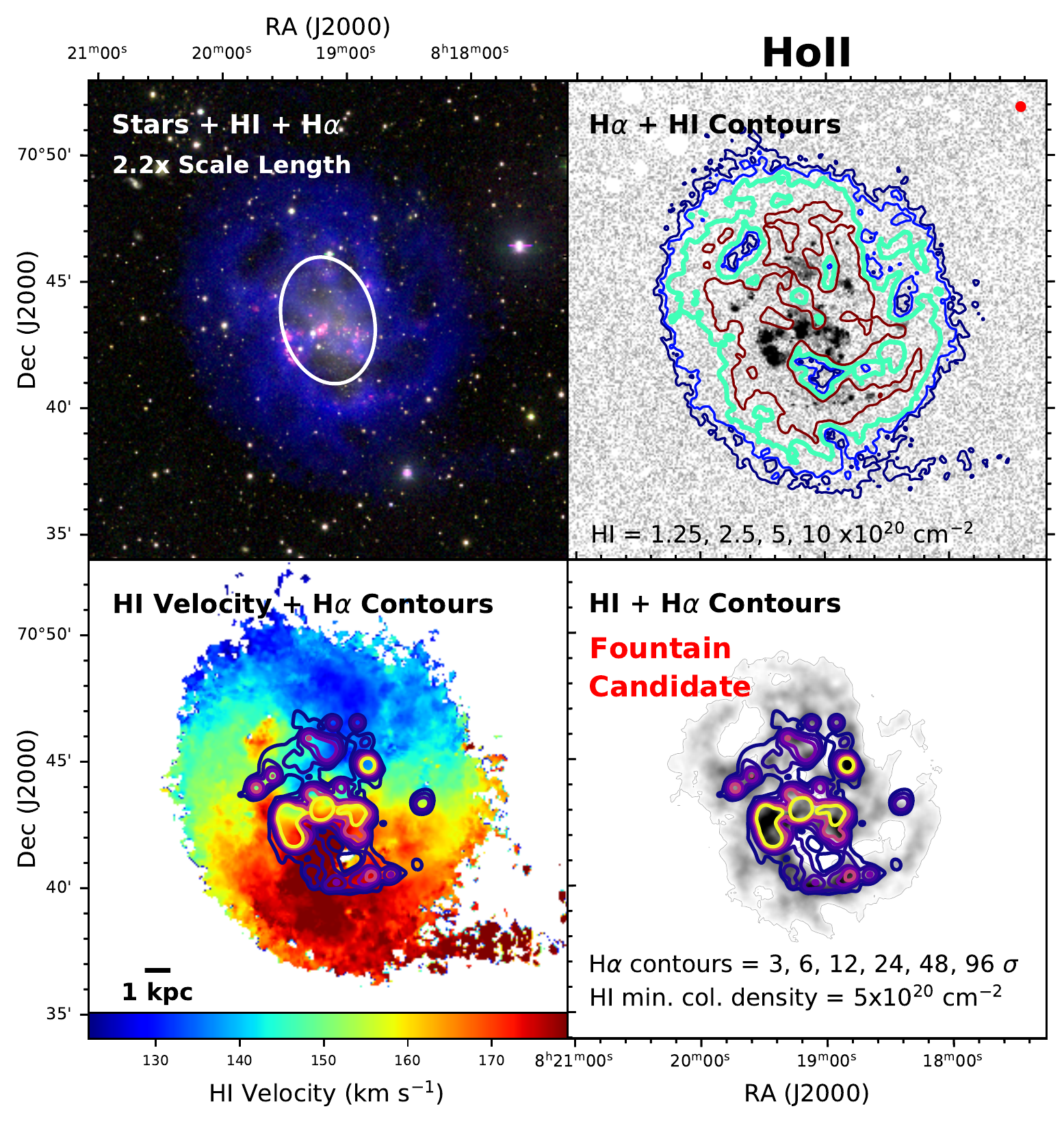}
\caption{Holmberg~II. Clockwise from the top left: combined \HI, B-band, \ha, R-band images of the galaxy, \ha\ map with \HI\ contours, \HI\ map with \ha\ contours, \HI\ velocity field with \ha\ contours. See description at the beginning of Appendix~\ref{sec:atlas} for details.}
\label{fig:hoii}
\end{figure*}

\subsection{NGC~4068}
In Figure~\ref{fig:ngc4068}, NGC~4068 has high surface brightness \ha\ emission tracing the location of recent star formation. This emission is spatially coincident with the peaks in the \HI\ density along the plane of the disk and is dominated by bubble-like structures typical of \HII\ regions. Based on the relative concentration of BHeB and RGB stars (see Table~\ref{tab:star_formation}), the starburst is NGC~4068 is widely spatially distributed, in agreement with the distribution of \HII\ regions. Defined structures in the \ha\ emission are also seen including filaments and arcs of material, particularly in the north. NGC~4068 also has diffuse \ha\ detected throughout the large, lower-density neutral gas cavities of the galaxy. On the west side of the disk, the \ha\ emission reaches the edge of the denser ISM in two locations. The low surface brightness ionized gas does not have a sharp edge to its morphology (i.e., there does not appear to be a pressure-bound bubble or shell). To the northeast, a second extension of low surface brightness \ha\ emission is detected with the shape of an arc or expanding edge of a superbubble, with tendrils of ionized gas extending to the north outside of the structure (see also Figure~\ref{fig:full_fovb}. The fountain features are denoted by arrows in Figure~\ref{fig:ngc4068}. The \HI\ has a low column density extension to the south where no \ha\ is detected. 

\begin{figure*}
\includegraphics[width=\textwidth]{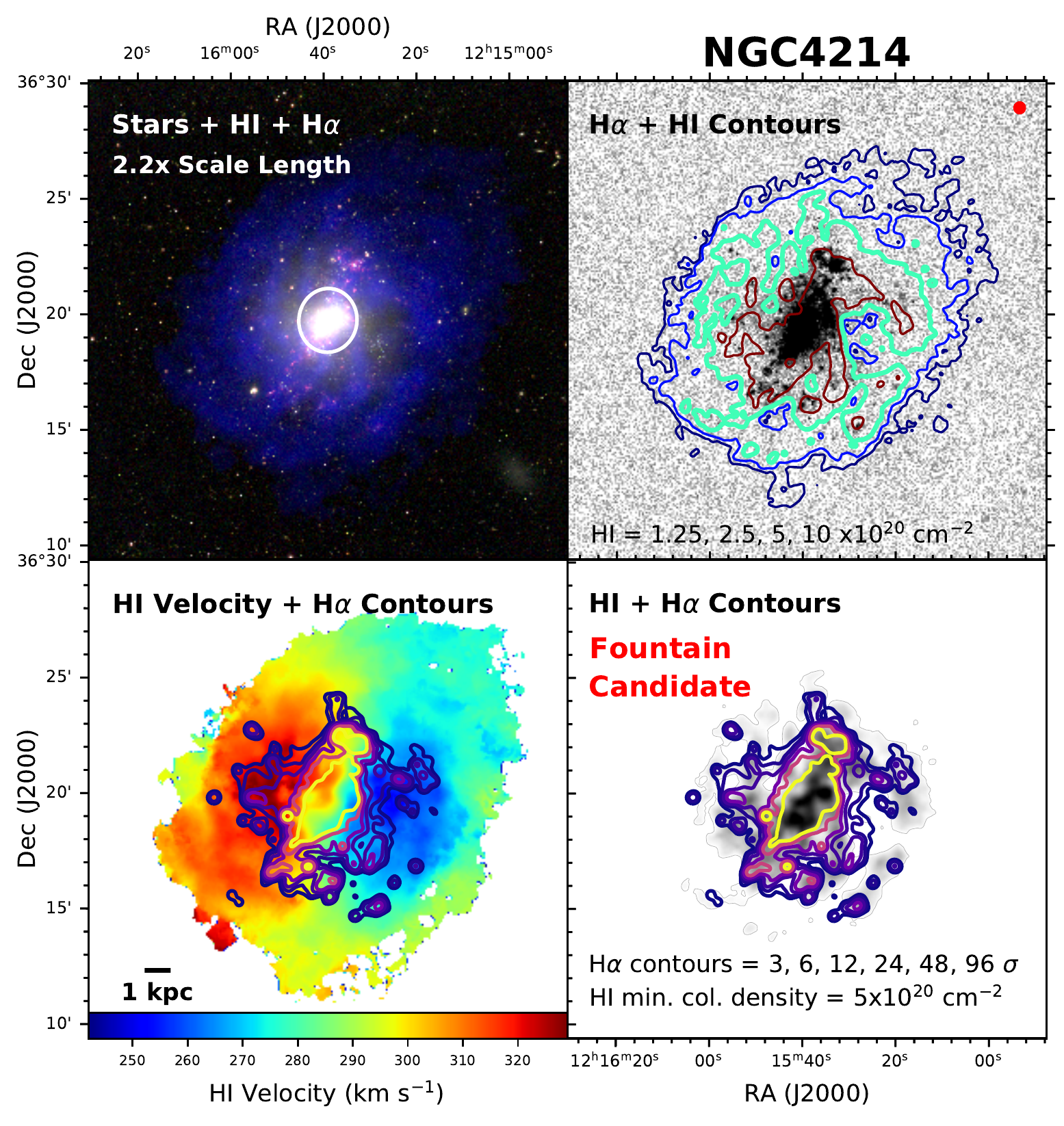}
\caption{NGC~4214. Clockwise from the top left: combined \HI, B-band, \ha, R-band images of the galaxy, \ha\ map with \HI\ contours, \HI\ map with \ha\ contours, \HI\ velocity field with \ha\ contours. See description at the beginning of Appendix~\ref{sec:atlas} for details.}
\label{fig:ngc4214}
\end{figure*}

\subsection{DDO~165} 
From Figure~\ref{fig:ddo165}, the \HI\ map shows the neutral gas has a central, low-density hole surrounding by a broken, ring-like structure of material. A previous study of the \HI\ concluded that the galaxy likely experienced a blow-out of the ISM in the recent past \citep{Cannon2011a}. The majority of \ha\ emission is coincident with this ring-like \HI\ distribution, and with a similar spatial extent. The higher surface brightness \ha\ emission regions have circular, bubble-like appearances and coincide with higher \HI\ densities. The majority of the detected \ha\ emission lies in the southern part of the galaxy where most of the recent star formation has occurred based on the identified location of the young BHeB stars \citep{McQuinn2012a}. The center of the ring-like feature has very low-surface brightness \ha\ detected. The galaxy is located in a region of Galactic cirrus which may be confused with the lowest \ha\ contour level shown.

\begin{figure*}
\includegraphics[width=\textwidth]{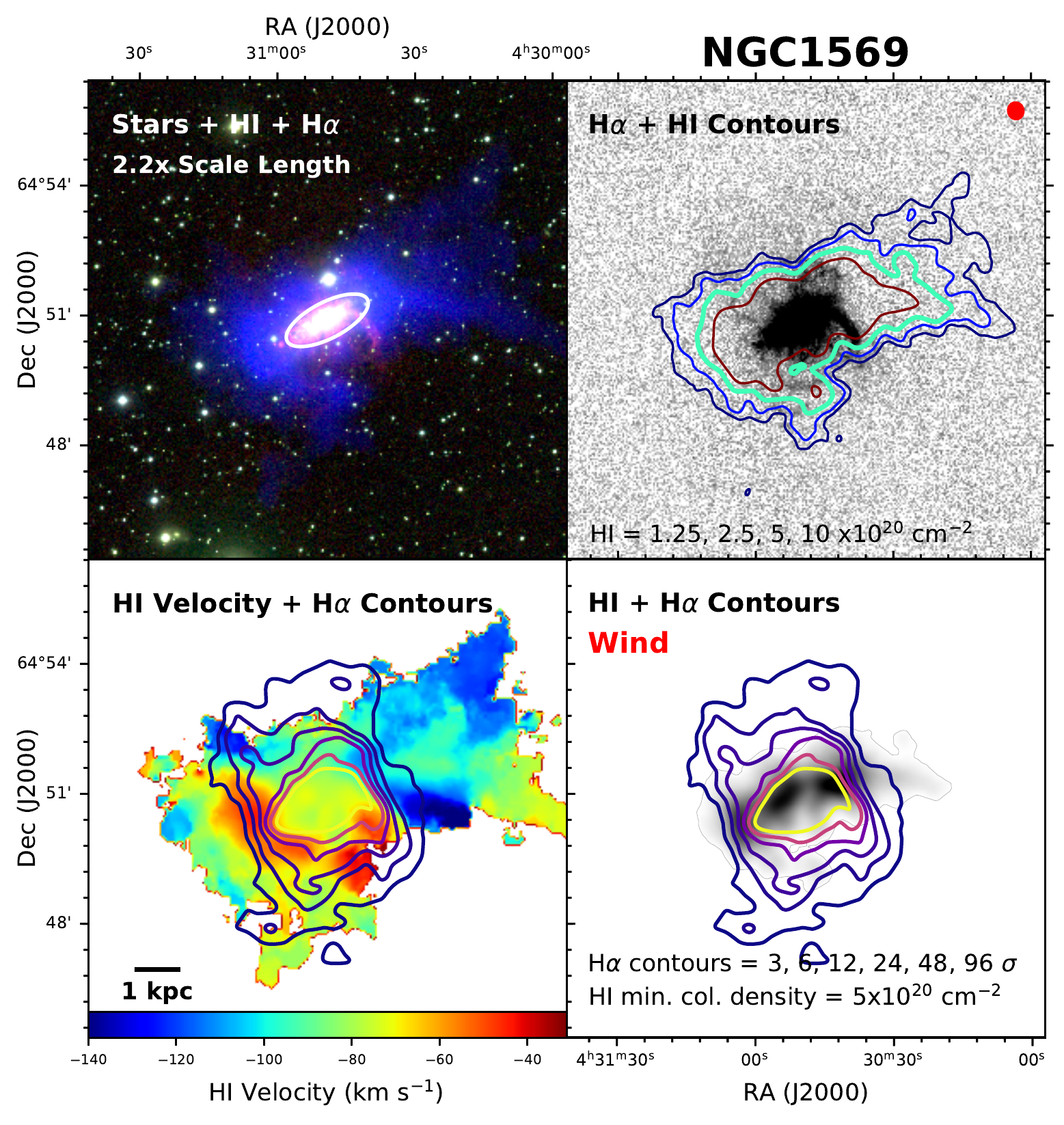}
\caption{NGC~4214. Clockwise from the top left: combined \HI, B-band, \ha, R-band images of the galaxy, \ha\ map with \HI\ contours, \HI\ map with \ha\ contours, \HI\ velocity field with \ha\ contours. See description at the beginning of Appendix~\ref{sec:atlas} for details.}\label{fig:ngc1569}
\end{figure*}

\subsection{NGC~2366} 
From Figure~\ref{fig:ngc2366}, NGC~2366 has two significant superbubbles of ionized gas traced by \ha\ emission in the south and several other slightly smaller bubbles in the north and central region of the galaxy. These regions are spatially coincident with the sites of BHeB stars with ages less than 100 Myr, and in the case of the northern region less than 25 Myr \citep{McQuinn2012a}. Spectroscopically derived radial \ha\ velocities using a Fabry-Per{\'o}t interferometer for the central $4\times4\arcmin$ southern and central \HII\ regions range from 35$-$80 km s$^{-1}$ to $90-125$ km s$^{-1}$ respectively \citep{Garrido2004}, compared to an \ha\ maximum circular velocity of 54 km s$^{-1}$. In contrast to the smoother distribution of diffuse ionized gas seen in, for example, NGC~4068 and NGC~4163, the \ha\ emission outside the bubbles shows significant structure with many arcs and filaments. The extended low-surface brightness ionized gas reaches the edge of the \HI\ disk in one place, marked by the arrows. Given that velocity of the \ha\ emission from this region approaches the maximum circular velocity \citep{Garrido2004}, this material may be outflowing from the galaxy but will likely still be bound. The northern \ha\ bubble lies at the edge of a lower density \HI\ cavity at the edge of the disk. Expanding bubbles of ionized gas from star-forming regions at large galactocentric radii such as this will mix the ISM and newly formed metals in outer regions and add turbulence to the gas.

\begin{figure*}
\includegraphics[width=\textwidth]{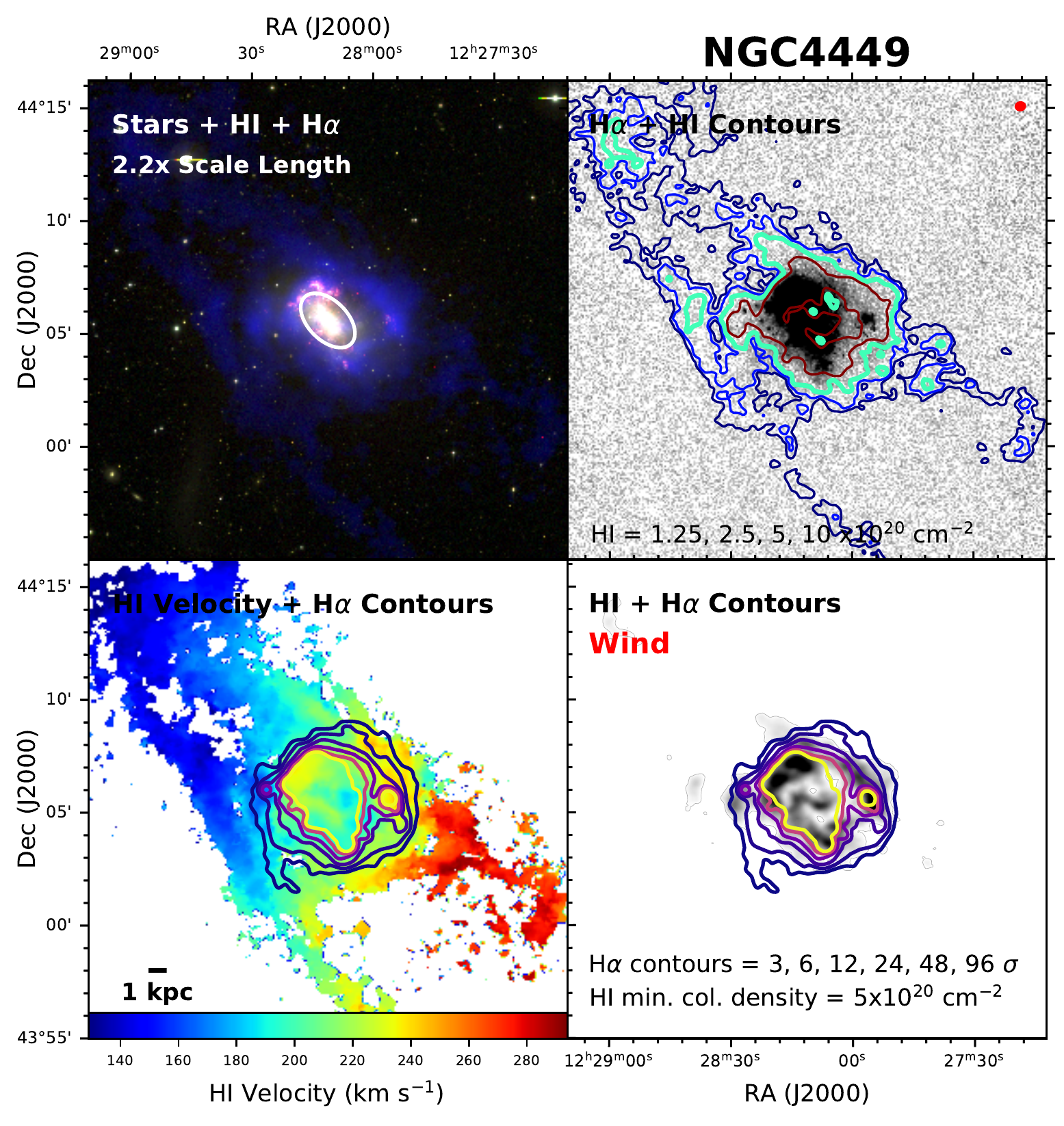}
\caption{NGC~4449. Clockwise from the top left: combined \HI, B-band, \ha, R-band images of the galaxy, \ha\ map with \HI\ contours, \HI\ map with \ha\ contours, \HI\ velocity field with \ha\ contours. See description at the beginning of Appendix~\ref{sec:atlas} for details.}
\label{fig:ngc4449}
\end{figure*}

\subsection{Holmberg~II} 
From Figure~\ref{fig:hoii}, the galaxy is inclined towards us with a unique ISM morphology including spiral structures with significant holes, visible in the extended \HI\ disk (seen in blue). The holes in the \HI\ are much larger than those found in more massive spiral galaxies, which has been attributed to the larger \HI\ scale height of 625 pc, rather than the more usual 120 pc \citep{Puche1992}. The larger scale height and implied lower gas density allows greater expansion of the hot ISM with less mass loading. There are numerous, high surface brightness \ha\ bubbles and structures throughout the galaxy with a large superbubble feature outlined in the southern half. Holmberg~II also has significant diffuse \ha\ emission throughout the disk that fills many of the \HI\ holes. Both the \ha\ higher surface brightness bubbles and the diffuse emission follow the spiral structure of the gas, seen most clearly along the northern and southern \HI\ spiral arms in the bottom right panel. Spectroscopically derived radial \ha\ velocities using a Fabry-Per{\'o}t interferometer for a $4\times4\arcmin$ field centered on the large \ha\ complex range from 120$-$145 km s$^{-1}$ \citep{Garrido2004}, compared to an \ha\ maximum circular velocity of $\sim$36 km s$^{-1}$. The large velocities in the emission co-located in projection over the \HI\ holes, particularly in the south, suggests that gas may be energetic enough to extend out of the gas disk, providing a vent for gas. 

\subsection{NGC~4214} From Figure~\ref{fig:ngc4214}, NGC~4214 has numerous high surface brightness \ha\ emission in the central region and extending along a Northwest to Southeast axis. The high surface brightness \ha\ emission in NGC~4214 presents a unique view of how well the \ha\ distribution traces the \HI\ in the spiral arm features of the galaxy. Many of the \ha\ high surface brightness regions in the top right panel clearly follow the peak \HI\ density structures, which are also the sites of recent star formation. Spectroscopically derived radial \ha\ velocities using a Fabry-Per{\'o}t interferometer range for the central $4\times4\arcmin$ field range from 260$-$310 km s$^{-1}$ and may be receding through the disk \citep{Garrido2004}, compared to a circular velocity of 79 km s$^{-1}$. The high surface brightness central region was also studied using an echelle spectrograph in order to measure the velocity components in the \ha\ emission \citep{Martin1998}. The bubbles in this area were found to be expanding, but the field of view was limited and, comparing their images to Figure~\ref{fig:full_fovb}, did not detect the lower surface brightness emission that nearly surrounds this central region and extends outward. This diffuse emission, also traced by the contours in the bottom right panel of Figure~\ref{fig:ngc4214}, extends on both sides of the central star-forming regions. This material appears to be expanding into the lower density surrounding ISM and out of the galaxy along our line of sight. This ionized gas lacks sharp boundaries and it is very possible that the warm gas we detect is part of a larger scale outflow.

\subsection{NGC~1569} 
From Figure~ \ref{fig:ngc1569}, the interior of the galaxy has numerous \HII\ regions and significant area covered with high surface brightness \ha\ emission. There are arcs and filaments of \ha\ emission along multiple directions from the galaxy and an ionized wind extending in the north and south from the minor axis. The \HI\ morphology is highly disturbed with a box-like morphology that is particularly stretched in the western direction. $HST$ \ha\ imaging of the well-studied dwarf starburst NGC~1569 traces a substantial outflow of ionized gas in a complicated network of filaments \citep{Grocholski2008}. Based on the expansion velocity of the warm gas, \citet{Westmoquette2008} conclude that even with the significant \ha\ emission and outflow, much of this gas may not escape to the IGM. 

\subsection{NGC~4449} 
From Figure~ \ref{fig:ngc4449}, the interior of the galaxy has numerous \HII\ regions of varying sizes seen in high-surface brightness \ha\ emission. Lower surface brightness \ha\ emission is also noted in a network of structure within the galaxy and more generally surrounding the entire galaxy, particularly evident around the northwest-southeast axis. The galaxy has a complicated recent history, with evidence of two minor mergers including one system that is currently being tidally disrupted \citep{Hunter1998, Martinez-Delgado2012}. The \HI\ disk is extremely large, with elongated structures extending outside the field of view in Figure~\ref{fig:ngc4449}. There is also a low-density cavity in the central \HI\ disk, similar to that noted in the starburst galaxy NGC~1569. Radial \ha\ velocities derived for the central $4\times4\arcmin$ field range from $165-210$ km s$^{-1}$ in a very chaotic velocity field and in the opposite sense from the neutral gas at larger radii \citep{Garrido2004}. On the one hand, it has been suggested that the large \HI\ halo may prevent a wind from venting material \citep{Summers2003}. However, the extended \HI\ has low column density and is extremely patchy. The current data suggest that ionized gas is able to effectively leave the main \HI\ disk of the galaxy, reaching regions with little neutral gas to impede its movement. The \ha\ emission has a curved morphology but lacks a defined edge indicating the ionized gas is not in pressure bound bubbles but free-flowing. Close inspection of the diffuse \ha\ emission shows a galactic wind present in nearly all directions seen in projection and dropping in intensity with radii.

\begin{figure}[b]
\includegraphics[width=0.63\textwidth]{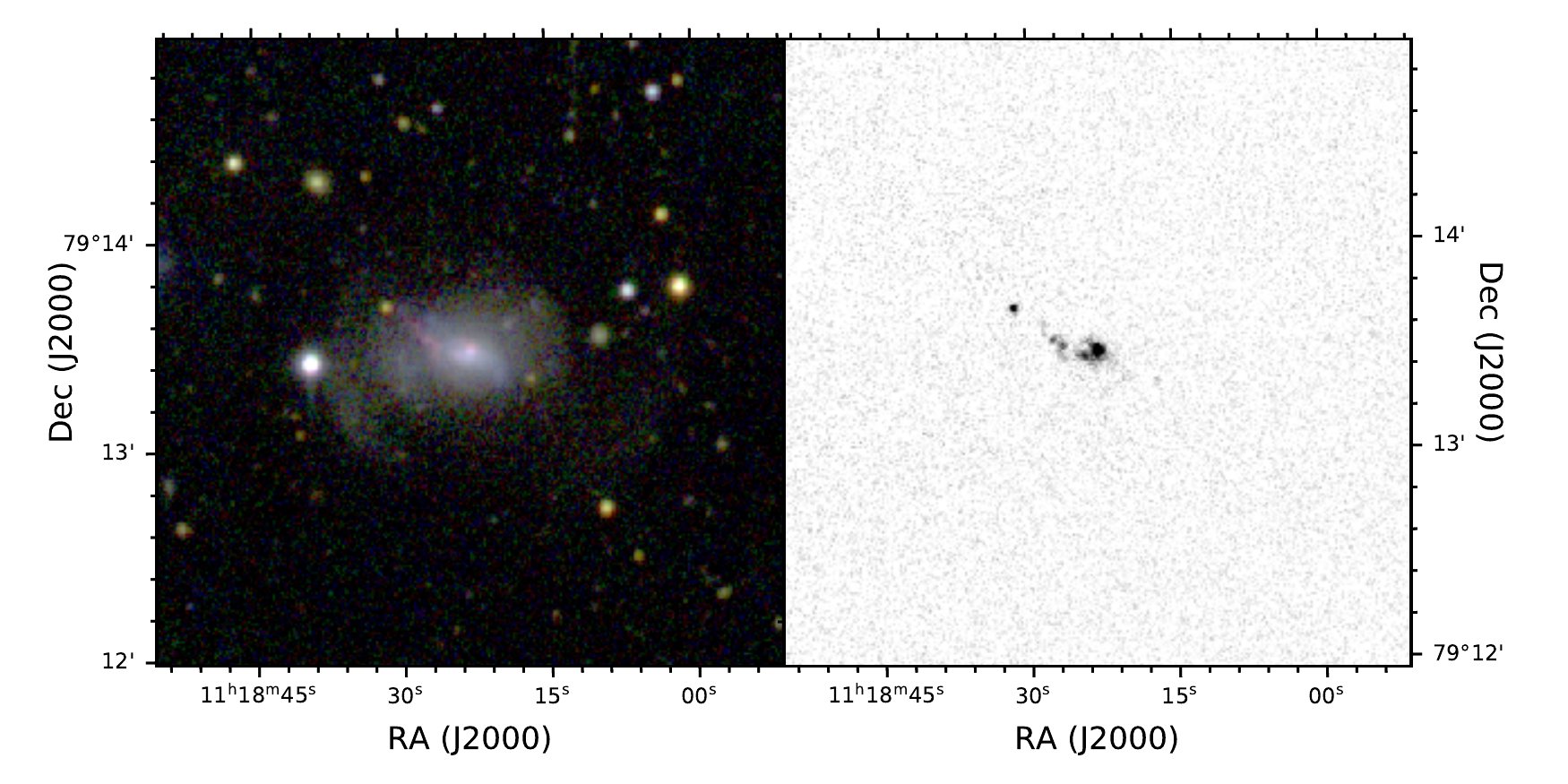}
\includegraphics[width=0.36\textwidth]{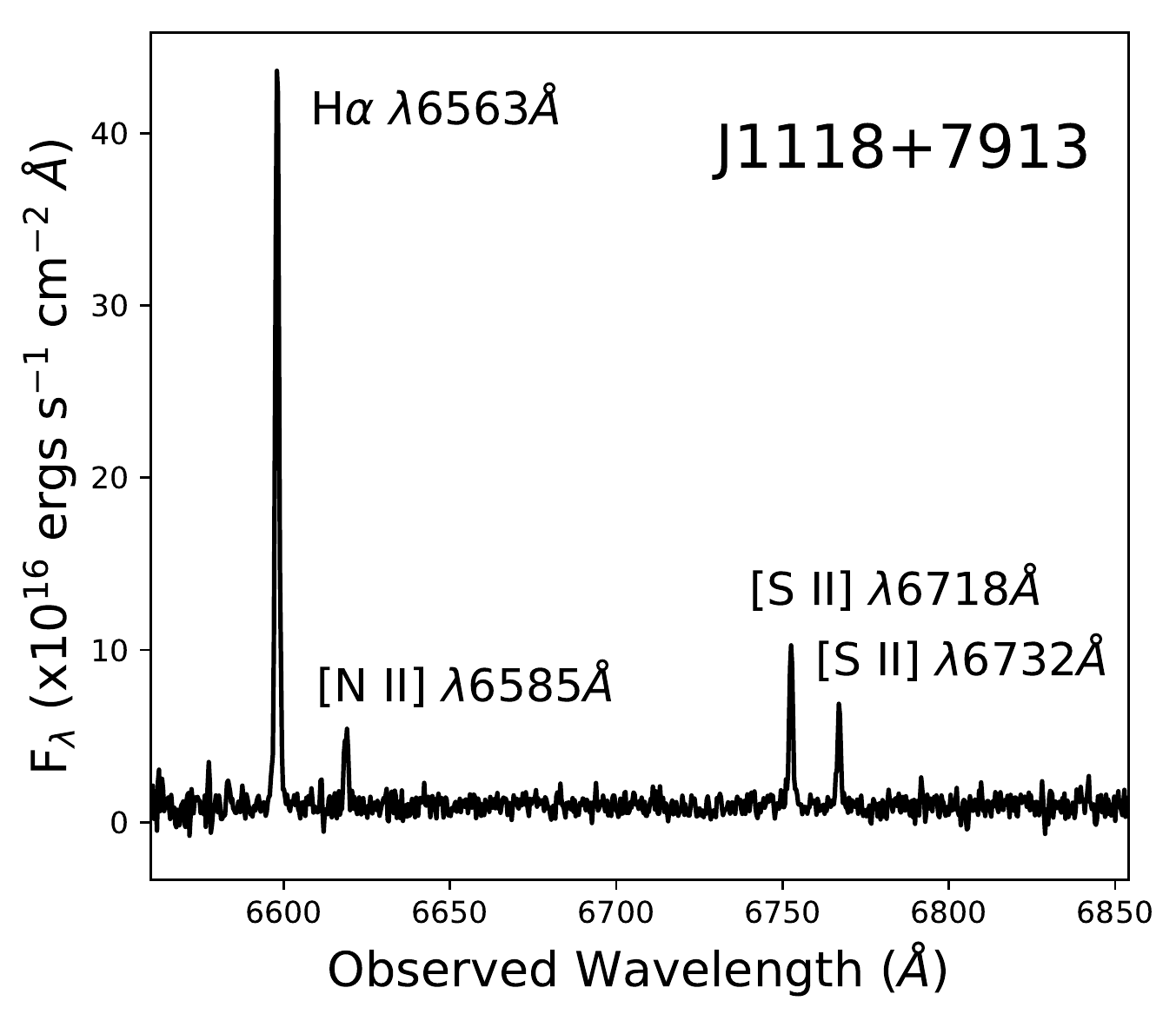}
\caption{Left: Three color image of a 3\arcmin\ field of view centered on the newly identified galaxy, J1118$+$7913, with North up and East left. Features are detected in the outer regions of the galaxy that appear to be stellar shells. Middle: Continuum subtracted \ha\ map of the same field of view. Diffuse ionized gas is detected in the central region of the galaxy. Right: Optical spectrum on the galaxy obtained on the WIYN 3.5m SparsePak instrument.}
\label{fig:newgal1}
\end{figure}

\section{J1118$+$7913: A Previously Uncatalogued Nearby Galaxy}\label{sec:j1118_7913}
We have identified a previously uncatalogued resolved galaxy, J1118$+$7913, located $\sim30$\arcmin\ from UGC~6456 in projection. The galaxy is identified in the GALEX all sky imaging survey, but not catalogued. In Figure~\ref{fig:newgal1} we present a 3-color image (B-band in blue, R-band in green, and \ha\ narrow band in red) and an \ha\ continuum subtracted image of the galaxy. The newly identified galaxy has a disturbed morphology including what appears to be multiple stellar shells outside the main disk visible even in integrated light. 

We obtained an optical spectrum of the galaxy, shown in the last panel of Figure~\ref{fig:newgal1}. J1118+7913 was observed with the SparsePak IFU on the WIYN 3.5m telescope\footnote{The WIYN Observatory is a joint facility of the University of Wisconsin-Madison, Indiana University, the National Optical Astronomy Observatory and the University of Missouri.} on 15 August 2017 for a total integration time of 300 seconds.  The instrumental set-up was similar to that described in \citet{Richards2016}, with a spectral resolution of 13.9 km s$^{-1}$ pixel$^{-1}$ and a wavelength range from
6580 \AA~to 6890 \AA. The SparsePak data were processed using standard tasks in the HYDRA package
within IRAF. The spectra were sky subtracted using the average intensity of all fibers not located
on the galaxy. The night was non-photometric but approximate flux calibration was obtained from 
observations of spectrophotometric standard stars \citep{Oke1990}. Emission lines from [NII], \ha\,
and [SII] were detected in nine of the fibers with relative velocity shifts indicative of ordered rotation.  
The derived systemic redshift of 1535 $\pm$ 2 km s$^{-1}$ is based on the average of heliocentric-corrected velocity 
centroids of the five emission lines in the spectrum of the center fiber, using vacuum 
wavelengths for these transitions as a reference.

\end{document}

%% file: tab1.tex
\begin{table*}
\begin{rotatetable*}
\setlength{\tabcolsep}{3pt}
\begin{center}
\begin{tabular}{lccccccccccrccc}
\hline
\hline 
			&  $\alpha$ 	&  $\delta$ 	& A$_B$	&  M$_{B,0}$	& Distance			& Ref.	& D$_{25}$ & h		 & R$_{\rm disk}$	& $1-b/a$ 		& PA		& {\it i} 		& 	12$+$		  & Target\\
Galaxy  		& (J2000)		& (J2000)		& (mag)   	& (mag)		& (Mpc)	  			& 		& (\arcsec)  &(\arcsec)& (kpc) 	& 	&  ($\degree$)	& ($\degree$)	& log(O/H) & Field \\
\colnumbers
\tableline
\multicolumn{12}{c}{Starburst Sample of Galaxies} \\
\tableline
UGC~9128        & 14:15:56.7s & $+$23:03:21s	&  0.084	& $-11.98\pm0.02$ & 2.21$\pm$0.07		& D09 	&  92.0	& 17.9 	&  0.42	& 0.63	& 39.3	& 68		& $7.75\pm0.05$ & ... \\	
UGC~4483        & 08:37:03.3s & $+$69:46:31s	&  0.123	&$-12.36\pm0.02$ & 3.41$\pm$0.12 	& D09 	&  66.0	& 11.8	&  0.43	& 0.55	& 160.3	& 63 		& $7.56\pm0.03$ & ... \\	
NGC~4163        & 12:12:09.0s & $+$36:10:10s	&  0.073	&$-13.63\pm0.02$ & 2.88$\pm$0.04 	& D09 	&  115.4 	& 16.7	&  0.51	& 0.66	& 5.4		& 70 		& $7.56\pm0.14$ & ... \\	
UGC~6456        & 11:27:57.4s & $+$78:59:39s	&  0.136	&$-13.70\pm0.03$ & 4.63$\pm$0.09		& T13	&  86.0	& 13.8	&  0.65	& 0.52	& 171.8	& 61 		& $7.73\pm0.05$ & ... \\	
NGC~4190	 & 12:13:44.7s & $+$36:38:09s	&  0.106	&$-14.02\pm0.02$  & 2.83$\pm$0.08	& T13	&  105.2	& 14.5	&  0.44	& 0.83	& 25.2	& 80		& \nodata	 	& NGC~4163 \\ 
NGC~4068        & 12:04:02.7s & $+$52:35:28s	&  0.078	&$-14.95\pm0.02$ & 4.38 $\pm$0.04	& T13	&  163.0	&  28.4	&  1.32	& 0.57	& 32.0	& 65 		& \nodata		& ...  \\		
DDO~165         	 & 13:06:25.5s & $+$67:42:27s	&  0.086	&$-15.17\pm0.03$ & 4.83$\pm$0.04		& T13	&  200.6 	&  38.5	&  1.98	& 0.49	& 94.0	& 59	 	& $7.80\pm0.06$ & ... \\	
NGC~2366        & 07:28:54.8s & $+$69:12:54s 	&  0.132	&$-15.77\pm0.04$ & 3.21$\pm0.05$ 	& D09 	&  475.2	& 94.2	&  3.22	& 0.34	& 26.4	& 49		& $7.91\pm0.05$ & ... \\	
Holmberg~II      & 08:19:09.3s & $+$70:43:30s 	&  0.115	&$-16.06\pm0.04$ & 3.38$\pm$0.05 	& D09 	&  375.8	& 69.2	&  2.49	& 0.72	& 13.6	& 74	 	& $7.92\pm0.10$ & ... \\	
NGC~4214        & 12:15:39.1s & $+$36:19:45s 	&  0.079	&$-17.13\pm0.02$ & 3.04$\pm$0.05 	& D09 	&  411.4	& 37.7	&  1.22	& 0.91	& 174.3	& 85		& $8.25\pm0.10$ & ... \\
NGC~1569	 & 04:30:49.1s & $+$64:50:54s	& 2.516	&$-17.86\pm0.02$ & 3.25$\pm$0.12		& T13	&  220.8	& 28.5	&  0.99	& 0.40	& -62.6	& 53		& 8.15 $\pm$ 0.05 & ... \\
NGC~4449        & 12:28:11.3s & $+$44:05:36s	&  0.070	&$-18.25\pm0.02$ & 4.26 $\pm$0.20	& T13	&  424.4	& 39.8	&  1.83	& 0.60	& 44.9	& 66 		& $8.32\pm0.03$ & ... \\
\hline
\multicolumn{12}{c}{Additional Galaxies in Observed Fields of View} \\
\hline

NGC~4102	& 12:06:23.0s & $+$52:42:39s	& 0.073	 & $-19.49\pm0.02$ 	& 19.5 mem 		&T13 	& 178.2	& 17.7	& 3.7		& 0.56	& 39.6	& 64		& \nodata	& NGC~4068 \\
UGC~7577	& 12:27:42.1s & $+$43:29:35s	& 0.074	 & $-13.86\pm0.02$ 	& 2.58$\pm$0.07 	& D09 	& 224.0	& 45.7	& 1.3		& 0.47	& 127.0	& 58		&7.97 $\pm$ 0.06 & NGC~4449 \\
UGC~7608	& 12:28:44.6s & $+$43:13:28s	& 0.062	 & $-15.25\pm0.03$ 	& 9.2 TF 			& T13 	& 102.6	& 31.0	& 3.1		& 0.92	& 60.7	& 85		& \nodata	& NGC~4449 \\
UGCA~276	& 12:14:58.3s & $+$36:13:08s	& 0.072	 & $-10.65\pm0.03$ 	& 2.95$\pm$0.07	& D09 	& 16.6	& 27.8	& 1.4		& 0.63	& 38.8	& 68		& \nodata	&NGC~4214 \\
UGC~7257	& 12:15:03.0s & $+$35:57:31s	& 0.057	 & $-5.96\pm0.02$	& 11.2 TF 			&T13 	& 77.0	& 9.9		& 0.7		& 0.62	& -13.3	& 68		& \nodata	&NGC~4214 \\
J1118$+$7913	& 11:18:24.5s & $+$79:13:28s	& 0.122  & ($16.73\pm0.03$) & \nodata 			& \nodata & 19.8	& 7.4		& \nodata	& 0.53	& 82.6	& 63		& \nodata	&UGC~6456 \\
\hline
\end{tabular}
\end{center}
\caption{{\bf The Galaxy Sample, Basic Properties, and Structural Parameters.} Galaxies are ordered by extinction corrected $M_B$ from faintest to brightest. Distances are based on the tip of the red giant branch method unless noted \citep[mem$=$group membership; TF$=$ Tully-Fisher;][]{Tully2013}. Col. 8 is the major-axis diameter measured at the B-band 25 mag arcsec$^{-2}$ isophote. Cols. 9, 10, 11, 12 are the scale-length (h) in arcsec, disk radius based on 2.2$\times$ scale length in units of kpc using the distances in Col. 5, ellipticity (1-b/a), and position angle measured from isophotal fitting. Col. 13 is the inclination angle determined from the ellipticity (cos {\em i}$=$b/a). Col. 14 is the gas-phase oxygen abundance based on the [O III] $\lambda$4363 auroral line. Col. 15 lists the primary STARBRIDS target whose field of view included the additional galaxy. References: D09$=$\citet{Dalcanton2009}; T13$=$\citet{Tully2013}. As there is no distance for the newly discovered galaxy J1118$=+$7913, we list the apparent magnitude parenthetically.}
\label{tab:galaxies}
\end{rotatetable*}
\end{table*}

%% file: tab2.tex
\begin{table}
\begin{center}
\setlength{\tabcolsep}{3pt}   
\begin{tabular}{lcccc}
\hline
\hline 
			& Peak SFR  		& 				&			&  \\
			& of Burst			& 				& sSFR		& Concentration	 \\
	  		&$\times10^{-3}$	& log				& $\times10^{-10}$& of Starburst	 \\
Galaxy  		&(\msun\ yr$^{-1}$)	& (M$_*$/	\msun)	& (yr$^{-1}$)	& (\%)	 \\
\hline
(1)                     & (2)                   	& (3)                  	& (4)                  & (5)                   \\
\hline
\hline 
UGC~9128        & 5$\pm$1		& 7.11$\pm$0.07	& 4			& 66 \\ 
UGC~4483        & 11$\pm$2		& 7.04$\pm$0.08	& 10			& 29 \\ 
NGC~4163        & 12$\pm$3		& 7.99$\pm$0.12	& 1			& 79 \\ 
UGC~6456        & 23$\pm$3		& 7.68$\pm$0.17	& 5			& 80 \\ 
NGC~4190	 & 14$^{+3}_{-1}$	& 7.49$\pm$0.34	& 5			& 75 \\ 
NGC~4068        & 46$\pm$3		& 8.34$\pm$0.07	& 2			& 03 \\ 
DDO~165         	 & 80$\pm$5		& 8.28$\pm$0.08	& 4			& 35 \\ 
NGC~2366        & 160$\pm$10		& 8.41$\pm$0.05	& 6			& 15 \\ 
Holmberg~II      & 180$\pm20$		& 8.48$\pm$0.06	& 6			& 06 \\ 
NGC~4214        & 130$\pm$40		& 8.99$\pm$0.03	& 1			& 30 \\ 
NGC~1569	& 240$\pm$10		& 8.85$\pm$0.04	& 3			& 85 \\ 
NGC~4449        & 970$\pm$70		& 9.32$\pm$0.07	& 5	 		& 22 \\ 
\hline
\end{tabular}
\end{center}
\caption{{\bf Star Formation Properties.} Peak SFR of the burst and stellar mass measurements are from \citep{McQuinn2010b, McQuinn2015d}. Stellar masses are derived from the SFHs assuming a 30\% recycling fraction \citep{Kennicutt1994}, with the exceptions of Holmberg~II and NGC~4214 whose spatial extent significantly extends outside the HST field of view. For these galaxies, stellar masses are based on 3.6$\micron$ fluxes from Spitzer Space Telescope observations and assuming a mass-to-light ratio of 0.47 \citep{Dale2009, Engelbracht2008, McGaugh2014}. Specific star formation rate is calculated from Cols. 2 and 3. Concentration of the starbursts is based on the relative distribution of young stars (traced by BHeB stars) to old stars (traced by RGB stars; 1 - (BHeB extent / RGB extent)) from \citet{McQuinn2012a, McQuinn2015d}. For Ho~II and NGC~4214, the areal coverage of the $HST$ imaging was insufficient to compare the spatial distribution of the BHeB and RGB stars. Thus, the concentration is calculated using 1 - ( radius of 90\% \ha\ flux / radius of stellar disk).  A high concentration value indicates a centrally concentrated burst while a low number indicates the recent star formation activity is spatially distributed. }
\label{tab:star_formation}
\end{table}

%% file: tab3.tex
\begin{table*}
\setlength{\tabcolsep}{4pt} 
\begin{center}
\begin{tabular}{llccccccccc}
\hline
\hline 
			&			& Obs.		& \ha			& R 		& B		& seeing 	& sky $\sigma$	& log F(\ha)	 	& S(\ha)				& EM	\\
Galaxy  		& Telescope 	& Date		& (min)		& (min)  	& (min)  	& (\arcsec)& (cts)	& erg/s/cm$^2$		& erg/s/cm$^2$/arcsec$^2$ & pc/cm$^6$\\
\hline
(1)			& (2)			& (3)			& (4)			& (5)		& (6)		& (7)		& (8)		& (9) 	& (10) 	& (11) 	\\
\hline
\hline 
UGC~9128	& KPNO 4m 	& 2013-02-13	& 100 (80) 	& 10		&  10		& 1.3		& 14.3	& $-13.79\pm0.11$	 & 3.2$\times10^{-18}$ 	& 1.6 \\
UGC~4483    	& Bok 2.3m	& 2013-03-15	& 120		& 7.5		& 7.5		& 1.6	 	& 7.9		& $-12.42\pm0.05$	 & 2.2$\times10^{-18}$ 	& 1.1 \\
NGC~4163     	& KPNO 4m 	& 2013-02-11	& 100		& 10		& 10	 	& 1.6		& 10.2	& $-12.74\pm0.05$	 & 1.8$\times10^{-18}$ 	& 0.9  \\
UGC~6456     	& Bok 2.3m	& 2013-03-15	& 120		& 7.5	 	& 7.5	 	& 1.8		& 6.6		& $-12.14\pm0.05$	 & 1.4$\times10^{-18}$ 	& 0.7  \\
NGC~4190	& KPNO 4m 	& 2013-02-11	& 100		&  10		& 10	 	& 1.6		& 10.2	& $-12.25\pm0.05$	 & 1.9$\times10^{-18}$ 	& 0.9 \\
NGC~4068     	& KPNO 4m	& 2013-02-11	& 140		& 10 		& 10	 	& 1.4		& 9.2		& $-12.08\pm0.05$	 & 1.6$\times10^{-18}$ 	& 0.8  \\
DDO~165      	& Bok 2.3m	& 2013-03-15	& 120		& 7.5 	& 7.5	 	& 1.6		& 6.4		& $-12.59\pm0.05$	 & 6.7$\times10^{-18}$ 	& 3.3 \\
NGC~2366     	& KPNO 4m	& 2013-02-09	& 120		& 10 		& 10	 	& 1.9		& 10.0	& $-10.95\pm0.05$	 & 3.1$\times10^{-18}$ 	& 1.5  \\
Holmberg~II   	& KPNO 4m	& 2013-02-11	& 120		& 10		& 10	 	& 1.5		& 9.1		& $-11.31\pm0.05$	 & 3.6$\times10^{-18}$ 	& 1.7  \\
NGC~4214     	& KPNO 4m	& 2013-02-13	& 100		& 10  	& 10  	& 1.5		& 8.4		& $-10.74\pm0.05$	 & 2.7$\times10^{-18}$ 	& 1.3  \\
NGC~1569	& WIYN 0.9m	& 2016-10-04	& 40			& 20		& 60 		& 1.2		& 4.25	& $-10.60\pm0.05$ & 1.8$\times10^{-17}$	& 8.6 \\
NGC~4449     	& Bok 2.3m	& 2013-03-17	& 100		& 7.5 	& 7.5	 	& 1.7 	& 4.7		& $-10.50\pm0.05$	 & 2.9$\times10^{-18}$ 	& 1.4  \\
\hline
\end{tabular}
\end{center}
\caption{ {\bf Summary of the Observations and \ha\ Measurements.} Cols. 4-6 list total integration times per filter. For the \ha -on and off filters, integration times were equal, except for UGC~9128 where we excluded one pointing in the \ha-off filter due to poor seeing conditions; total \ha -off integration time is listed parenthetically. Col. 7 and 8 are the seeing and mean sky noise for the continuum subtracted \ha\ image. Col. 9 is the total \ha\ flux for each system, including any extended, diffuse emission. Col. 10 lists the $3\sigma$ surface brightness limits based on the sky sigma measured from images smoothed with a 9\arcsec\ Gaussian kernel. Col. 11 is the Emission Measure calculated from the surface brightness limits and assuming Case B recombination.}
\label{tab:observations}
\end{table*}

%% file: tab4.tex
\begin{table*}
\begin{center}
\begin{tabular}{lllll}
\hline
\hline 
			& 					& 								& 		&\\
Galaxy  		& Beam Size			& Source of \HI\ data 				& PID	&  VLA Configuration\\
\hline
(1)			& (2)					& (3)								& (4)		& (5)	\\
\hline
\hline 
UGC~9128	& 12.4$\times$11.0		& VLA-ANGST; \citet{Ott2012} 				& AO215		& B$+$C$+$D\\
			&					& LITTLE THINGS; \citet{Hunter2012} 	& AH0927 	& \\
UGC~4483    	&  10.8$\times$8.9		& \citet{vanZee1998b} 				& AZ090		& B$+$C\\
NGC~4163     	&  21.3$\times$20.1		& VLA-ANGST; \citet{Ott2012}				& AO215		& B$+$C$+$D \\
			&					& LITTLE THINGS; \citet{Hunter2012}	& AH0927 	& \\
UGC~6456     	&  27.8$\times$20.2		& PI Huang; PI Hunter 				& AH453; AH623& C$+$D\\
NGC~4190	& 10.5$\times8.9$		& VLA-ANGST; \citet{Ott2012}				& AO215		& B$+$C$+$D \\
NGC~4068     	& 18.7$\times$16.7		& \citet{Richards2016}		   		& 16A-013	& C \\
DDO~165      	&  20.9$\times$20.2		& PI Patterson 						& AP198		& C\\
NGC~2366     	& 41.5$\times$41.0		& THINGS Survey; \citet{Walter2008} 	& AW605; AW539& B$+$C$+$D\\
Holmberg~II   	&  20.3$\times$19.7		& \citet{Puche1992}; 					& AP196		& B$+$C$+$D \\
NGC~4214     	& 28.3$\times$27.2		& THINGS Survey; \citet{Walter2008} 	& AM418		& B$+$C$+$D\\
NGC~1569     	& 23.4$\times$21.3		& LITTLE THINGS; \citet{Hunter2012}	& AW325; AW605& B$+$C$+$D\\ 
NGC~4449     	& 24.46$\times$20.8		& \citet{Hunter1999b}				& AH375; AH359 	& B$+$C$+$D \\
			&					&								& AH513; AH540	& \\
\hline
\end{tabular}
\end{center}
\caption{{\bf Summary of the \HI\ Observations from the literature.} VLA-ANGST $=$ VLA ACS Nearby Galaxy Survey Treasury; THINGS $=$ The \HI\ Nearby Galaxy Survey; LITTLE THINGS $=$ Local Irregulars That Trace Luminosity Extremes in THINGS. }
\label{tab:HI_observations}
\end{table*}

%% file: tab5.tex
\begin{table*}
\begin{center}
\end{center}
\setlength{\tabcolsep}{2pt}
\begin{center}
\begin{tabular}{l | cccc | ccc | ll | cC}
\hline
\hline 
Galaxy	& log			& V$_{\rm circ}$	& \HI\ R$_{\rm max}$& log   & log	& v$_{\rm esc}$	& R$_{\rm vir}$& SFR($\tau_{\rm{v}=25}$) & SFR($\tau_{\rm{v}=50}$) & $<\dot{M}>$ & $<\eta>$\\
		& (M$_*$/\msun) & km~s$^{-1}$ 		& (kpc)	& (M$_{\rm d}$/\msun) &(M$_{\rm h}$/\msun) & (km~s$^{-1}$) & (kpc) & (\msun~yr$^{-1}$)	& (\msun~yr$^{-1}$)	& (\msun~yr$^{-1}$)	&  \\
\hline
\hline 
UGC~9128       	& 7.11$\pm$0.7		&12.1	& 1.18	& 7.6		& 10.2	& 39		& 81		& 2.5$\times10^{-4}$ (c) 	& 2.5$\times10^{-4}$ (c) 	& $5.5\times10^{-4}$	 & 2.2 \\
UGC~4483       	& 7.04$\pm$0.8	& 22.8	& 1.82	& 8.3		& 10.1	& 38		& 78		& 4.1$\times10^{-3}$ (c) 	& 4.1$\times10^{-3}$ (c) 	& $2.5\times10^{-3}$	 & 1.0\\
NGC~4163       	& 7.99$\pm$0.12	& 18.1	& 1.54	& 8.1		& 10.6	& 55		& 113	& 4.6$\times10^{-3}$  	& 3.8$\times10^{-3}$ 	& $2.9\times10^{-2}$ & 7.1 \\
UGC~6456       	& 7.68$\pm$0.17	&  20.5	& 2.47	& 8.4		& 10.4	& 49		& 100	& 6.4$\times10^{-3}$  	& 2.1$\times10^{-2}$ 	& $2.9\times10^{-2}$ & 2.4 \\
NGC~4190	& 7.49$\pm$0.34	& 39.3	& 1.78	& 8.8		& 10.3	& 45		& 93		& 1.3$\times10^{-2}$  	& 1.7$\times10^{-2}$ 	& $2.1\times10^{-2}$ & 1.4 \\
NGC~4068       & 8.34$\pm$0.07	& 35.4	& 3.82	& 9.1		& 10.8	& 63		& 129	& 2.2$\times10^{-2}$ (c)  	& 2.4$\times10^{-2}$ 	(c) & $8.4\times10^{-3}$  & 0.6 \\
DDO~165      	& 8.28$\pm$0.08	& \nodata	& \nodata	& \nodata	& 10.7	& 64		& 126	& \nodata				& \nodata				& \nodata & \nodata \\
NGC~2366       	& 8.41$\pm$0.05	& 53.7	& 7.78	& 9.7		& 10.9	& 67		& 138	& 1.7$\times10^{-2}$ (c)  	& 6.3$\times10^{-2}$ (c) 	& 7.0$\times10^{-2}$ & 0.2 \\
Holmberg~II    	& 8.48$\pm$0.06	& 36.1	& 5.57	& 9.2		& 10.8	& 66		& 133	& 7.7$\times10^{-2}$ 	& 12.4$\times10^{-2}$ 	& 8.1$\times10^{-2}$ & 0.8 \\
NGC~4214      	& 8.99$\pm$0.03	& 79$^{**}$& 2.8	& 10.		& 11.1	& 80		& 166	& 8.4$\times10^{-2}$ 	&  6.4$\times10^{-2}$ 	& 8.0$\times10^{-2}$ & 1.2 \\
NGC~1569	& 8.85$\pm$0.04	& 50$^{**}$& 2.7	& 9.3		& 11.0	& 76		& 157	& 2.4$\times10^{-1}$ 	& 2.4$\times10^{-1}$		& 9.0$\times10^{-1}$ & 3.4 \\
NGC~4449       	& 9.32$\pm$0.07	& 75$^{**}$& 12.	& 10.		& 11.3	& 91		& 188	& 4.4$\times10^{-1}$ 	& 4.1$\times10^{-1}$ 	& 6.9$\times10^{-1}$ & 1.6 \\
\hline
\end{tabular}
\end{center}
\caption{Mass loading measurements. Stellar mass measurements are repeated from Table~\ref{tab:star_formation} to enable an easier comparison with other mass values. V$_{\rm circ}$ is the rotational velocity measured from the \HI\ rotation curves at the radius R$_{\rm max}$. Three galaxies marked with $^{**}$ had complicated kinematics; we adopt V$_{\rm circ}$ derived from the detailed dynamical modelling of \citet{Lelli2014b} for NGC~4214 and NGC~1569 measured at the optical radius, and of \citet{Hunter1998} for NGC~4449 measured at the edge of main gaseous disc. M$_{\rm d}$ is the dynamical mass based on the rotation velocities and the radial extent of the neutral gas. M$_{\rm h}$ is the halo mass based the stellar mass and assuming the abundance matching scaling relation from \citet{Moster2010}. V$_{\rm esc}$ is the escape velocity calculated based on the halo mass. R$_{\rm vir}$ is the virial radius based on analysis from \citet{Bordoloi2014} and provides context for the extent of the diffuse \ha\ emission shown in Figure~\ref{fig:SB_histograms}. SFRs are from \citet{McQuinn2010a} at two look-back times ($\tau$) corresponding to the amount of time needed for the outflows to reach the furthermost radius traveling at 25 and 50 km s$^{-1}$ respectively. For the galaxies with fountain flows, the SFRs were measured from the central regions (rather than a global average) from \citet{McQuinn2012a} and are delineated in the table with a (c). The mass loss rates ($\dot{\rm M}$) are the average rates based on outflow velocities of 25 $-$ 50 km s$^{-1}$. Similarly, the mass loading factors ($\eta$) represent an average of the mass loss rates calculated assuming two wind velocities divided by the SFRs at the corresponding look-back times.}
\label{tab:mass_load}
\end{table*}